\documentclass[10pt,journal,final]{IEEEtran}
\IEEEoverridecommandlockouts

\usepackage{amsfonts,amssymb}
\usepackage{ifpdf}
\usepackage{cite}

\usepackage[hyphens]{url}
\usepackage{stfloats}
\usepackage{bm}
\usepackage{color,xcolor}
\usepackage{subfigure}
\usepackage{caption}
\usepackage{multirow}
\usepackage{booktabs}

\ifCLASSINFOpdf
   \usepackage[pdftex]{graphicx}
  
   \graphicspath{{../pdf/}{../jpeg/}}
   \DeclareGraphicsExtensions{.pdf,.jpeg,.png}
\else
   \usepackage[dvips]{graphicx}

   \graphicspath{{../eps/}}

   \DeclareGraphicsExtensions{.eps}
\fi

\usepackage[cmex10]{amsmath}

\usepackage{algorithm}
\usepackage{algorithmic}
\ifCLASSOPTIONcompsoc
\else
  \usepackage[caption=false,font=footnotesize]{subfig}
\fi

\usepackage{makecell}
\begin{document}
\title{3D Multi-Target Localization Via  Intelligent Reflecting Surface: Protocol  and Analysis}
\author{Meng Hua,~\IEEEmembership{Member, IEEE},
Guangji Chen,
Kaitao Meng,~\IEEEmembership{Member,~IEEE},
Shaodan Ma,~\IEEEmembership{Senior Member,~IEEE},
Chau Yuen,~\IEEEmembership{Fellow,~IEEE},
and Hing Cheung So,~\IEEEmembership{Fellow,~IEEE}

\thanks{M. Hua is with the Department of Electrical Engineering, City University of Hong Kong, Hong Kong, 999077, China, and also with the Department of Electrical and Electronic Engineering, Imperial College London,
	London SW7 2AZ, UK (e-mail: m.hua@imperial.ac.uk).}

\thanks{G. Chen is with the School of Electrical and Optical
	Engineering, Nanjing University of Science and Technology, Nanjing 210094,
	China (e-mail: guangjichen@njust.edu.cn).}

\thanks{K. Meng is with the Department of Electronic and Electrical Engineering, University College London, WC1E 6BT, UK (email: kaitao.meng@ucl.ac.uk).}

\thanks{S. Ma is with  the State Key Laboratory of Internet of Things for Smart City, University of Macau, Macau, 999078, China (email:shaodanma@um.edu.mo).}

\thanks{C. Yuen is with the School of
	Electrical and Electronics Engineering, Nanyang Technological University,  639798, Singapore  (e-mail: chau.yuen@ntu.edu.sg).}

\thanks{ H. C. So is  with  the Department of Electrical Engineering, City University of Hong Kong, Hong Kong, 999077, China (e-mail: h.c.so@cityu.edu.hk).}

}
\vspace{-2cm}
\maketitle
\begin{abstract}
With the emerging environment-aware applications, ubiquitous sensing is expected to play a key role in future networks. In this paper, we study a 3-dimensional (3D) multi-target localization system where multiple intelligent reflecting surfaces (IRSs) are applied to create virtual line-of-sight (LoS) links that bypass the base station (BS) and targets. To fully unveil the fundamental limit of IRS for sensing, we first study a single-target-single-IRS case and propose a novel \textit{two-stage localization protocol} by controlling the on/off state of IRS. To be specific, in the IRS-off stage, we derive the  Cram\'{e}r-Rao bound (CRB) of the azimuth/elevation direction-of-arrival (DoA) of the BS-target link and design a  DoA estimator based on the  MUSIC algorithm. In the IRS-on stage,  the  CRB of the azimuth/elevation DoA of the IRS-target link is derived and a simple DoA estimator based on the on-grid IRS beam scanning method is proposed. Particularly, the impact of echo signals reflected by IRS from different paths on sensing performance is analyzed and we show that only the signal passing through the  BS-IRS-target link is required while that of the BS-target link can be neglected provided that the number of BS antennas is sufficiently large and the dedicated sensing beam at the BS is aligned with the departure transmit array response from the BS to the IRS. Moreover, we prove that the single-beam of the IRS is not capable of sensing, but it can be achieved with \textit{multi-beam}. Based on the  two obtained DoAs,  the 3D single-target  location is constructed.  We then extend to the multi-target-multi-IRS case and propose an \textit{IRS-adaptive sensing protocol} by controlling the on/off state of multiple IRSs, and a multi-target localization algorithm is developed.  Simulation results demonstrate the effectiveness of our scheme and show that  sub-meter-level positioning accuracy can be achieved.

\end{abstract}

\section{Introduction}
The future sixth-generation (6G) wireless networks are expected to not only provide high-data rate and low-latency communication services but also require ubiquitous sensing for supporting environment-aware applications such as smart transportation, autonomous driving, pedestrian/animal intrusion detection, and unmanned aerial vehicle trajectory tracking \cite{Liu2022survey,liu2020joint,meng2023UAV}.   This thus gives rise to the emerging research area of integrated sensing and communication, where the sensing function is integrated into the conventional communication base station (BS) to share a common platform but consisting of two functionalities \cite{zhang2021overview}. Recently, the 3rd Generation Partnership Project (3GPP) has released a latest document on sensing and communication including key performance indicators, sensing applications, regulation, and privacy, etc \cite{onlineweb}. Since the physical information/features of the target can be extracted from received echo signals reflected by objects in the surrounding environments, it is anticipated that wireless sensing will play a key role in the next generation of wireless networks. 

However, the current cellular BS for sensing faces three challenges. First, the direct line-of-sight (LoS) link between the BS and the target is frequently blocked due to the complex environments such as buildings and trees. Second, the received power of echo signals at the BS is typically very small due to the high  path loss caused by the round-trip propagation (i.e.,  BS-target-BS link)  and absorption by the target. Third, the cellular BSs are generally deployed sparsely in the cellular networks because of the practical deployment costs and huge energy consumption, which indicates that there are many ``dead zones''  where the physical information/features in these places cannot be analyzed by using wireless sensing techniques.

Recently, the emerging technology, namely intelligent reflecting surface (IRS), which offers a new paradigm to wireless communications \cite{pan2021overview,liu2021reconfigurable,WU2020towards},  is attracting considerable attention. The IRS is a planar surface consisting of a large number of passive reflecting elements, each of which is able to adjust the phase/amplitude of the signal impinging on the IRS, thereby changing the wireless signal propagation. Besides, the IRS is passive, its hardware cost and power consumption are low compared to the  BS with radio-frequency chains. Due to these advantages, there are substantial works paying attention to the research of integrating the IRS into wireless communication systems with various applications such as unmanned aerial vehicle\cite{9400768,9293155,9690481} and mobile edge computing \cite{9887822,9279326,9982493}. In fact, the IRS is also appealing for sensing due to the following three reasons.  First, the IRS is able to create a virtual link between the BS and the target even when its direct path is blocked. Second, the IRS can enhance the received power of echo signals reflected by the IRS due to the passive beamforming gain. Third, the IRS can be widely deployed in the region to achieve ubiquitous signal coverage due to its low hardware cost. Note that there have been  substantial works studying the IRS for sensing, see e.g.,  \cite{9913311,meng2022Intelligent,liu2021dual,9416177,10143420}. However, they focus on \textit{beamforming design} by assuming that the target information is known \textit{a priori}, which is  applicable  to the target tracking case. 
 
So far, there are only a handful of works focusing on IRS for sensing from signal processing perspectives and its research is still in its infancy. 
In \cite{shao2022target,hu2023IRS,song2023intelligent,li2023beam,yang2021wireless,hua2023intelligentsensing}, estimation of target parameters such as time-of-arrival (ToA), direction-of-arrival (DoA), and location by using the IRS, is studied. Particularly,  \cite{hua2023intelligentsensing} unveils the relationships between the  Cram\'{e}r-Rao bound (CRB) for DoA/ToA with the number of IRS reflecting elements $N_r$ and the number of active sensors $N_s$, and shows the CRB of DoA is inversely proportional to ${\cal O}(N_r^2N_s^3)$ and that of ToA is inversely proportional to ${\cal O}(N_r^2N_s)$. However, the above works all assume that the direct link between the BS and the target is absent. When the direct link exists, the BS will receive four different echo signals, namely,   BS$ \to $target$ \to $BS, BS$ \to $IRS$ \to $target$ \to $IRS$ \to $BS, BS$ \to $IRS$ \to $target$ \to $BS, and  BS$ \to $target$ \to $IRS$ \to $BS, which are difficult to distinguish
from the combined signals. Several studies, see \cite{Elzanaty2021Reconfigurable,Keykhosravi2021SISO,Keykhosravi2021semi}, consider both direct link and the reflected link by IRS for sensing/localization, and analysis for ToA-based target sensing/localization is performed.  However, as the IRS is close to the target, the time differences between any different links are significantly small, and  ultra-wideband signals are needed to distinguish them, which, however, is at the cost of hardware implementation.  
Besides,  ToA-based ranging requires perfect clock synchronization between the BS and the target for  active sensing, and its ranging performance is affected by the clock drift and offset \cite{Aditya2018survey}. Therefore,  \cite{Elzanaty2021Reconfigurable,Keykhosravi2021SISO,Keykhosravi2021semi} focusing on the \textit{bistatic radar systems} assisted by the IRS would meet inevitably time-mismatch issue. In fact, the \textit{monostatic radar system} assisted by the IRS may be more appealing due to the following reasons. First, the hardware cost of the monostatic radar system is much cheaper than that of the bistatic radar system. Second, the clock offset does not exist in the monostatic radar system, which simplifies the hardware design. Third, the cellular monostatic BS has strong signal processing abilities/computation resources to estimate the information of the target (e.g.,  its location)  via analyzing the echo signals received at the BS.  To the best of our knowledge, there is no work considering the monostatic radar system assisted by the IRS for target localization.  How to estimate target location assisted by the IRS in the monostatic radar system remains unknown.

\begin{figure}[!t]
	\centerline{\includegraphics[width=3.2in]{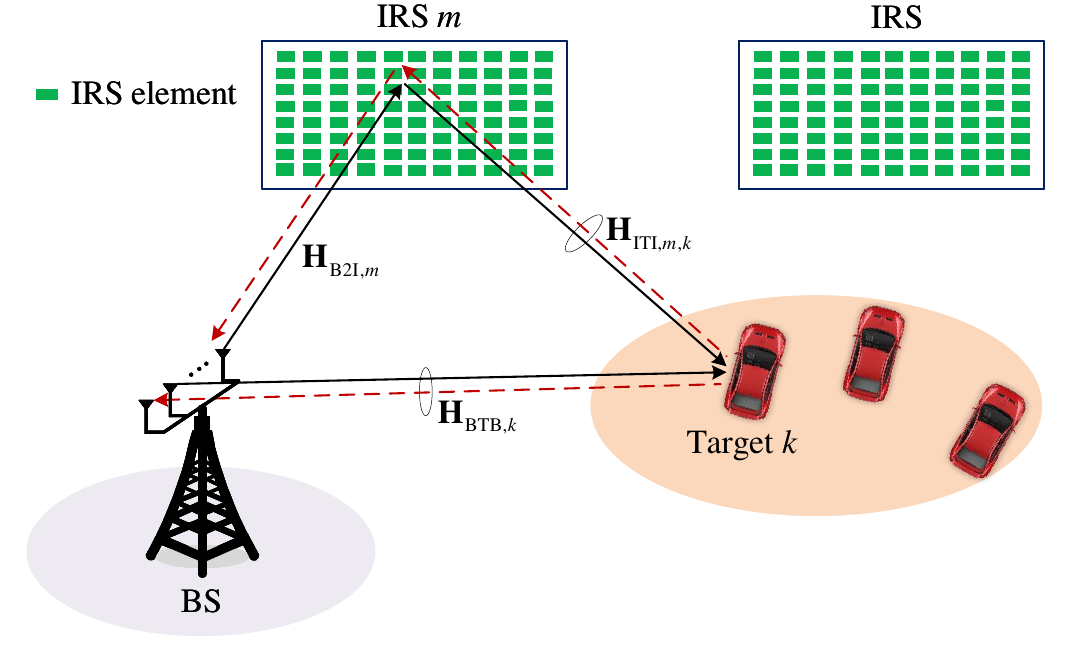}}
	\caption{System model of multi-IRS-assisted 3D multi-target  localization.} \label{fig1}
	\vspace{-0.5cm}
\end{figure}
To fill the gap in this research direction, we consider a \textit{multi-IRS-assisted 3-dimensional (3D)  multi-target}  localization system as shown in Fig.~\ref{fig1}, where the monostatic BS sends the probing signals and then receives the echo signals for estimating 3D target locations. Particularly, we tackle the following practical challenges in this paper. First, we propose an efficient scheme to address the impact of different echo path signals constructed at the BS. Second, we develop an \textit{IRS-adaptive sensing} protocol to efficiently estimate the target location. Third, we prove that the single-beam of the IRS is not able to estimate the target location but at least three different IRS beams is required.  Fourth, to realize multi-target localization, an effective target match algorithm is devised. 
Our main contributions are summarized as follows:
\begin{itemize}
	\item  We first study an IRS-aided single-target case and propose an \textit{localization protocol} which consists of two stages. In the first stage, the IRS is turned off, and then the BS estimates the azimuth/elevation DoA of the BS-target link. Specifically, the MUSIC algorithm is applied for azimuth/elevation DoA estimation and then the CRB of the azimuth/elevation DoA is analytically derived. Particularly, we prove that a \textit{spatially white } probing signal is an optimal waveform to achieve the minimum CRB. In the second stage, the IRS is turned on, and the BS first removes the signal reflected by the direct link based on the former stage. To mitigate the different path echo signals received at the BS, we design the BS beamformer directly towards the IRS and meanwhile consider the BS applied a large number of antennas. Based on this design, 
		the powers of  signals passing through  the  BS$ \to $IRS$ \to $target$ \to $IRS$ \to $BS link is significantly higher than that of  the   BS$ \to $IRS$ \to $target$ \to $BS link and  the BS$ \to $target$ \to $IRS$ \to $BS link, and thus only  former one is considered. Then, we propose an on-grid IRS beam scanning method to estimate the azimuth/elevation DoA of the IRS-target link. Besides, the CRB of its DoA is also analytically derived. We prove that if the single beam of the IRS is adopted, the BS cannot estimate the location of the target, instead, the \textit{multi-beam of the IRS} (at least three different IRS beams) is needed. Based on the DoAs obtained from the two stages, we propose an efficient  3D target location construction algorithm.
	
	\item Next, we extend the single-target localization case to multiple targets where multiple IRSs are employed. To distinguish the composite signals received at the BS from different paths reflected by different IRSs and suppress the echo signals reflected by different IRSs, 	we propose an \textit{IRS-adaptive sensing}  protocol where one IRS is turned on while the other IRSs are turned off at any time frames. Then, at each time frame,  our localization algorithm in the single-target localization case can be directly applied. To associate the multiple locations accurately, an effective target matching algorithm is further proposed. 
	\item Finally, our simulation results verify the effectiveness of the proposed 3D localization scheme and show that  sub-meter  accuracy can be achieved.
\end{itemize} 

The rest of this paper is organized as follows. Section II introduces the channel model and the IRS-aided localization protocol. In Section III, estimation algorithm for DoA of the BS-target link in the single-target case is proposed. In Section IV, estimation algorithm for DoA of the IRS-target link in  the single-target case is proposed. In Section V, 3D localization is developed for the single-target case.  In Section VI, the protocol and algorithm for the multi-target localization assisted by multi-IRS are proposed.
Numerical results are provided in Section VII and the paper is concluded in Section VIII.

\emph{Notations}: Boldface upper-case and lower-case letters denote matrices and vectors, respectively.   ${\left(  \cdot  \right)^{-1}}, {\left(  \cdot  \right)^T},{\left(  \cdot  \right)^*}$, and ${\left(  \cdot  \right)^H}$ stand for the inverse, transpose,  conjugate,  conjugate transpose, and pseudo-inverse operators, respectively. ${\mathbb E}\left\{  \cdot  \right\}$ is the expectation operator. ${\mathop{\rm Re}\nolimits} \left\{ x \right\}$ stands for the real part of $x$, $\left| x \right|$ represents the absolute value of $x$, and  ${\rm{diag}}\left( {\bf{x}} \right)$ stands for a diagonal matrix with the diagonal entries specified by  $\bf x$. ${\bf x} \sim {\cal CN}\left( {{{\bm \mu }},{{\sigma^2{\bf{I}} }}} \right)$ denotes a circularly symmetric complex Gaussian (CSCG) vector $\bf x$ with mean of  $ \bm {\mu}$ and  covariance matrix of ${\sigma ^2}{\bf{I}}$. $ \otimes $ denotes the Kronecker product operator and  $K! = K \times \left( {K - 1} \right) \times  \cdots  \times 1$.


\section{System Model}
As shown in Fig.~\ref{fig1},  the sensing system consists of one BS, $M$ IRSs, and $K$ targets. We consider a   3D Cartesian coordinate system where the locations of the BS,  IRS $m$, and  target $k$ are denoted by ${{\bf{q}}_{{\rm{BS}}}} = {\left[ {{x_{{\rm{BS}}}},{y_{{\rm{BS}}}},{z_{{\rm{BS}}}}} \right]^T}$, ${{\bf{q}}_{{\rm{I,}}m}} = {\left[ {{x_{{\rm{I,}}m}},{y_{{\rm{I,}}m}},{z_{{\rm{I,}}m}}} \right]^T}$, and  ${{\bf{q}}_{{\rm{T,}}k}} = {\left[ {{x_{{\rm{T,}}k}},{y_{{\rm{T,}}k}},{z_{{\rm{T,}}k}}} \right]^T}$, respectively, $k = 1, \ldots ,K, m = 1, \ldots ,M$. Each IRS is deployed to create a virtual LoS link between the BS and the target so that the target-related parameters can be sensed by the BS via the IRS.  
We assume that the IRS location is known\textit{ a priori} by the BS since the IRS is always attached to facades of buildings, while the target location is unknown and needs to be estimated. To estimate the 3D location of the target, we consider that the BS is equipped with a uniform planar array (UPA) with  $N_{{\rm{BS}}}= N_{{\rm{BS}}}^y \times N_{{\rm{BS}}}^z$ elements,  while  IRS $m$ is also a UPA with ${N_{{\rm{r,}}m}} = N_{{\rm{r,}}m}^y \times N_{{\rm{r,}}m}^z$ elements. In this paper, we first focus on the analysis of a 3D  single-target  location estimation via one IRS and then extend it to the general case for multiple targets via multiple IRSs discussed in Section VI. 
\subsection{Channel Model}
As shown in Fig.~\ref{fig1}, the BS receives composite signals reflected by the target from four links: 1) BS$ \to $target$ \to $BS; 2) BS$ \to $IRS$ \to $target$ \to $IRS$ \to $BS; 3) BS$ \to $IRS$ \to $target$ \to $BS; 4) BS$ \to $target$ \to $IRS$ \to $BS. These channels are modelled as follows.

To facilitate the channel description,  let ${\bf{u}}\left( {\phi ,N} \right)$ be the steering vector of the form: 
\begin{align}
{\bf{u}}\left( {\phi ,N} \right) = {\left[ {{e^{ - \frac{{j\left( {N - 1} \right)\pi \phi }}{2}}}, \ldots ,{e^{\frac{{j\left( {N - 1} \right)\pi \phi }}{2}}}} \right]^T}, \label{response_vector}
\end{align}
where $N$ denotes the number of elements of a uniform linear array and $\phi $ represents the steering vector direction. Note that the array origin is set at the array centroid in this paper.

In general, the IRS  is deployed with a high altitude and the LoS links can be established to both the BS and the target. In addition, we  assume that the direct link between the BS and the target exists and is dominated by LoS. 
Following \cite{shao2022target,pang2023cellular,Elzanaty2021Reconfigurable,xing2021location},
  the BS$ \to $target $k$ $ \to $BS channel is modeled as 
\begin{align}
\!\!\!\!{{\bf{H}}_{{{\rm BTB},k}}} = 
{\beta _{{{\rm BTB},k}}}{\bf{a}}\left( {\mu _{{\rm{B2T}},k}^{\rm{D}},\nu _{{\rm{B2T}},k}^{\rm{D}}} \right){{\bf{a}}^T}\left( {\mu _{{\rm{B2T}},k}^{\rm{D}},\nu _{{\rm{B2T}},k}^{\rm{D}}} \right), \label{BS_IRS_channel}
\end{align}
where ${\beta _{{\rm{BTB}},k}} = \sqrt {\frac{{{\lambda ^2}{\kappa _{{{\rm RCS},m}}}}}{{64{\pi ^3}d_{{\rm{B2T}},k}^4}}} {e^{ - j\frac{{4\pi }}{\lambda }{d_{{{ \rm B2T},k}}}}}$, $\lambda $ stands for the carrier wavelength, ${d_{{\rm{B2T}},k}}$ is the distance between the BS and  target $k$, ${{\kappa _{{\rm{RCS}},k}}}$ denotes the radar-cross section of target $k$. In addition, ${\bf{a}}\left( {\mu _{{\rm{B2T}},k}^{\rm{D}},\nu _{{\rm{B2T}},k}^{\rm{D}}} \right)$ denotes the transmit array response vector from the BS to  target $k$, which can be expressed as 
\begin{align}
\!\!\!\!{\bf{a}}\left( {\mu _{{\rm{B2T}},k}^{\rm{D}},\nu _{{\rm{B2T}},k}^{\rm{D}}} \right)={\bf{u}}\left( {\mu _{{\rm{B2T}},k}^{\rm{D}},N_{{\rm{BS}}}^y} \right) \otimes {\bf{u}}\left( {\nu _{{\rm{B2T}},k}^{\rm{D}},N_{{\rm{BS}}}^z} \right). \label{transmit_array-vector}
\end{align}
Note that ${\mu _{{\rm{B2T}},k}^{\rm{D}}}$ and ${\nu _{{\rm{B2T}},k}^{\rm{D}}}$ denote the spatial azimuth angle-of-departure (AoD)  and  spatial elevation AoD from the BS to target $k$, respectively. The relationship between the azimuth/elevation spatial AoD and physical  azimuth/elevation AoD is given by 
\begin{align}
&\mu _{{\rm{B2T}},k}^{\rm{D}} = \frac{{2{d_{{{\rm BS}}}}}}{\lambda }\cos \left( {\phi _{{\rm{B2T}},k}^{\rm{D}}} \right)\sin \left( {\varphi _{{\rm{B2T}},k}^{\rm{D}}} \right),\\
&\nu _{{\rm{B2T}},k}^{\rm{D}} = \frac{{2{d_{{\rm{BS}}}}}}{\lambda }\sin \left( {\phi _{{\rm{B2T}},k}^{\rm{D}}} \right),
\end{align}
where $d_{\rm BS}$ stands for the  distance between two adjacent BS antennas, ${\varphi _{{\rm{B2T}},k}^{\rm{D}}}$ and ${\phi _{{\rm{B2T}},k}^{\rm{D}}}$ denote the corresponding physical azimuth and elevation AoDs, respectively.

The BS$\to$IRS $m$ channel is modelled as  
\begin{align}
\!\!\!\!{{\bf{H}}_{{\rm{B2I}},m}} =
 {\beta _{{\rm{B2I}},m}}{{\bf{b}}_{\rm{r}}}\left( {\mu _{{\rm{B2I}},m}^{\rm{A}},\nu _{{\rm{B2I}},m}^{\rm{A}}} \right){{\bf{a}}^T}\left( {\mu _{{\rm{B2I}},m}^{\rm{D}},\nu _{{\rm{B2I}},m}^{\rm{D}}} \right),
\end{align}
where ${\beta _{{\rm{B2I}},m}}{\rm{ = }}\sqrt {\frac{{{\lambda ^2}}}{{16{\pi ^2}d_{{\rm{B2I}},m}^2}}} {e^{ - j\frac{{2\pi }}{\lambda }{d_{{{ \rm B2I},m}}}}}$, ${{d_{{\rm{B2I}},m}}}$ is the distance between the BS and IRS $m$.  In addition, ${\bf{a}}\left( {\mu _{{\rm{B2I}},m}^{\rm{D}},\nu _{{\rm{B2I}},m}^{\rm{D}}} \right)$ is the transmit array response vector from the BS to  IRS $m$ with spatial azimuth  and  elevation AoDs denoted by $\mu _{{\rm{B2I}},m}^{\rm{D}} = \frac{{2{d_{{\rm{BS}}}}}}{\lambda }\cos \left( {\phi _{{\rm{B2I}},m}^{\rm{D}}} \right)\sin \left( {\varphi _{{\rm{B2I}},m}^{\rm{D}}} \right)$ and $\nu _{{\rm{B2I}},m}^{\rm{D}} = \frac{{2{d_{{\rm{BS}}}}}}{\lambda }\sin \left( {\phi _{{\rm{B2I}},m}^{\rm{D}}} \right)$, which is similarly defined in \eqref{transmit_array-vector}, and ${{\bf{b}}_{\rm{r}}}\left( {\mu _{{\rm{B2I}},m}^{\rm{A}},\nu _{{\rm{B2I}},m}^{\rm{A}}} \right)={\bf{u}}\left( {\mu _{{\rm{B2I}},m}^{\rm{A}},N_{\rm{r}}^y} \right) \otimes {\bf{u}}\left( {\nu _{{\rm{B2I}},m}^{\rm{A}},N_{\rm{r}}^z} \right)$ stands for the receive array response vector from the BS to  IRS $m$ with spatial azimuth  and  elevation angle-of-arrivals (AoAs)  denoted by   $\mu _{{\rm{B2I}},m}^{\rm{A}}{\rm{ = }}\frac{{2{d_{{\rm{IRS}}}}}}{\lambda }\cos \left( {\phi _{{\rm{B2I}},m}^{\rm{A}}} \right)\sin \left( {\varphi _{{\rm{B2I}},m}^{\rm{A}}} \right)$ and $\nu _{{\rm{B2I}},m}^{\rm{A}} = \frac{{2{d_{{\rm{IRS}}}}}}{\lambda }\sin \left( {\phi _{{\rm{B2I}},m}^{\rm{A}}} \right)$, where ${{d_{{\rm{IRS}}}}}$ is the inter-element spacing.

The IRS $m$$\to$target $k$$\to$IRS $m$ channel can be  modeled as
\begin{align}
&{{\bf{H}}_{{\rm{ITI}},m,k}} = \notag\\
&{\beta _{{\rm{ITI}},m,k}}{{\bf{b}}_{\rm{r}}}\left( {\mu _{{\rm{I2T}},m,k}^{\rm{D}},\nu _{{\rm{I2T}},m,k}^{\rm{D}}} \right){\bf{b}}_{\rm{r}}^T\left( {\mu _{{\rm{I2T}},m,k}^{\rm{D}},\nu _{{\rm{I2T}},m,k}^{\rm{D}}} \right),
\end{align}
where ${\beta _{{\rm{ITI}},m,k}} = \sqrt {\frac{{{\lambda ^2}{\kappa _{{\rm{RCS}},k}}}}{{64{\pi ^3}d_{{\rm{I2T}},m,k}^4}}} {e^{ - j\frac{{4\pi }}{\lambda }{d_{{\rm{I2T}},m,k}}}}$, ${d_{{\rm{I2T}},m,k}}$ is the distance between  IRS $m$ and  target $k$, ${{\bf{b}}_{\rm{r}}}\left( {\mu _{{\rm{I2T}},m,k}^{\rm{D}},\nu _{{\rm{I2T}},m,k}^{\rm{D}}} \right)={\bf{u}}\left( {\mu _{{\rm{I2T}},m,k}^{\rm{D}},N_{\rm{r}}^y} \right) \otimes {\bf{u}}\left( {\nu _{{\rm{I2T}},m,k}^{\rm{D}},N_{\rm{r}}^z} \right)$ denotes the transmit array response vector from IRS $m$ to  target $k$ with spatial azimuth  and  elevation AoDs  denoted by   $\mu _{{\rm{I2T}},m,k}^{\rm{D}}=\frac{{2{d_{{\rm{IRS}}}}}}{\lambda }\cos \left( {\phi _{{\rm{I2T}},m,k}^{\rm{D}}} \right)\sin \left( {\varphi _{{\rm{I2T}},m,k}^{\rm{D}}} \right)$ and $\nu _{{\rm{I2T}},m,k}^{\rm{D}} = \frac{{2{d_{{\rm{IRS}}}}}}{\lambda }\sin \left( {\phi _{{\rm{I2T}},m,k}^{\rm{D}}} \right)$.

Similarly, the BS$\to$target $k$$\to $IRS $m$ channel can be modeled as 
\begin{align}
&{{\bf{H}}_{{\rm{BTI}},m,k}} = \notag\\
&\quad{\beta _{{\rm{BTI}},m,k}}{{\bf{b}}_{\rm{r}}}\left( {\mu _{{\rm{I2T}},m,k}^{\rm{D}},\nu _{{\rm{I2T}},m,k}^{\rm{D}}} \right){{\bf{a}}^T}\left( {\mu _{{\rm{B2T}},k}^{\rm{D}},\nu _{{\rm{B2T}},k}^{\rm{D}}} \right),
\end{align}
 where ${\beta _{{\rm{BTI}},m,k}} = \sqrt {\frac{{{\lambda ^2}{\kappa _{{{\rm RCS},k}}}}}{{64{\pi ^3}d_{{\rm{B2T}},k}^2d_{{\rm{I2T}},m,k}^2}}} {e^{ - j\frac{{2\pi }}{\lambda }\left( {{d_{{{\rm B2T},k}}}+{d_{{\rm{I2T}},m,k}}} \right)}}$.

\begin{figure}[!t]
	\centerline{\includegraphics[width=3in]{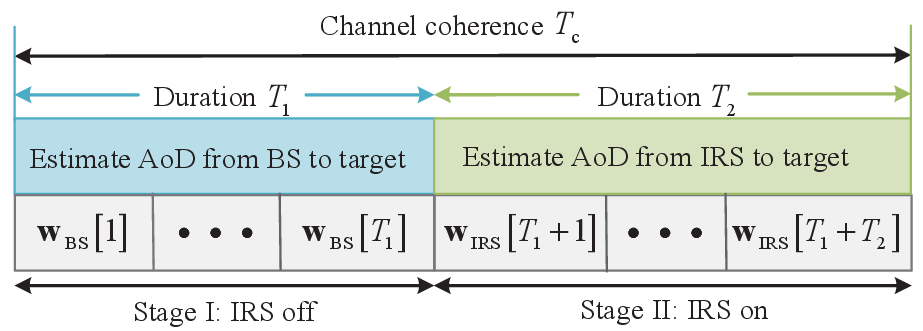}}
	\caption{Localization protocol for a single target assisted by a single IRS.} \label{fig2}
	\vspace{-0.5cm}
\end{figure}
\subsection{IRS-aided Localization Protocol}
It is difficult to estimate the 3D target  location via directly analyzing received signals reflected by the target due to the following two reasons. First, the BS receives two types of echo signals: one is directly from the target and the other is reflected by the IRS,  and these two types of signals are coherently combined at the BS. Second, the effective angle of the BS-IRS-target channel cannot be estimated by using traditional estimation approaches such as  MUSIC \cite{cheney2001linear}. To address these, we decouple the target location estimation into two stages, namely stage I and stage II, as shown in  Fig.~\ref{fig2}. Without loss of generality, let $T_{\rm c}$, $T_1$, and $T_2$  be the duration of channel coherence, stage I, and stage II, respectively, satisfying    $T_{\rm c}=T_1+T_2$. In stage I, we turn off the IRS, the system model in Fig.~\ref{fig1} is simplified to the traditional radar system, then the BS sends different beams with different directions over time to scan the target and the target information such as the  AoD and channel coefficient can be estimated. In stage II, we turn on the IRS and then subtract the echo signals that are directly reflected by the IRS from composite signals and then adjust different IRS beams over time to illuminate the target.
Note that in stage II, the direct channel information between the BS and the target is assumed to be known \textit{a  priori} via either estimation in stage I or offline estimation.
\section{Single Target: BS-target DoA Estimation}
In this section, we show how to estimate the azimuth/elevation of the BS-target DoA.
As shown in Fig.~\ref{fig2}, the IRS is turned off in stage I and  $T_1$ samples of transmit probing signals are used for  target estimation. In this case, we drop index $k$ and $m$ when describing the channel for notational simplicity. Denote  the set of samples in stage I as ${{\cal T}_1} = \left\{ {  1, \ldots ,{T_1}} \right\}$.
Let ${{\bf{w}}_{{\rm{BS}}}}\left[ i \right] \in {{\mathbb C}^{{N_{{\rm{BS}}}} \times 1}}, i \in {{\cal T}_1}$ and  $ {P_{{\rm{BS}}}}$ denote the BS transmit probing signals at the $i$-th sample and available power budget at the BS, respectively. Define ${\bf{W}}_{{\rm{BS}}}^{\rm{I}} = \left[ {{{\bf{w}}_{{\rm{BS}}}}\left[ 1 \right], \ldots,{{\bf{w}}_{{\rm{BS}}}}\left[ {{T_1}} \right]} \right]$ as the collection of $T_1$ transmit probing signals over  $T_1$ samples, the  sample coherence matrix of the  transmit probing signals is given by 
\begin{align}
	{{\bf{R}}_{{{\bf{w}}_{{\rm{BS}}}}}} = \frac{1}{{{T_1}}}{\bf{W}}_{{\rm{BS}}}^{\rm{I}}{\bf{W}}_{{\rm{BS}}}^{{\rm{I,}}H} = \frac{1}{{{T_1}}}\sum\limits_{t{\rm{ = }}1}^{{T_1}} {{{\bf{w}}_{{\rm{BS}}}}\left[ t \right]{\bf{w}}_{{\rm{BS}}}^H\left[ t \right]},
\end{align}
which satisfies  ${\rm{tr}}\left( {{{\bf{R}}_{{{\bf{w}}_{{\rm{BS}}}}}}} \right) = {P_{{\rm{BS}}}}$.

The echo signal received  at the BS through the BS-target link at the $i$-th sample is 
\begin{align}
{\bf{y}}_{{\rm{BS}}}^{\rm{I}}\left[ i \right] = {{\bf{H}}_{{\rm{BTB}}}}{{\bf{w}}_{{\rm{BS}}}}\left[ i \right] + {\bf{n}}_{{\rm{BS}}}^{\rm{I}}\left[ i \right],i \in {\cal T}_1
\end{align}
where ${\bf{n}}_{{\rm{BS}}}^{\rm{I}}\left[ i \right] \sim {\cal CN}\left( {{\bf 0},{\sigma ^2}{{\bf{I}}_{{N_{{\rm{BS}}}}}}} \right)$ is the received noise with ${{\sigma ^2}}$ denoting the noise power at each BS antenna.

\subsection{BS-target DoA Estimation}

Upon collecting $T_1$ received samples at the BS, we can write them into a matrix form given by 
\begin{align}
{\bf{Y}}_{{\rm{BS}}}^{\rm{I}}=\left[ {{\bf{y}}_{{\rm{BS}}}^{\rm{I}}\left[ 1 \right], \ldots ,{\bf{y}}_{{\rm{BS}}}^{\rm{I}}\left[ {{T_1}} \right]} \right]. \label{StageI: total_signals}
\end{align}
To obtain a high-resolution direction estimate, the MUSIC algorithm is applied \cite{cheney2001linear}. The main key idea behind it is that the  signal subspace over the auto-correlation matrix of the received signals is orthogonal to the noise subspace. To be specific, based on \eqref{StageI: total_signals}, its auto-correlation matrix is given by 
\begin{align}
{{\bf{R}}_{{\bf{Y}}_{{\rm{BS}}}^{\rm{I}}}}{\rm{ = }}\frac{1}{{{T_1}}}{\bf{Y}}_{{\rm{BS}}}^{\rm{I}}{\bf{Y}}_{{\rm{BS}}}^{{\rm{I,}}H}.
\end{align}
Since there is only one target, the dimension of the signal subspace is $1$ while the dimension of the noise subspace is $N_{\rm BS}-1$. Thus, we can perform eigenvalue decomposition of  ${{\bf{R}}_{{\bf{Y}}_{{\rm{BS}}}^{\rm{I}}}}$ as 
\begin{align}
{{\bf{R}}_{{\bf{Y}}_{{\rm{BS}}}^{\rm{I}}}}{\rm{ = }}\left[ {{{\bf{U}}_{\rm{s}}}{\kern 1pt} {\kern 1pt} {\kern 1pt} {{\bf{U}}_{\rm{n}}}} \right]\left[ {\begin{array}{*{20}{c}}
		{{\Sigma _{\rm{s}}}{\kern 1pt} {\kern 1pt} {\kern 1pt} }&{}\\
		{}&{{{\bf{\Sigma }}_{\rm{n}}}{\kern 1pt} {\kern 1pt} }
\end{array}} \right]{\left[ {{{\bf{U}}_{\rm{s}}}{\kern 1pt} {\kern 1pt} {\kern 1pt} {{\bf{U}}_{\rm{n}}}} \right]^H},
\end{align}
where ${{\bf{U}}_{\rm{s}}}{\kern 1pt}  \in {{\mathbb C}^{{N_{{\rm{BS}}}} \times 1}}$ and ${{\bf{U}}_{\rm{n}}} \in {{\mathbb C}^{{N_{{\rm{BS}}}} \times \left( {{N_{{\rm{BS}}}} - 1} \right)}}$.
Following \cite{cheney2001linear}, the DoA of the BS-target channel  can be obtained via the following  MUSIC spectrum function:
\begin{align}
&\left( {\hat \mu _{{\rm{B2T}}}^{\rm{D}},\hat \nu _{{\rm{B2T}}}^{\rm{D}}} \right)\;=\notag\\
&\quad~~ \mathop {\arg \max }\limits_{\mu _{{\rm{B2T}}}^{\rm{D}},\nu _{{\rm{B2T}}}^{\rm{D}}} \frac{1}{{{{\bf{a}}^H}\left( {\mu _{{\rm{B2T}}}^{\rm{D}},\nu _{{\rm{B2T}}}^{\rm{D}}} \right){{\bf{U}}_{\rm{n}}}{\bf{U}}_{\rm{n}}^H{\bf{a}}\left( {\mu _{{\rm{B2T}}}^{\rm{D}},\nu _{{\rm{B2T}}}^{\rm{D}}} \right)}}, \label{MUSIC_stageI}
\end{align}
which can be solved by a two-dimensional search.
\subsection{CRB for BS-target DoA Estimation}
Since the DoA estimation mean square error  (MSE) of the MUSIC method is difficult to quantify, we resort to its analysis by using the well-known CRB method, which serves as a lower bound of any unbiased estimators. Let ${\bm{\eta }}_{{\rm{BS}}}^{\rm{I}}{\rm{ = }}{\left[ {\mu _{{\rm{B2T}}}^{\rm{D}},\nu _{{\rm{B2T}}}^{\rm{D}},{\bm{\beta }}_{{\rm{BTB}}}^T} \right]^T}$  denote the collection of unknown parameters to be estimated in stage I, including the target's azimuth/elevation DoA  and channel coefficient, where    ${{\bm{\beta }}_{{\rm{BTB}}}} = {\left[ {{\mathop{\rm Re}\nolimits} \left\{ {{\beta _{{\rm{BTB}}}}} \right\}{\kern 1pt} {\kern 1pt} {\kern 1pt} {\kern 1pt} {\kern 1pt} {\mathop{\rm Im}\nolimits} \left\{ {{\beta _{{\rm{BTB}}}}} \right\}{\kern 1pt} } \right]^T}$. We rewrite  \eqref{StageI: total_signals} into a more tractable form given by 
\begin{align}
{\bf{Y}}_{{\rm{BS}}}^{\rm{I}}={\beta _{{\rm{BTB}}}}{\bf{A}}\left( {\mu _{{\rm{B2T}}}^{\rm{D}},\nu _{{\rm{B2T}}}^{\rm{D}}} \right){\bf{W}}_{{\rm{BS}}}^{\rm{I}} + {\bf{N}}_{{\rm{BS}}}^{\rm{I}}, \label{tageI: total_signals_v1}
\end{align}
where ${\bf{A}}\left( {\mu _{{\rm{B2T}}}^{\rm{D}},\nu _{{\rm{B2T}}}^{\rm{D}}} \right) = {\bf{a}}\left( {\mu _{{\rm{B2T}}}^{\rm{D}},\nu _{{\rm{B2T}}}^{\rm{D}}} \right){{\bf{a}}^T}\left( {\mu _{{\rm{B2T}}}^{\rm{D}},\nu _{{\rm{B2T}}}^{\rm{D}}} \right)$ and ${\bf{N}}_{{\rm{BS}}}^{\rm{I}} = \left[ {{\bf{n}}_{{\rm{BS}}}^{\rm{I}}\left[ 1 \right], \ldots ,{\bf{n}}_{{\rm{BS}}}^{\rm{I}}\left[ {{T_1}} \right]} \right]$.  

Define the vector form of ${\bf{Y}}_{{\rm{BS}}}^{\rm{I}}$ as ${\bf{y}}_{{\rm{BS}}}^{\rm{I}} = {\rm{vec}}\left( {{\bf{Y}}_{{\rm{BS}}}^{\rm{I}}} \right) \in {{\mathbb C}^{{N_{{\rm{BS}}}}T_1 \times 1}}$. The mean and covariance matrix of ${\bf{y}}_{{\rm{BS}}}^{\rm{I}}$ can be readily obtained as 
\begin{align}
&{\bf{u}}_{{\rm{BS}}}^{\rm{I}} = {\mathbb E}\left\{ {{\bf{y}}_{{\rm{BS}}}^{\rm{I}}} \right\} = {\beta _{{\rm{BTB}}}}{\rm{vec}}\left( {{\bf{A}}\left( {\mu _{{\rm{B2T}}}^{\rm{D}},\nu _{{\rm{B2T}}}^{\rm{D}}} \right){\bf{W}}_{{\rm{BS}}}^{\rm{I}}} \right),\\
&{\bf{R}}_{{\rm{BS}}}^{\rm{I}} = {\mathbb E}\left\{ {\left( {{\bf{y}}_{{\rm{BS}}}^{\rm{I}} - {\bf{u}}_{{\rm{BS}}}^{\rm{I}}} \right){{\left( {{\bf{y}}_{{\rm{BS}}}^{\rm{I}} - {\bf{u}}_{{\rm{BS}}}^{\rm{I}}} \right)}^H}} \right\} = {\sigma ^2}{{\bf{I}}_{{N_{{\rm{BS}}}}{T_1}}}. \label{covariance}
\end{align}
Since the noise is Gaussian distributed, the FIM can be calculated as 
\begin{align}
{{\bf{F}}_{{\bm{\eta }}_{{\rm{BS}}}^{\rm{I}}}}=\left[ {\begin{array}{*{20}{c}}
		{f_{\mu _{{\rm{B2T}}}^{\rm{D}}\mu _{{\rm{B2T}}}^{\rm{D}}}^{\rm{I}}}&{f_{\mu _{{\rm{B2T}}}^{\rm{D}}\nu _{{\rm{B2T}}}^{\rm{D}}}^{\rm{I}}}&{{\bf{f}}_{\mu _{{\rm{B2T}}}^{\rm{D}}{\bm{\beta }}_{{\rm{BTB}}}^T}^{\rm{I}}}\\
		{f_{\mu _{{\rm{B2T}}}^{\rm{D}}\nu _{{\rm{B2T}}}^{\rm{D}}}^{{\rm{I,}}T}}&{f_{\nu _{{\rm{B2T}}}^{\rm{D}}\nu _{{\rm{B2T}}}^{\rm{D}}}^{\rm{I}}}&{{\bf{f}}_{\nu _{{\rm{B2T}}}^{\rm{D}}{\bm{\beta }}_{{\rm{BTB}}}^T}^{\rm{I}}}\\
		{{\bf{f}}_{\mu _{{\rm{B2T}}}^{\rm{D}}{\bm{\beta }}_{{\rm{BTB}}}^T}^{{\rm{I,}}T}}&{{\bf{f}}_{\nu _{{\rm{B2T}}}^{\rm{D}}{\bm{\beta }}_{{\rm{BTB}}}^T}^{{\rm{I,}}T}}&{{\bf{F}}_{{\bm{\beta }}_{{\rm{BTB}}}^T{\bm{\beta }}_{{\rm{BTB}}}^T}^{{\rm{I}}}}
\end{array}} \right], \label{fishermatrix_stageI}
\end{align}
where the $\left( {i,j} \right)$ entry of ${{\bf{F}}_{{\bm{\eta }}_{{\rm{BS}}}^{\rm{I}}}}$ is given by \cite{stoica2005spectral} 
\begin{align}
{\left[ {{{\bf{F}}_{{\bm{\eta }}_{{\rm{BS}}}^{\rm{I}}}}} \right]_{ij}} &={\rm{ tr}}\left( {{\bf{R}}_{{\rm{BS}}}^{{\rm{I,}} - 1}\frac{{\partial {\bf{R}}_{{\rm{BS}}}^{\rm{I}}}}{{\partial {\bm{\eta }}_{{\rm{BS,}}i}^{\rm{I}}}}{\bf{R}}_{{\rm{BS}}}^{{\rm{I,}} - 1}\frac{{\partial {\bf{R}}_{{\rm{BS}}}^{\rm{I}}}}{{\partial {\bm{\eta }}_{{\rm{BS,}}j}^{\rm{I}}}}} \right)  \notag\\
&+2{\mathop{\rm Re}\nolimits} \left\{ {\frac{{\partial {\bf{u}}_{{\rm{BS}}}^H}}{{\partial {\bm{\eta }}_{{\rm{BS,}}i}^{\rm{I}}}}{\bf{R}}_{{\rm{BS}}}^{{\rm{I,}} - 1}\frac{{\partial {\bf{u}}_{{\rm{BS}}}^{\rm{I}}}}{{\partial {\bm{\eta }}_{{\rm{BS,}}j}^{\rm{I}}}}} \right\}. \label{Fisherexpresion}
\end{align}
The following notation is defined for later use: $\frac{{\partial {\bf{A}}\left( {\mu _{{\rm{B2T}}}^{\rm{D}},\nu _{{\rm{B2T}}}^{\rm{D}}} \right)}}{{\partial \mu _{{\rm{B2T}}}^{\rm{D}}}} = {{\bf{\dot A}}_{\mu _{{\rm{B2T}}}^{\rm{D}}}}$ and $\frac{{\partial {\bf{A}}\left( {\mu _{{\rm{B2T}}}^{\rm{D}},\nu _{{\rm{B2T}}}^{\rm{D}}} \right)}}{{\partial \nu _{{\rm{B2T}}}^{\rm{D}}}} = {{\bf{\dot A}}_{\nu _{{\rm{B2T}}}^{\rm{D}}}}$.

\textbf{\textbf{\underline{\textit{Lemma}}} 1:} The derivation of FIM ${{\bf{F}}_{{\bm{\eta }}_{{\rm{BS}}}^{\rm{I}}}}$ is given by 
\begin{align}
&f_{\mu _{{\rm{B2T}}}^{\rm{D}}\mu _{{\rm{B2T}}}^{\rm{D}}}^{\rm{I}} = \frac{{2{{\left| {{\beta _{{\rm{BTB}}}}} \right|}^2}{T_1}}}{{{\sigma ^2}}}{\rm{tr}}\left( {{{{\bf{\dot A}}}_{\mu _{{\rm{B2T}}}^{\rm{D}}}}{{\bf{R}}_{{{\bf{w}}_{{\rm{BS}}}}}}{\bf{\dot A}}_{\mu _{{\rm{B2T}}}^{\rm{D}}}^H} \right),\label{FIM_1}\\ 
&f_{\mu _{{\rm{B2T}}}^{\rm{D}}\nu _{{\rm{B2T}}}^{\rm{D}}}^{\rm{I}} = \frac{{2{{\left| {{\beta _{{\rm{BTB}}}}} \right|}^2}}}{{{\sigma ^2}}}{\mathop{\rm Re}\nolimits} \left\{ {{\rm{tr}}\left( {{{{\bf{\dot A}}}_{\nu _{{\rm{B2T}}}^{\rm{D}}}}{{\bf{R}}_{{{\bf{w}}_{{\rm{BS}}}}}}{\bf{\dot A}}_{\mu _{{\rm{B2T}}}^{\rm{D}}}^H} \right)} \right\},\label{FIM_2}\\
&{\bf{f}}_{\mu _{{\rm{B2T}}}^{\rm{D}}{\bm{\beta }}_{{\rm{BTB}}}^T}^{\rm{I}} = \frac{{2{T_1}}}{{{\sigma ^2}}}{\mathop{\rm Re}\nolimits} \left\{ {\beta _{{\rm{BTB}}}^H\left[ {1{\kern 1pt} {\kern 1pt} {\kern 1pt} {\kern 1pt} {\kern 1pt} {\kern 1pt} j} \right]{\rm{tr}}\left( {{\bf{A}}{{\bf{R}}_{{{\bf{w}}_{{\rm{BS}}}}}}{\bf{\dot A}}_{\mu _{{\rm{B2T}}}^{\rm{D}}}^H} \right)} \right\},\label{FIM_3}\\
&f_{\nu _{{\rm{B2T}}}^{\rm{D}}\nu _{{\rm{B2T}}}^{\rm{D}}}^{\rm{I}} = \frac{{2{{\left| {{\beta _{{\rm{BTB}}}}} \right|}^2}{T_1}}}{{{\sigma ^2}}}{\rm{tr}}\left( {{{{\bf{\dot A}}}_{\nu _{{\rm{B2T}}}^{\rm{D}}}}{{\bf{R}}_{{{\bf{w}}_{{\rm{BS}}}}}}{\bf{\dot A}}_{\nu _{{\rm{B2T}}}^{\rm{D}}}^H} \right),\label{FIM_4}\\
&{\bf{f}}_{\nu _{{\rm{B2T}}}^{\rm{D}}{\bm{\beta }}_{{\rm{BTB}}}^T}^{\rm{I}}{\rm{ = }}\frac{{2{T_1}}}{{{\sigma ^2}}}{\mathop{\rm Re}\nolimits} \left\{ {\beta _{{\rm{BTB}}}^H\left[ {1{\kern 1pt} {\kern 1pt} {\kern 1pt} {\kern 1pt} {\kern 1pt} {\kern 1pt} j} \right]{\rm{tr}}\left( {{\bf{A}}{{\bf{R}}_{{{\bf{w}}_{{\rm{BS}}}}}}{\bf{\dot A}}_{\nu _{{\rm{B2T}}}^{\rm{D}}}^H} \right)} \right\},\\
&{\bf{F}}_{{\bm{\beta }}_{{\rm{BTB}}}^T{\bm{\beta }}_{{\rm{BTB}}}^T} = \frac{{2{T_1}}}{{{\sigma ^2}}}{\rm{tr}}\left( {{\bf{A}}{{\bf{R}}_{{{\bf{w}}_{{\rm{BS}}}}}}{{\bf{A}}^H}} \right){{\bf{I}}_2}. \label{FIM_6}
\end{align}
\emph{Proof:}  Please refer to Appendix~\ref{appendix_lemma1}. \hfill\rule{2.7mm}{2.7mm}

\subsection{Performance Analysis}
The CRB of ${{\bm{\eta }}_{{\rm{BS}}}^{\rm{I}}}$ is the inverse of FIM ${{\bf{F}}_{{\bm{\eta }}_{{\rm{BS}}}^{\rm{I}}}}$, which is given by 
\begin{align}
{\rm{CRB}}\left( {{\bm{\eta }}_{{\rm{BS}}}^{\rm{I}}} \right) = {\bf{F}}_{{\bm \eta }_{{\rm{BS}}}^{\rm{I}}}^{ - 1}.
\end{align}
The corresponding CRB minimization optimization problem can be formulated as 
\begin{subequations} \label{problem1}
\begin{align}
&\mathop {\min }\limits_{{{\bf{R}}_{{{\bf{w}}_{{\rm{BS}}}}}}} {\rm{tr}}\left( {{\rm{CRB}}\left( {{\bm \eta} _{{\rm{BS}}}^{\rm{I}}} \right)} \right)\\\label{problem1_obj}
&{\rm{s}}{\rm{.t.}}~{\rm{tr}}\left( {{{\bf{R}}_{{{\bf{w}}_{{\rm{BS}}}}}}} \right) = {P_{{\rm{BS}}}}.
\end{align}
\end{subequations}
It is difficult to  solve  \eqref{problem1} since the objective function ${\rm{tr}}\left( {{\rm{CRB}}\left( {{\bm \eta} _{{\rm{BS}}}^{\rm{I}}} \right)} \right)$ cannot be expressed in an analytical function of ${{{\bf{R}}_{{{\bf{w}}_{{\rm{BS}}}}}}}$. To solve it, the following lemma is applied.

\textbf{\textbf{\underline{\textit{Lemma}}} 2:}   Let ${\bf{Q}} \in {{\mathbb C}^{N \times N}}$ be a positive-definite Hermitian matrix, the following inequality holds \cite{Yang2007MIMO}:
\begin{align}
{\rm{tr}}\left( {{{\bf{Q}}^{ - 1}}} \right) \ge \sum\limits_{n = 1}^N {\frac{1}{{{{\bf{Q}}_{ii}}}}},
\end{align}
where the equality holds if and only if ${\bf{Q}}$ is a diagonal   matrix.

\textbf{\textbf{\underline{\textit{Theorem}}} 1:}   The optimal solution to problem \eqref{problem1} is given by 
\begin{align}
{{\bf{R}}_{{{\bf{w}}_{{\rm{BS}}}}}} = \frac{{{P_{{\rm{BS}}}}}}{{{N_{{\rm{BS}}}}}}{{\bf{I}}_{{N_{{\rm{BS}}}}}}. \label{BS_beam}
\end{align}
\emph{Proof:} Please refer to Appendix~\ref{appendix_Theorem1}. \hfill\rule{2.7mm}{2.7mm}

This solution is intuitive since there is no prior information on the target location, the BS will transmit a \textit{spatially white} probing signal to give a constant power at any direction. 
Based on  Theorem 1, we can adopt the discrete Fourier transform (DFT)-based codebook for BS beamformer as ${\left[ {{\bf{W}}_{{\rm{BS}}}^{\rm{I}}} \right]_{n,t}} = \sqrt {\frac{{{P_{{\rm{BS}}}}}}{{{N_{{\rm{BS}}}}}}} {e^{ - j\frac{{2\pi \left( {t - 1} \right)\left( {n - 1} \right)}}{{{T_1}}}}}$.
%
%
%

%

\section{Single Target: IRS-target DoA Estimation}
In stage II, the IRS is turned on. Denote the set of samples in stage II as ${{\cal T}_2} = \left\{ {{T_1} + 1, \ldots ,{T_1} + {T_2}} \right\}$.
Let  ${\bf{\Theta }}\left[ t \right] = {\rm{diag}}\left( {{{\bf{w}}_{{\rm{IRS}}}}\left[ t \right]} \right)$, where ${{\bf{w}}_{{\rm{IRS}}}}\left[ t \right] = {\left[ {{e^{j{\theta _{t,1}}}}, \ldots ,{e^{j{\theta _{t,{N_{\rm{r}}}}}}}} \right]^T}$  represents the IRS phase shift vector at the $t$-th sample, $t \in {\cal T}_2$.

The echo signal received  at the BS at the $t$-th ($t \in {{\cal T}_2}$) sample is given by 
\begin{align}
&{\bf{y}}_{{\rm{BS}}}^{{\rm{II}}}\left[ t \right] = \left( {{{\bf{H}}_{{\rm{BTB}}}}{\rm{ + }}{\bf{H}}_{{\rm{B2I}}}^T{\bf{\Theta }}\left[ t \right]{{\bf{H}}_{{\rm{ITI}}}}{\bf{\Theta }}\left[ t \right]{\bf{H}}_{{\rm{B2I}}} + {\bf{H}}_{{\rm{B2I}}}^T{\bf{\Theta }}\left[ t \right]} \right.\times\notag\\
&\!{{\bf{H}}_{{\rm{BTI}}}} \left. { + {\bf{H}}_{{\rm{BTI}}}^T{\bf{\Theta }}\left[ t \right]{\bf{H}}_{{\rm{B2I}}}^{}{\rm{ + }}{\bf{H}}_{{\rm{B2I}}}^T{\bf{H}}_{{\rm{B2I}}}^{}} \right){{\bf{w}}_{{\rm{BS}}}}\left[ t \right] + {\bf{n}}_{\rm BS}^{\rm II}\left[ t \right], \label{stageII_signal}
\end{align}
where ${\bf{n}}_{{\rm{BS}}}^{{\rm{II}}}\left[ t \right] \sim {\cal CN}\left( {0,{\sigma ^2}{{\bf{I}}_{{N_{{\rm{BS}}}}}}} \right)$.
Since the IRS location is known \textit{ a priori}, the channel information between the BS and the IRS, i.e., ${{\bf{H}}_{{\rm{B2I}}}^T{\bf{H}}_{{\rm{B2I}}}^{}}$,  can be directly calculated based on the channel model in \eqref{BS_IRS_channel}. In addition, the direct link between the BS and the target, i.e., ${{{\bf{H}}_{{\rm{BTB}}}}}$, can  be estimated in an offline stage or stage I.  Therefore,  removing ${{\bf{H}}_{{\rm{B2I}}}^T{\bf{H}}_{{\rm{B2I}}}^{}}$ and ${{{\bf{H}}_{{\rm{BTB}}}}}$ from \eqref{stageII_signal} yields
\begin{align}
&{\bf{\tilde y}}_{{\rm{BS}}}^{{\rm{II}}}\left[ t \right] = \left( {{\bf{H}}_{{\rm{B2I}}}^T{\bf{\Theta }}\left[ t \right]{{\bf{H}}_{{\rm{ITI}}}}{\bf{\Theta }}\left[ t \right]{\bf{H}}_{{\rm{B2I}}}^{} + {\bf{H}}_{{\rm{B2I}}}^T{\bf{\Theta }}\left[ t \right]{{\bf{H}}_{{\rm{BTI}}}}} \right.\notag\\
&\qquad~~ \left. { + {\bf{H}}_{{\rm{BTI}}}^T{\bf{\Theta }}\left[ t \right]{\bf{H}}_{{\rm{B2I}}}^{}} \right){{\bf{w}}_{{\rm{BS}}}}\left[ t \right] + {\bf{ n}}_{{\rm{BS}}}^{{\rm{II}}}\left[ t \right]\notag\\
&=\Big( {{\beta _{{\rm{ITI}}}}\beta _{{\rm{B2I}}}^2{q^2}\left[ t \right]{\bf{a}}\left( {\mu _{{\rm{B2I}}}^{\rm{D}},\nu _{{\rm{B2I}}}^{\rm{D}}} \right){{\bf{a}}^T}\left( {\mu _{{\rm{B2I}}}^{\rm{D}},\nu _{{\rm{B2I}}}^{\rm{D}}} \right){\rm{ + }}{\beta _{{\rm{BTI}}}}} \notag\\
&   \times {\beta _{{\rm{B2I}}}}q\left[ t \right]{\bf{a}}\left( {\mu _{{\rm{B2I}}}^{\rm{D}},\nu _{{\rm{B2I}}}^{\rm{D}}} \right){{\bf{a}}^T}\left( {\mu _{{\rm{B2T}}}^{\rm{D}},\nu _{{\rm{B2T}}}^{\rm{D}}} \right) + {\beta _{{\rm{BTI}}}}{\beta _{{\rm{B2I}}}}q\left[ t \right]\notag\\
&\times {{\bf{a}}\left( {\mu _{{\rm{B2T}}}^{\rm{D}},\nu _{{\rm{B2T}}}^{\rm{D}}} \right){{\bf{a}}^T}\left( {\mu _{{\rm{B2I}}}^{\rm{D}},\nu _{{\rm{B2I}}}^{\rm{D}}} \right)} \Big){{\bf{w}}_{{\rm{BS}}}}\left[ t \right] + {\bf{n}}_{{\rm{BS}}}^{{\rm{II}}}\left[ t \right], \label{stageII_signal_1}
\end{align}
where $q\left[ t \right] = {\bf{b}}_{\rm{r}}^T\left( {\mu _{{\rm{B2I}}}^{\rm{A}},\nu _{{\rm{B2I}}}^{\rm{A}}} \right){\bf{\Theta }}\left[ t \right]{{\bf{b}}_{\rm{r}}}\left( {\mu _{{\rm{I2T}}}^{\rm{D}},\nu _{{\rm{I2T}}}^{\rm{D}}} \right)$. 
\begin{figure}[!t]
	\centerline{\includegraphics[width=3in]{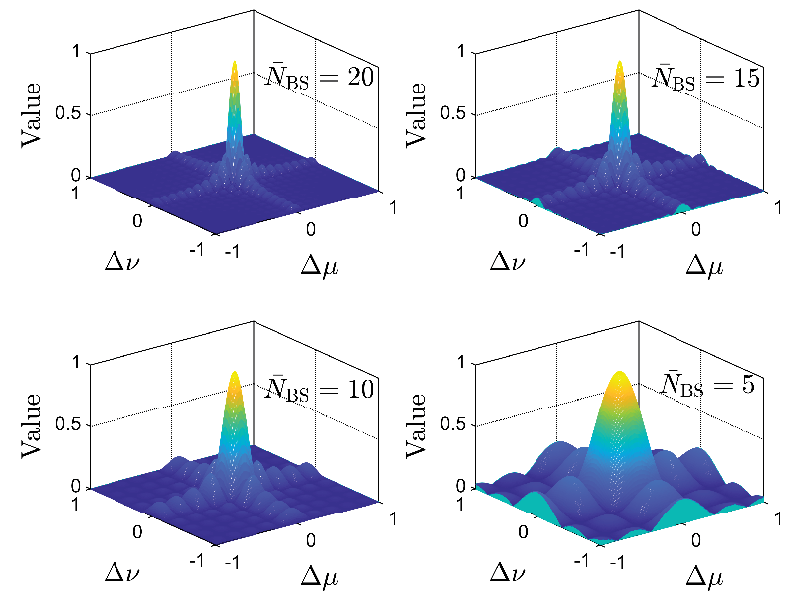}}
	\caption{Normalized beam power distribution versus azimuth and elevation directions under different ${\bar N}_{\rm BS}$.} \label{energyleakage}
\end{figure}
\begin{figure}[!t]
	\centerline{\includegraphics[width=3in]{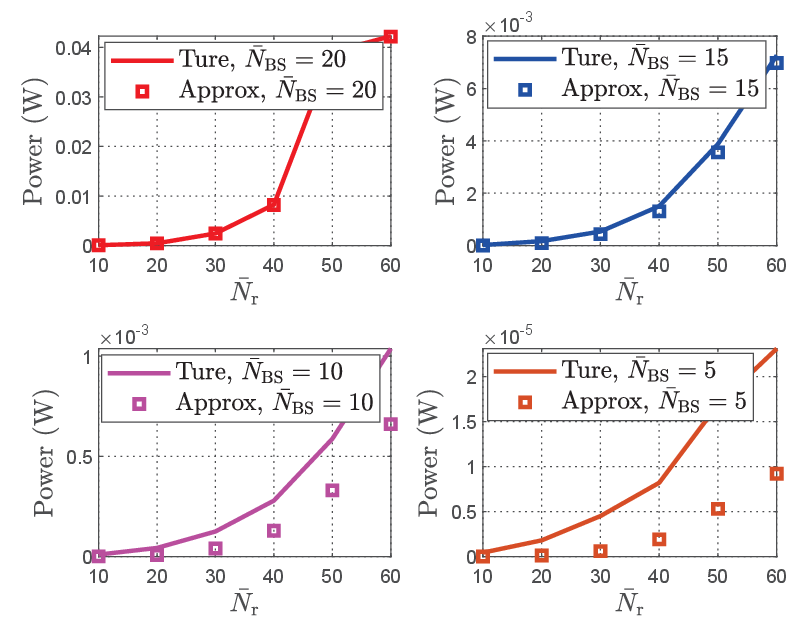}}
	\caption{True value versus approximate value in terms of average received power.}  \label{trueVSappro}
\end{figure}

It has been shown in \cite{bekkerman2006Target} that the localization  MSE tends to decrease with the increase of average received power.  Motivated by this, a reasonable approach for designing BS transmit beamformer is to align it with the steering direction from the BS to the IRS,  i.e., ${{\bf{w}}_{{\rm{BS}}}}\left[ t \right] = \sqrt {\frac{{{P_{{\rm{BS}}}}}}{{{N_{{\rm{BS}}}}}}} {{\bf{a}}^{\rm{*}}}\left( {\mu _{{\rm{B2I}}}^{\rm{D}},\nu _{{\rm{B2I}}}^{\rm{D}}} \right)$. This approach has two advantages. On the one hand,  the power of signals going through the BS-IRS link would be enhanced. On the other hand,  the energy leakage of the BS to other directions is confined to a small value especially when $N_{\rm BS}$ is large, which indicates that the received power of signal passing through the BS-target link at the BS would be small and the IRS will play a dominating role for sensing. Let  ${\bf{w}}_{{\rm{BS}}}^{{\rm{filter,}}T}\left[ t \right] = {{\bf{a}}^H}\left( {\mu _{{\rm{B2I}}}^{\rm{D}},\nu _{{\rm{B2I}}}^{\rm{D}}} \right)$ denote the matched filter at the BS, then the received signal after the matched filtering can be expressed  as 
\begin{align}
{{\tilde y}}_{{\rm{BS}}}^{{\rm{II}}}\left[ t \right] = {\bf{w}}_{{\rm{BS}}}^{{\rm{filter,}}T}\left[ t \right]{\bf{\tilde y}}_{{\rm{BS}}}^{{\rm{II}}}\left[ t \right]. \label{stageII_signal_2}
\end{align}
We note that since the target location and the channel state information are unknown, the signals passing through three different paths, i.e., BS-target-IRS-BS link, BS-IRS-target-BS link, and BS-IRS-target-IRS-BS link, will be coherently added, implying that the received power may be decreased.
To distinguish the BS-IRS-target-IRS-BS link and BS-target-IRS-BS link (or BS-IRS-target-BS link), two factors should be considered, namely, $N_{\rm BS}$ and $N_{\rm r}$, where a larger $N_{\rm BS}$ would lead a smaller recevied power of BS-target-IRS-BS link and a larger $N_{\rm r}$ would lead 
the BS-IRS-target-IRS-BS link dominated over the BS-target-IRS-BS link.
To better illuminate it, we define the normalized beam gain of ${\bf{a}}\left( {{\mu _1},{\nu _1}} \right)$ in the direction $\left( {{\mu _2},{\nu _2}} \right)$ as 
\begin{align}
&f\left( {\Delta \mu ,\Delta \nu } \right) = \frac{1}{{\bar N_{{\rm{BS}}}^{2}}}\left| {{{\bf{a}}^H}\left( {{\mu _1},{\nu _1}} \right){\bf{a}}\left( {{\mu _2},{\nu _2}} \right)} \right|\notag\\
& = \frac{1}{{\bar N_{{\rm{BS}}}^2}}\left| {\sum\limits_{i = 1}^{\bar N_{{\rm{BS}}}^{}} {{e^{\frac{{j\left( {\bar N_{{\rm{BS}}}^{} - 2i + 1} \right)\pi \Delta \mu }}{2}}}} } \right|\left| {\sum\limits_{i = 1}^{\bar N_{{\rm{BS}}}^{}} {{e^{\frac{{j\left( {\bar N_{{\rm{BS}}}^{} - 2i + 1} \right)\pi \Delta \nu }}{2}}}} } \right|,
\end{align}
where $\bar N_{{\rm{BS}}}^{} = N_{{\rm{BS}}}^y = N_{{\rm{BS}}}^z$ and $\Delta \mu  = {\mu _1} - \mu $,$\Delta \nu  = {\nu _1} - {\nu _2}$. In Fig.~\ref{energyleakage},  we  plot $f\left( {\Delta \mu ,\Delta \nu } \right)$ versus $\Delta \mu $ and $\Delta \nu $ under different $\bar N_{{\rm{BS}}}^{}$.
It can be seen that a larger $\bar N_{{\rm{BS}}}$ leads to a smaller sidelobe
region, which indicates that the most of energy is focused on one dedicated direction and little energy is leaked to other directions. Apparently,  the energy leaked to the BS-target link is neglected as  $\bar N_{{\rm{BS}}}=20$, but cannot be ignored for $\bar N_{{\rm{BS}}}=5$ due to a large sidelobe region.

\textbf{\textbf{\underline{\textit{Lemma}}} 3:}  With the IRS phase shifts given in \eqref{problem2}, i.e., $q\left[ t \right] = {N_{\rm{r}}}$, the power of the BS-IRS-target-IRS-BS link exceeds that of the BS-target-IRS-BS link and the  BS-IRS-target-BS link when
\begin{align}
{N_{\rm{r}}} \ge \frac{{\sqrt 2 }}{{\beta _{{\rm{B2I}}}^{}}} \label{No.:element}
\end{align}

\emph{Proof:}  For the  BS-IRS-target-IRS-BS link, its power is given by 
\begin{align}
{P_1}& = \left\| {{\bf{H}}_{{\rm{B2I}}}^T{\bf{\Theta }}\left[ t \right]{{\bf{H}}_{{\rm{ITI}}}}{\bf{\Theta }}\left[ t \right]{\bf{H}}_{{\rm{B2I}}}^{}} \right\|_F^2\notag\\
& = {\left| {{\beta _{{\rm{ITI}}}}\beta _{{\rm{B2I}}}^2} \right|^2}N_{{\rm{BS}}}^2N_{\rm{r}}^4. \label{power1}
\end{align}
Next, for the  BS-target-IRS-BS link and the  BS-IRS-target-BS link, the power sum  of the two links is given by 
\begin{align}
{P_2} &= 2\left\| {{\bf{H}}_{{\rm{B2I}}}^T{\bf{\Theta }}\left[ t \right]{{\bf{H}}_{{\rm{BTI}}}}} \right\|_F^2\notag\\
&= 2{\left| {{\beta _{{\rm{ITI}}}}\beta _{{\rm{B2I}}}^{}} \right|^2}N_{{\rm{BS}}}^2N_{\rm{r}}^2. \label{power2}
\end{align}
Based on  \eqref{power1} and \eqref{power2}, \eqref{No.:element} can be directly obtained.

In Fig.~\ref{trueVSappro}, we further compare the power of ${{\tilde y}}_{{\rm{BS}}}^{{\rm{II}}}\left[ t \right]$ (denoted by `True') with the power of the BS-IRS-target-IRS-BS link (denoted by `Approx'), i.e., ${\left| {{\beta _{{\rm{ITI}}}}\beta _{{\rm{B2I}}}^2{N_{{\rm{BS}}}}\sqrt {{P_{{\rm{BS}}}}{N_{{\rm{BS}}}}} {q^2}\left[ t \right]} \right|^2}$. It is observed that as $\bar N_{{\rm{BS}}}$ is large, e.g., $\bar N_{{\rm{BS}}}=15$ and $\bar N_{{\rm{BS}}}=20$, the `Approx' approaches `True', which implies that the BS-IRS-target-IRS-BS link is dominated over the BS-target-IRS-BS link (or BS-IRS-target-BS link), and the latter link can be ignored. On the other hand, if both $\bar N_{{\rm{BS}}}$ and $\bar N_{{\rm{r}}}$ (${{\bar N}_{\rm{r}}} = N_{\rm{r}}^y = N_{\rm{r}}^z$) are small, e.g., $\bar N_{{\rm{BS}}}=5$ and $\bar N_{{\rm{r}}}=20$, the power of `Approx' can be  neglected,  indicating that the BS-target-IRS-BS link (or BS-IRS-target-BS link) is dominated over the  BS-IRS-target-IRS-BS link.\footnote{ We should point out that even $\bar N_{{\rm{BS}}}$ is small, the power of  BS-IRS-target-IRS-BS link will be comparable with that of the BS-target-IRS-BS link when  $\bar N_{{\rm{r}}}$ is large due to the passive beamforming gain provided by the IRS, and the signals from different links maybe coherently added or unconstructively added at the BS.  In this case, the sensing performance cannot be guaranteed and is thus omitted here for discussion.} In the following, we discuss two cases, namely,  Case 1 where $\bar N_{{\rm{BS}}}$ is large and  Case 2 where both $\bar N_{{\rm{BS}}}$ and $\bar N_{{\rm{r}}}$ are small.
\subsection{Case 1: $\bar N_{{\rm{BS}}}$ is Large}
In this case, the BS-target-IRS-BS link (or BS-IRS-target-BS link) can be neglected, and  \eqref{stageII_signal_2} is approximated as 
\begin{align}
{{\tilde y}}_{{\rm{BS}}}^{{\rm{II}}}\left[ t \right] \approx {\beta _{{\rm{ITI}}}}\beta _{{\rm{B2I}}}^2{N_{{\rm{BS}}}}\sqrt {{P_{{\rm{BS}}}}{N_{{\rm{BS}}}}} {q^2}\left[ t \right] + \tilde n_{{\rm{BS}}}^{{\rm{II}}}\left[ t \right],\label{stageII_receivedsignal}
\end{align}
where $\tilde n_{{\rm{BS}}}^{{\rm{II}}}\left[ t \right]  \sim {\cal CN}\left( {0,{N_{{\rm{BS}}}}{\sigma ^2}} \right)$.
\subsubsection{IRS-target DoA Estimation}
To maximize the average received power, i.e., $P_{{\rm{BS}}}^{{\rm{II}}}\left[ t \right] = {\mathbb E}\left\{ {{{\left| {{{\tilde y}}_{{\rm{BS}}}^{{\rm{II}}}\left[ t \right]} \right|}^2}} \right\}$,   we next design ${\bf{\Theta }}\left[ t \right]$ to maximize $\left| {q\left[ t \right]} \right|$.  Recall that ${{\bf{b}}_{\rm{r}}}\left( {\mu _{{\rm{B2I}}}^{\rm{A}},\nu _{{\rm{B2I}}}^{\rm{A}}} \right){\rm{ = }}{\bf{u}}\left( {\mu _{{\rm{B2I}}}^{\rm{A}},N_{\rm{r}}^y} \right) \otimes {\bf{u}}\left( {\nu _{{\rm{B2I}}}^{\rm{A}},N_{\rm{r}}^z} \right)$ and ${{\bf{b}}_{\rm{r}}}\left( {\mu _{{\rm{I2T}}}^{\rm{D}},\nu _{{\rm{I2T}}}^{\rm{D}}} \right){\rm{ = }}{\bf{u}}\left( {\mu _{{\rm{I2T}}}^{\rm{D}},N_{\rm{r}}^y} \right) \otimes {\bf{u}}\left( {\nu _{{\rm{I2T}}}^{\rm{D}},N_{\rm{r}}^z} \right)$, we then have 
\begin{align}
&q\left[ t \right]={\bf{b}}_{\rm{r}}^T\left( {\mu _{{\rm{B2I}}}^{\rm{A}},\nu _{{\rm{B2I}}}^{\rm{A}}} \right){\bf{\Theta }}\left[ t \right]{{\bf{b}}_{\rm{r}}}\left( {\mu _{{\rm{I2T}}}^{\rm{D}},\nu _{{\rm{I2T}}}^{\rm{D}}} \right)\notag\\
&\quad ~{\kern 1pt}{\kern 1pt}{\kern 1pt}= \left( {{{\bf{u}}^T}\left( {\mu _{{\rm{B2I}}}^{\rm{A}},N_{\rm{r}}^y} \right) \otimes {{\bf{u}}^T}\left( {\nu _{{\rm{B2I}}}^{\rm{A}},N_{\rm{r}}^z} \right)} \right)  \notag\\
&\qquad~~\odot\left( {{{\bf{u}}^T}\left( {\mu _{{\rm{I2T}}}^{\rm{D}},N_{\rm{r}}^y} \right) \otimes {{\bf{u}}^T}\left( {\nu _{{\rm{I2T}}}^{\rm{D}},N_{\rm{r}}^z} \right)} \right){\bf{w}}_{{\rm{IRS}}}^{}\left[ t \right]\notag\\
&\quad ~{\kern 1pt}{\kern 1pt}{\kern 1pt}= \left( {{{\bf{u}}^T}\left( {\mu _{{\rm{B2I}}}^{\rm{A}},N_{\rm{r}}^y} \right) \odot {{\bf{u}}^T}\left( {\mu _{{\rm{I2T}}}^{\rm{D}},N_{\rm{r}}^y} \right)} \right)\notag\\
& \qquad~~\otimes \left( {{{\bf{u}}^T}\left( {\nu _{{\rm{B2I}}}^{\rm{A}},N_{\rm{r}}^z} \right) \odot {{\bf{u}}^T}\left( {\nu _{{\rm{I2T}}}^{\rm{D}},N_{\rm{r}}^z} \right)} \right){\bf{w}}_{{\rm{IRS}}}^{}\left[ t \right]\notag\\
& \quad ~{\kern 1pt}{\kern 1pt}{\kern 1pt}= \left( {{{\bf{u}}^T}\left( {\mu ,N_{\rm{r}}^y} \right) \otimes {{\bf{u}}^T}\left( {\nu ,N_{\rm{r}}^z} \right)} \right){{\bf{w}}_{{\rm{IRS}}}}\left[ t \right]\notag\\
& \quad ~{\kern 1pt}{\kern 1pt}{\kern 1pt}\overset{(a)}{=} \left( {{{\bf{u}}^T}\left( {\mu ,N_{\rm{r}}^y} \right){\bf{w}}_{{\rm{IRS}}}^y\left[ t \right]} \right)\left( {{{\bf{u}}^T}\left( {\nu ,N_{\rm{r}}^z} \right){\bf{w}}_{{\rm{IRS}}}^z\left[ t \right]} \right), \label{IRSphaseshiftfunction}
\end{align}
where $\mu  = \mu _{{\rm{B2I}}}^{\rm{A}}{\rm{ + }}\mu _{{\rm{I2T}}}^{\rm{D}}$  and $\nu  = \nu _{{\rm{B2I}}}^{\rm{A}}{\rm{ + }}\nu _{{\rm{I2T}}}^{\rm{D}}$,  (a) holds  since ${{\bf{w}}_{{\rm{IRS}}}}\left[ t \right]{\rm{ = }}{\bf{w}}_{{\rm{IRS}}}^y\left[ t \right] \otimes {\bf{w}}_{{\rm{IRS}}}^z\left[ t \right]$, ${\bf{w}}_{{\rm{IRS}}}^y\left[ t \right] \in {{\mathbb C}^{N_{\rm{r}}^y \times 1}}$, and ${\bf{w}}_{{\rm{IRS}}}^z\left[ t \right] \in {{\mathbb C}^{N_{\rm{r}}^z \times 1}}$. 

Substituting \eqref{IRSphaseshiftfunction} into \eqref{stageII_receivedsignal}  yields 
\begin{align}
 {{\tilde y}}_{{\rm{BS}}}^{{\rm{II}}}\left[ t \right] &= {\alpha }{\left( {{{\bf{u}}^T}\left( {\mu ,N_{\rm{r}}^y} \right){\bf{w}}_{{\rm{IRS}}}^y\left[ t \right]} \right)^2}{\left( {{{\bf{u}}^T}\left( {\nu ,N_{\rm{r}}^z} \right){\bf{w}}_{{\rm{IRS}}}^z\left[ t \right]} \right)^2}\notag\\
 &+ {\tilde{n}}_{{\rm{BS}}}^{{\rm{II}}}\left[ t \right],  \label{stageII_sigal_3}
\end{align}
where ${\alpha } = \sqrt {{P_{{\rm{BS}}}}{N_{{\rm{BS}}}}} N_{\rm BS} \beta _{{\rm{B2I}}}^2{\beta _{{\rm{ITI}}}}$. Note that the traditional DoA estimation methods such as the MUSIC algorithm cannot be adopted in stage II since the direct LoS link does not exist. To estimate $\mu$ from \eqref{stageII_sigal_3}, a codebook of IRS beam is applied. Before designing the IRS codebook, we consider the following optimization problem 
 \begin{subequations} \label{problem2}
	\begin{align}
		&\mathop {\max }\limits_{{\bf{z}} \in {{\mathbb C}^{N \times 1}}} {\left| {{{\bf{u}}^T}\left( {\mu ,N} \right){\bf{z}}} \right|^2}\\
		&{\rm{s}}{\rm{.t}}{\rm{.}}~ \left| {{{\left[ {\bf{z}} \right]}_i}} \right| = 1,i = 1, \ldots ,N.
	\end{align}
\end{subequations} 
By employing  Cauchy-Schwarz inequality, it is not difficult to show that the optimal solution to problem  \eqref{problem2} is given by ${\bf{z}} = {e^{ - j\arg \left( {{\bf{u}}\left( {\mu ,N} \right)} \right)}} = {{\bf{u}}^*}\left( {\mu ,N} \right)$. This indicates that the IRS beam should be set to align with one steering direction. Inspired by this, the IRS codebook ${\cal W}_{{\rm{IRS}}}^y = \left\{ {{\bf{w}}_{{\rm{IRS}}}^y\left[ {{T_1} + 1} \right], \ldots ,{\bf{w}}_{{\rm{IRS}}}^y\left[ {{T_1} + {T_2}} \right]} \right\}$ is designed as follows.  We note that  $\mu$ is limited to $\left[ { - 2,2} \right]$ (we assume that the IRS reflecting element spacing is equal to half of carrier wavelength), however, the frequency ambiguity cannot be inevitably incurred due to ${\bf{u}}\left( {\mu ,N} \right) = -  {\bf{u}}\left( {\mu  \pm 2,N} \right)$ for even $N$ and ${\bf{u}}\left( {\mu ,N} \right) =   {\bf{u}}\left( {\mu  \pm 2,N} \right)$ for odd ${ N}$. To circumvent it, a simple way is to carefully deploy the IRS so that the LoS azimuth direction is perpendicular to its BS, namely,  $\mu _{{\rm{B2I}}}^{\rm{A}}=0$.\footnote{Note that we cannot set  $\nu _{{\rm{B2I}}}^{\rm{A}}=0$ in this case since the unique 3D target location cannot be constructed based on  \eqref{angletolocation}. In addition, the  IRS is higher than the target, so we have $\nu _{{\rm{I2T}}}^{\rm{D}}\le 0$, which indicates that $\nu _{{\rm{I2T}}}^{\rm{D}}$ can be uniquely determined.} Thus, we equally  partition the search  space $\left[ { - 1,1} \right]$ into $T_2-1$ subspaces with $T_2$ narrow beams. Then, the IRS beam for each sample  is designed as 
\begin{align}
&{\bf{w}}_{{\rm{IRS}}}^y\left[ i \right] = {{\bf{u}}^*}\left( { - 1 + \frac{{2\left( {i - {T_1}} \right)}}{{T_2^y}},N_{\rm{r}}^y} \right),  i\in {{\cal T}^y_2},\\
&{\bf{w}}_{{\rm{IRS}}}^z\left[ j \right] = {{\bf{u}}^*}\left( { - 1 + \frac{{2\left( {j - T_2^y - {T_1}} \right)}}{{T_2^z}},N_{\rm{r}}^z} \right),  j\in {{\cal T}^z_2},
\end{align}
where ${\cal T}_2^y = \left\{ {{T_1} + 1, \ldots ,{T_1} + T_2^y} \right\}$ and ${\cal T}_2^z = \left\{ {{T_1} + T_2^y + 1, \ldots ,{T_1} + T_2^y + T_2^z} \right\}$, and $T_2^{} = T_2^y+T_2^z$.
Thus, the best beam direction can achieve the largest of $\left| {q\left[ t \right]} \right|$ or equivalently the maximum received power ${\left| {{{\hat y}}_{{\rm{BS}}}^{{\rm{II}}}\left[ t \right]} \right|^2}$. Therefore, we can compare   different received powers under different IRS beams and choose the best one according to the following criteria:\footnote{Note that we do not consider the joint azimuth and elevation DoA estimation as   $\left( {t_{{\rm{index}}}^{y,{\rm{opt}}},t_{{\rm{index}}}^{z,{\rm{opt}}}} \right) = \mathop {\arg \max }\limits_{i \in T_2^y,j \in T_2^z} {\left| {\tilde y_{{\rm{BS}}}^{{\rm{II}}}\left[ {i,j} \right]} \right|^2}$ since this requires substantial samples, i.e., $T_2 = T_2^yT_2^z$.}
\begin{align}
\left\{ \begin{array}{l}
t_{{\rm{index}}}^{y,{\rm{opt}}} = \mathop {\arg \max }\limits_{i \in T_2^y} {\left| {\tilde y_{{\rm{BS}}}^{{\rm{II}}}\left[ {i,j} \right]} \right|^2},\\
t_{{\rm{index}}}^{z,{\rm{opt}}} = \mathop {\arg \max }\limits_{j \in T_2^z} {\left| {\tilde y_{{\rm{BS}}}^{{\rm{II}}}\left[ {i,j} \right]} \right|^2},
\end{array} \right.\label{direction_stageII}
\end{align}
where 
\begin{align}
{{\tilde y}}_{{\rm{BS}}}^{{\rm{II}}}\left[ i,j \right] &= {\alpha }{\left( {{{\bf{u}}^T}\left( {\mu ,N_{\rm{r}}^y} \right){\bf{w}}_{{\rm{IRS}}}^y\left[ i \right]} \right)^2}{\left( {{{\bf{u}}^T}\left( {\nu ,N_{\rm{r}}^z} \right){\bf{w}}_{{\rm{IRS}}}^z\left[ j \right]} \right)^2}\notag\\
&+ {\tilde{n}}_{{\rm{BS}}}^{{\rm{II}}}\left[ t \right].
\end{align}
Accordingly, the estimated spatial azimuth and elevation  angles are respectively  given by 
\begin{align}
&\hat \mu  =  - 1 + \frac{2}{{T_2^y - 1}}\left( {t_{{\rm{index}}}^{y{\rm{,opt}}} - {T_1} - 1} \right),\\
&\hat \nu  =  - 1 + \frac{2}{{T_2^z - 1}}\left( {t_{{\rm{index}}}^{z{\rm{,opt}}} -T_2^y- {T_1} - 1} \right).
\end{align} 
Then, the estimated DoA of the IRS-target link is given by $\hat \mu _{{\rm{I2T}}}^{\rm{D}} = \hat \mu  - \mu _{{\rm{B2I}}}^{\rm{A}}$ and $\hat \nu _{{\rm{I2T}}}^{\rm{D}} = \hat \nu  - \nu _{{\rm{B2I}}}^{\rm{A}}$.
\subsubsection{CRB for IRS-target DoA Estimation}
Let ${\bm{\eta }}_{{\rm{IRS}}}^{{\rm{II}}}={\left[ {\mu ,\nu ,{{\bm{\alpha }}^T}} \right]^T}$  denote the collection of unknown parameters to be estimated in stage II, including the target's azimuth/elevation DoA and channel coefficient, where    ${\bm{\alpha }} = \left[ {{\mathop{\rm Re}\nolimits} \left\{ \alpha  \right\}{\kern 1pt} {\kern 1pt} {\kern 1pt} {\mathop{\rm Im}\nolimits} \left\{ \alpha  \right\}} \right]^T$.\footnote{The FIMs of the parameter vectors ${\bm{\eta }}_{{\rm{IRS}}}^{{\rm{II}}}={\left[ {\mu ,\nu ,{{\bm{\alpha }}^T}} \right]^T}$ and ${\bm \eta} _{{\rm{IRS}}}^{{\rm{II}}} = {\left[ {\mu _{{\rm{I2T}}}^{\rm{D}},\nu _{{\rm{I2T}}}^{\rm{D}},{{\bm \alpha} ^T}} \right]^T}$ are the same, which can be readily checked via chain rule. Thus, we use the former for calculating  FIM.}  We first rewrite \eqref{stageII_sigal_3} as 
\begin{align}
\tilde y_{{\rm{BS}}}^{{\rm{II}}}\left[ t \right] = \alpha {\bf{w}}_{{\rm{IRS}}}^T\left[ t \right]{\bf{q}}{{\bf{q}}^T}{{\bf{w}}_{{\rm{IRS}}}}\left[ t \right] + \tilde n_{{\rm{BS}}}^{{\rm{II}}}\left[ t \right],
\end{align}
where ${\bf{q}} = {\bf{u}}\left( {\mu ,N_{\rm{r}}^y} \right) \otimes {\bf{u}}\left( {\nu ,N_{\rm{r}}^z} \right)$.
Then, we collect  $T_2$ samples in a vector form given by 
\begin{align}
{\bf{\tilde y}}_{{\rm{BS}}}^{{\rm{II}}} = {\left[ {\tilde y_{{\rm{BS}}}^{{\rm{II}}}\left[ {{T_1} + 1} \right], \ldots ,\tilde y_{{\rm{BS}}}^{{\rm{II}}}\left[ {{T_1} + {T_2}} \right]} \right]^T},
\end{align}
where its mean and covariance matrix are respectively given by 
\begin{align}
&{\bf{u}}_{{\rm{IRS}}}^{{\rm{II}}} = \alpha \left[ {{\bf{w}}_{{\rm{IRS}}}^T\left[ {{T_1} + 1} \right]{\bf{q}}{{\bf{q}}^T}{{\bf{w}}_{{\rm{IRS}}}}\left[ {{T_1} + 1} \right]} \right., \ldots ,\notag\\
&\qquad\quad \left. {{\bf{w}}_{{\rm{IRS}}}^T\left[ {{T_1} + {T_2}} \right]{\bf{q}}{{\bf{q}}^T}{{\bf{w}}_{{\rm{IRS}}}}\left[ {{T_1} + {T_2}} \right]} \right]^T,\\
&{\bf{R}}_{{\rm{IRS}}}^{{\rm{II}}} = {N_{{\rm{BS}}}}{\sigma ^2}{{\bf{I}}_{{T_2}}}.
\end{align}
Similar to \eqref{fishermatrix_stageI}, the FIM of ${\bm{\eta }}_{{\rm{IRS}}}^{{\rm{II}}}={\left[ {\mu ,\nu ,{{\bm{\alpha }}^T}} \right]^T}$ is given by 
\begin{align}
{\bf{F}}_{{\bm{\eta }}_{{\rm{IRS}}}^{{\rm{II}}}}=\left[ {\begin{array}{*{20}{c}}
		{f_{\mu \mu }^{{\rm{II}}}}&{f_{\mu \nu }^{{\rm{II}}}}&{{\bf{f}}_{\mu {{\bm{\alpha }}^T}}^{{\rm{II}}}}\\
		{f_{\mu \nu }^{{\rm{II,}}T}}&{f_{\nu \nu }^{{\rm{II}}}}&{{\bf{f}}_{\nu {{\bm{\alpha }}^T}}^{{\rm{II}}}}\\
		{{\bf{f}}_{\mu {{\bm{\alpha}}^T}}^{{\rm{II,}}T}}&{{\bf{f}}_{\nu {{\bm{\alpha }}^T}}^{{\rm{II,}}T}}&{{\bf{F}}_{{{\bm{\alpha }}^T}{{\bm{\alpha }}^T}}^{{\rm{II}}}}
\end{array}} \right],
\end{align}
with $(i,j)$ entry  is
\begin{align}
{\left[ {{\bf{F}}_{{\bm{\eta }}_{{\rm{IRS}}}^{{\rm{II}}}}^{}} \right]_{ij}}{\rm{ = }}\frac{2}{{{\sigma ^2}{N_{{\rm{BS}}}}}}{\mathop{\rm Re}\nolimits} \left\{ {\frac{{\partial {\bf{u}}_{{\rm{IRS}}}^{{\rm{II,}}H}}}{{\partial {\bm{\eta }}_{{\rm{IRS,}}i}^{{\rm{II}}}}}\frac{{\partial {\bf{u}}_{{\rm{IRS}}}^{{\rm{II}}}}}{{\partial {\bm{\eta }}_{{\rm{IRS,}}j}^{{\rm{II}}}}}} \right\}.
\end{align}

\textbf{\textbf{\underline{\textit{Lemma}}} 4:} Each entry of   FIM ${\bf{F}}_{{\bm{\eta }}_{{\rm{IRS}}}^{{\rm{II}}}}$ is derived in  \eqref{fisher_matrix_stageII_1}-\eqref{fisher_matrix_stageII_6}.  

\emph{Proof:}  Please refer to Appendix~\ref{appendix_lemma3}. \hfill\rule{2.7mm}{2.7mm}

\noindent Accordingly, the CRB of ${\bm{\eta }}_{{\rm{IRS}}}^{{\rm{II}}}={\left[ {\mu ,\nu ,{{\bm{\alpha }}^T}} \right]^T}$ is given by 
\begin{align}
{\rm{CRB}}\left( {{\bm \eta} _{{\rm{IRS}}}^{{\rm{II}}}} \right) = {\bf{F}}_{{\bm \eta} _{{\rm{IRS}}}^{{\rm{II}}}}^{ - 1}.
\end{align}

\textbf{\textbf{\underline{\textit{Theorem}}} 2:}   If the IRS phase shift vector is only allowed to adjust once, i.e., ${{\bf{w}}_{{\rm{IRS}}}}\left[ {{T_1} + 1} \right] = , \ldots ,= {{\bf{w}}_{{\rm{IRS}}}}\left[ {{T_1} + {T_2}} \right]$,   ${\rm{CRB}}\left( {{\bm \eta} _{{\rm{IRS}}}^{{\rm{II}}}} \right)=+\infty$.

\emph{Proof:} Please refer to Appendix~\ref{appendix_Theorem2}. \hfill\rule{2.7mm}{2.7mm}

Theorem 2 indicates that for   purely LoS in  both BS-IRS and IRS-target links, the static IRS configuration  (i.e., the IRS phase shift vector only adjusts once within a coherence channel time) cannot perform sensing, and the dynamic IRS configuration is needed for sensing. 

\begin{table*}[!t]
	\centering
	\caption{CRBs of $\mu$ and $\nu$ under different number of  IRS beams.}
	\begin{tabular}{ c c c c c c  c c }
	\toprule[1.2pt]
		Number of different IRS beams & 1                              & 2                               & 3 & 4 & 5 & 6 & $T_2$ \\ \hline	\\[-12pt]	\addlinespace
		CRB($\mu$)     & $1.60\times10^{13}$ & 4.16 & $3.78\times10^{-4}$ & $1.23\times10^{-4}$ & $6.89\times10^{-5}$ & $5.04\times10^{-5}$ & $1.67\times10^{-5}$  \\ 
		CRB($\nu$)     &$3.71\times10^{12}$ & 1.23 &$3.26\times10^{-4}$ & $1.03\times10^{-4}$ & $6.54\times10^{-5}$ & $5.12\times10^{-5}$ & $1.69\times10^{-5}$  \\ \toprule[1.2pt]
	\end{tabular}
\end{table*}

\textbf{\textbf{\underline{\textit{Remark}}} 1:}  To enable sensing assisted by the IRS, the minimum number of different IRS beams required is $3$.  To clearly show it,  we plot the average CRBs of $\mu$ and $\nu$ versus the number of different IRS beams under $T_2=60$ in Table I. One can observe that CRBs of $\mu$ and $\nu$ are very small when the number of different IRS beams exceeds $3$.
This is expected since there are two variables in  \eqref{stageII_receivedsignal} and at least $3$ nonlinear equations are required to obtain the unique solution. It should be pointed out that for $2$ nonlinear equations, there may incur multiple feasible solutions, and thus the performance is still bad for $2$ different IRS beams.

%

\textbf{\textbf{\underline{\textit{Remark}}} 2:} For only one IRS beam, target tracking is infeasible since its CRB is infinite according to Theorem 2. This result is somewhat counter-intuitive compared to conventional radar tracking in which one radar beam, e.g., phased-array radar, is feasible. The reason is that the BS-IRS-target link is not a direct LoS link, the degrees of freedom for resolving the channel coefficients and the target direction are limited, which can be mathematically seen in Theorem 2. Thus, different from conventional radar systems where only one radar beam is feasible for target tracking, it is necessary to provide multi-beam for the IRS-assisted sensing architecture. 
\subsection{Case 2: Both $\bar N_{{\rm{BS}}}$ and $\bar N_{{\rm{r}}}$ are Small}  In this case, the  BS-IRS-target-IRS-BS link can be neglected, and  \eqref{stageII_signal_2} is approximated as 
\begin{align}
\tilde y_{{\rm{BS}}}^{{\rm{II}}}\left[ t \right] \approx \tilde \alpha bq\left[ t \right] + \tilde n_{{\rm{BS}}}^{{\rm{II}}}\left[ t \right], \label{approx_2}
\end{align}
where $\tilde \alpha  = 2\sqrt {{P_{{\rm{BS}}}}{N_{{\rm{BS}}}}} \beta _{{\rm{B2I}}}^{}{\beta _{{\rm{BTI}}}}$ and $b = {{\bf{a}}^H}\left( {\mu _{{\rm{B2I}}}^{\rm{D}},\nu _{{\rm{B2I}}}^{\rm{D}}} \right){\bf{a}}\left( {\mu _{{\rm{B2T}}}^{\rm{D}},\nu _{{\rm{B2T}}}^{\rm{D}}} \right)$.
\subsubsection{IRS-target DoA Estimation} It can be readily verified that maximizing the average received power of $\tilde y_{{\rm{BS}}}^{{\rm{II}}}\left[ t \right]$ is equivalent to maximizing ${\left| {q\left( t \right)} \right|^2}$, which implies that the proposed method in Case 1 is also applicable to Case 2  for IRS-target DoA estimation.
\subsubsection{CRB for IRS-target DoA Estimation} Comparing \eqref{approx_2} with \eqref{stageII_receivedsignal}, the expressions are different and the CRB of Case 2 should be recalculated. The  mean and covariance matrix after collecting  $T_2$ samples from  \eqref{approx_2} can be respectively given by 
\begin{align}
&{\bf{\hat u}}_{{\rm{IRS}}}^{{\rm{II}}} = \tilde \alpha b{\bf{W}}_{{\rm{IRS}}}^T{\bf{q}},\\
&{\bf{\hat R}}_{{\rm{IRS}}}^{{\rm{II}}} = {N_{{\rm{BS}}}}{\sigma ^2}{{\bf{I}}_{{T_2}}}
\end{align}
where ${\bf{W}}_{{\rm{IRS}}}^{} = \left[ {{\bf{w}}_{{\rm{IRS}}}^{}\left[ {{T_1} + 1} \right], \ldots ,{\bf{w}}_{{\rm{IRS}}}^{}\left[ {{T_1} + {T_2}} \right]} \right]$. Let ${\bm{\hat \eta }}_{{\rm{IRS}}}^{{\rm{II}}} = {\left[ {\mu _{{\rm{I2T}}}^{\rm{D}},\nu _{{\rm{I2T}}}^{\rm{D}},\mu _{{\rm{B2T}}}^{\rm{D}},\nu _{{\rm{B2T}}}^{\rm{D}},{{{\bm{\tilde \alpha }}}^T}} \right]^T}$, where ${\bm{\tilde \alpha }} = \left[ {{\mathop{\rm Re}\nolimits} \left\{  \tilde \alpha  \right\}{\kern 1pt} {\kern 1pt} {\kern 1pt} {\mathop{\rm Im}\nolimits} \left\{ \tilde \alpha  \right\}} \right]^T$  denotes the unknown  parameters needed to be estimated and  ${\bf{\hat F}}_{{\bm{\eta }}_{{\rm{IRS}}}^{{\rm{II}}}}$ stands for its FIM  in Case 2. 
We can readily obtain the following identities:
\begin{align}
&\frac{{\partial {\bf{\hat u}}_{{\rm{IRS}}}^{{\rm{II}}}}}{{\partial \mu _{{\rm{I2T}}}^{\rm{D}}}} = \tilde \alpha b{\bf{W}}_{{\rm{IRS}}}^T{{{\bf{\dot q}}}_\mu },{\kern 1pt} {\kern 1pt} {\kern 1pt} \frac{{\partial {\bf{\hat u}}_{{\rm{IRS}}}^{{\rm{II}}}}}{{\partial \nu _{{\rm{I2T}}}^{\rm{D}}}} = \tilde \alpha b{\bf{W}}_{{\rm{IRS}}}^T{{{\bf{\dot q}}}_\nu },\\
&\frac{{\partial {\bf{\hat u}}_{{\rm{IRS}}}^{{\rm{II}}}}}{{\partial \mu _{{\rm{B2T}}}^{\rm{D}}}} = \tilde \alpha {{\dot b}_{\mu _{{\rm{B2T}}}^{\rm{D}}}}{\bf{W}}_{{\rm{IRS}}}^T{\bf{q}},{\kern 1pt} {\kern 1pt} \frac{{\partial {\bf{\hat u}}_{{\rm{IRS}}}^{{\rm{II}}}}}{{\partial \nu _{{\rm{B2T}}}^{\rm{D}}}} = \tilde \alpha {{\dot b}_{\nu _{{\rm{B2T}}}^{\rm{D}}}}{\bf{W}}_{{\rm{IRS}}}^T{\bf{q}},\\
&\frac{{\partial {\bf{\hat u}}_{{\rm{IRS}}}^{{\rm{II}}}}}{{\partial {{{\bf{\tilde {\bm \alpha} }}}^T}}} = \left[ {1{\kern 1pt} {\kern 1pt} {\kern 1pt} j} \right] \otimes b{\bf{W}}_{{\rm{IRS}}}^T{\bf{q}},
\end{align}
where ${{\dot b}_{\mu _{{\rm{B2T}}}^{\rm{D}}}} = {{\bf{a}}^H}\left( {\mu _{{\rm{B2I}}}^{\rm{D}},\nu _{{\rm{B2I}}}^{\rm{D}}} \right){{{\bf{\dot a}}}_{\mu _{{\rm{B2T}}}^{\rm{D}}}}\left( {\mu _{{\rm{B2T}}}^{\rm{D}},\nu _{{\rm{B2T}}}^{\rm{D}}} \right)$ and ${{\dot b}_{\nu _{{\rm{B2T}}}^{\rm{D}}}} = {{\bf{a}}^H}\left( {\mu _{{\rm{B2I}}}^{\rm{D}},\nu _{{\rm{B2I}}}^{\rm{D}}} \right){{{\bf{\dot a}}}_{\nu _{{\rm{B2T}}}^{\rm{D}}}}\left( {\mu _{{\rm{B2T}}}^{\rm{D}},\nu _{{\rm{B2T}}}^{\rm{D}}} \right)$.
As a result,  the FIM  can be directly obtained based on ${\left[ {{\bf{F}}_{{\bm{\hat \eta }}_{{\rm{IRS}}}^{{\rm{II}}}}^{}} \right]_{ij}}{\rm{ = }}\frac{2}{{{\sigma ^2}{N_{{\rm{BS}}}}}}{\rm{Re}}\left\{ {\frac{{\partial {\bf{\hat u}}_{{\rm{IRS}}}^{{\rm{II}},H}}}{{\partial {\bm{\hat \eta }}_{{\rm{IRS}},i}^{{\rm{II}}}}}\frac{{\partial {\bf{\hat u}}_{{\rm{IRS}}}^{{\rm{II}}}}}{{\partial {\bm{\hat \eta }}_{{\rm{IRS}},j}^{{\rm{II}}}}}} \right\}$. Thus, its CRB can also be obtained. 

Note that for Case 2, similar insights in Theorem 2 as well as Remarks 1 and 2 in Case 1 can be observed since it also has two variables in \eqref{approx_2}, which are not repetitively discussed here for brevity.


\section{Single Target: 3D Localization Construction}
After obtaining the estimated angles $\hat \mu _{{\rm{B2T}}}^{\rm{D}}$, $\hat \nu _{{\rm{B2T}}}^{\rm{D}}$, $\hat \mu _{{\rm{I2T}}}^{\rm{D}}$, and $\hat \nu _{{\rm{I2T}}}^{\rm{D}}$ from Sections III and IV, we next calculate the target's 3D location. It follows from the geometric relationship that 
\begin{align}
	\left\{ \begin{array}{l}
		\hat \mu _{{\rm{B2T}}}^{\rm{D}} = \frac{{{y_{\rm{T}}} - {y_{{\rm{BS}}}}}}{{{d_{{\rm{B2T}}}}}} = \frac{{{y_{\rm{T}}} - {y_{{\rm{BS}}}}}}{{\sqrt {{{\left( {{x_{\rm{T}}} - {x_{{\rm{BS}}}}} \right)}^2}{\rm{ + }}{{\left( {{y_{\rm{T}}} - {y_{{\rm{BS}}}}} \right)}^2} + {{\left( {{z_{\rm{T}}} - {z_{{\rm{BS}}}}} \right)}^2}} }},\\
		\hat \nu _{{\rm{B2T}}}^{\rm{D}} = \frac{{{z_{\rm{T}}} - {z_{{\rm{BS}}}}}}{{{d_{{\rm{B2T}}}}}} = \frac{{{z_{\rm{T}}} - {z_{{\rm{BS}}}}}}{{\sqrt {{{\left( {{x_{\rm{T}}} - {x_{{\rm{BS}}}}} \right)}^2}{\rm{ + }}{{\left( {{y_{\rm{T}}} - {y_{{\rm{BS}}}}} \right)}^2} + {{\left( {{z_{\rm{T}}} - {z_{{\rm{BS}}}}} \right)}^2}} }},\\
		\hat \mu _{{\rm{I2T}}}^{\rm{D}} = \frac{{{y_{\rm{T}}} - {y_{\rm{I}}}}}{{{d_{{\rm{I2T}}}}}} = \frac{{{y_{\rm{T}}} - {y_{\rm{I}}}}}{{\sqrt {{{\left( {{x_{\rm{T}}} - {x_{\rm{I}}}} \right)}^2}{\rm{ + }}{{\left( {{y_{\rm{T}}} - {y_{\rm{I}}}} \right)}^2} + {{\left( {{z_{\rm{T}}} - {z_{\rm{I}}}} \right)}^2}} }},\\
		\hat \nu _{{\rm{I2T}}}^{\rm{D}} = \frac{{{z_{\rm{T}}} - {z_{\rm{I}}}}}{{{d_{{\rm{I2T}}}}}} = \frac{{{z_{\rm{T}}} - {z_{\rm{I}}}}}{{\sqrt {{{\left( {{x_{\rm{T}}} - {x_{\rm{I}}}} \right)}^2}{\rm{ + }}{{\left( {{y_{\rm{T}}} - {y_{\rm{I}}}} \right)}^2} + {{\left( {{z_{\rm{T}}} - {z_{\rm{I}}}} \right)}^2}} }},
	\end{array} \right. \label{angletolocation}
\end{align}
After some algebraic operations, \eqref{angletolocation} can be transformed into 
\begin{align}
\left\{ {\begin{array}{*{20}{l}}
	{{d_{{\rm{B2T}}}} = \frac{{\hat \nu _{{\rm{I2T}}}^{\rm{D}}\left( {{y_{{\rm{BS}}}} - {y_{\rm{I}}}} \right) - \hat \mu _{{\rm{I2T}}}^{\rm{D}}\left( {{z_{{\rm{BS}}}} - {z_{\rm{I}}}} \right)}}{{\hat \mu _{{\rm{I2T}}}^{\rm{D}}\hat \nu _{{\rm{B2T}}}^{\rm{D}} - \hat \mu _{{\rm{B2T}}}^{\rm{D}}\hat \nu _{{\rm{I2T}}}^{\rm{D}}}}}\\
	{{d_{{\rm{I2T}}}} = \frac{{\hat \nu _{{\rm{B2T}}}^{\rm{D}}\left( {{y_{\rm{I}}} - {y_{{\rm{BS}}}}} \right) - \hat \mu _{{\rm{B2T}}}^{\rm{D}}\left( {{z_{\rm{I}}} - {z_{{\rm{BS}}}}} \right)}}{{\hat \mu _{{\rm{B2T}}}^{\rm{D}}\hat \nu _{{\rm{I2T}}}^{\rm{D}} - \hat \mu _{{\rm{I2T}}}^{\rm{D}}\hat \nu _{{\rm{B2T}}}^{\rm{D}}}}}\\
	{{y_{\rm{T}}} = {y_{{\rm{BS}}}} + \hat \mu _{{\rm{B2T}}}^{\rm{D}}{d_{{\rm{B2T}}}}}\\
	{{z_{\rm{T}}} = {z_{{\rm{BS}}}} + \hat \nu _{{\rm{B2T}}}^{\rm{D}}{d_{{\rm{B2T}}}}}
	\end{array}} \right.
\end{align}
After determining the solutions ${d_{{\rm{B2T}}}}$, ${d_{{\rm{I2T}}}}$, ${y_{\rm{T}}}$, and ${z_{\rm{T}}}$, we next compute ${x_{\rm{T}}}$.  Note that ${x_{\rm{T}}}$ should satisfy the following two equations:
\begin{align}
\left\{ \begin{array}{l}
	{\left( {{x_{\rm{T}}} - {x_{{\rm{BS}}}}} \right)^2} = d_{{\rm{B2T}}}^2 - {\left( {{y_{\rm{T}}} - {y_{{\rm{BS}}}}} \right)^2} - {\left( {{z_{\rm{T}}} - {z_{{\rm{BS}}}}} \right)^2},\\
	{\left( {{x_{\rm{T}}} - {x_{\rm{I}}}} \right)^2} = d_{{\rm{I2T}}}^2 - {\left( {{y_{\rm{T}}} - {y_{\rm{I}}}} \right)^2} - {\left( {{z_{\rm{T}}} - {z_{\rm{I}}}} \right)^2}.
\end{array} \right. \label{equation_x}
\end{align}
By considering the fact that only the target residing in the front half-space of IRS can be illuminated, which indicates that $x_{\rm T} \ge x_{\rm I}$. Based on this, by solving  \eqref{equation_x}, we can obtain
\begin{align}
\left\{ \begin{array}{l}
	{x_{\rm{T}}} = {x_{{\rm{BS}}}} \pm \sqrt {d_{{\rm{B2T}}}^2 - {{\left( {{y_{\rm{T}}} - {y_{{\rm{BS}}}}} \right)}^2} - {{\left( {{z_{\rm{T}}} - {z_{{\rm{BS}}}}} \right)}^2}}, \\
	{x_{\rm{T}}} = {x_{\rm{I}}} + \sqrt {d_{{\rm{I2T}}}^2 - {{\left( {{y_{\rm{T}}} - {y_{\rm{I}}}} \right)}^2} - {{\left( {{z_{\rm{T}}} - {z_{\rm{I}}}} \right)}^2}}. 
\end{array} \right.
\end{align}
Next, we can compare two points that are the closest to a given point and choose one as the estimated $x_{\rm T}$. To be specific, we adopt the following least squares  criterion: 
\begin{align}
&{{ x}_{\rm{T}}} = \notag\\
&\mathop {\arg \min }\limits_{{x_{\rm{T}}}} \left\{ {\left| {{x_{\rm{T}}} - {x_{\rm{I}}} - \sqrt {d_{{\rm{I2T}}}^2 - {{\left( {{y_{\rm{T}}} - {y_{\rm{I}}}} \right)}^2} - {{\left( {{z_{\rm{T}}} - {z_{\rm{I}}}} \right)}^2}} } \right|} \right\}, \label{single_location}
\end{align}
where ${x_{\rm{T}}} \in \left\{ {{x_{{\rm{T,1}}}},{x_{{\rm{T,2}}}}} \right\}$, ${x_{{\rm{T,1}}}} = {x_{{\rm{BS}}}} + \sqrt {d_{{\rm{B2T}}}^2 - {{\left( {{y_{\rm{T}}} - {y_{{\rm{BS}}}}} \right)}^2} - {{\left( {{z_{\rm{T}}} - {z_{{\rm{BS}}}}} \right)}^2}} $, and ${x_{{\rm{T,2}}}} = {x_{{\rm{BS}}}} - \sqrt {d_{{\rm{B2T}}}^2 - {{\left( {{y_{\rm{T}}} - {y_{{\rm{BS}}}}} \right)}^2} - {{\left( {{z_{\rm{T}}} - {z_{{\rm{BS}}}}} \right)}^2}} $.
\begin{figure}[!t]
	\centerline{\includegraphics[width=3.2in]{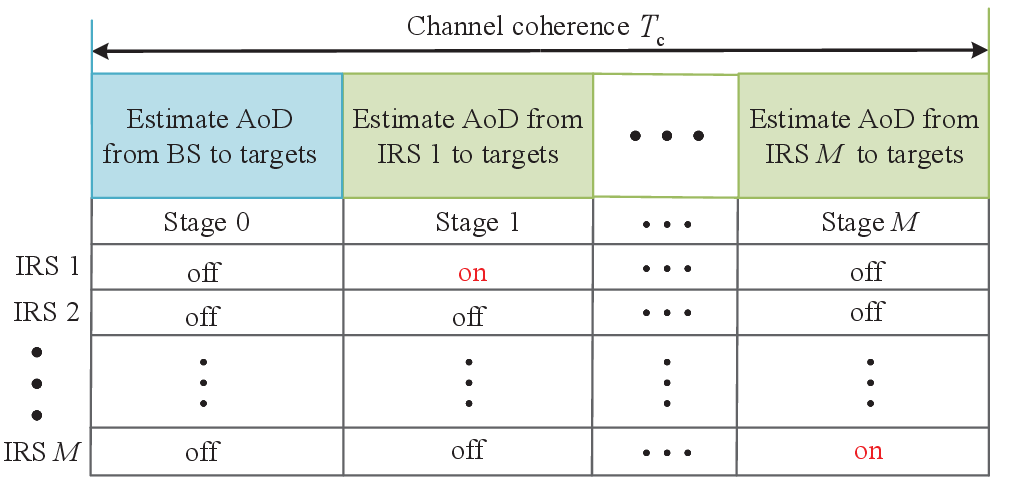}}
	\caption{Illustration of localization protocol for multi-target with multiple IRSs.} \label{fig4}
		\vspace{-0.5cm}
\end{figure}
\section{Multiple Target Localization}
For multi-target localization, one IRS is not feasible since there are $K!$ pairs of locations for $K$ targets. To estimate multiple targets, multiple IRSs are required as shown in Fig.~\ref{fig1}.  To distinguish the composite signals received at the BS from different paths reflected by different IRSs and suppress the echo signals reflected by different IRSs, an \textit{IRS adaptive sensing protocol} is proposed in Fig.~\ref{fig4}. In Fig.~\ref{fig4}, the channel coherence time $T$ is divided into $M+1$ stages, where all the IRSs are turned off in stage 0, while only the  $m$-th IRS is turned on in stage $m$.  Based on this protocol, the interplay between multiple IRSs does not exist, which significantly simplifies the signal processing design for target localization. As a result, the proposed algorithm for one target location estimation can also be applicable to the case with multiple targets with some moderate modifications. These steps  are described  as follows: 
\subsubsection{BS-target DoA Estimation} 
Since the BS beams are orthogonal over time in \eqref{BS_beam}, it indicates that the signals reflected by different targets are not coherent provided that the number of samples is larger than that of BS antennas.  Thus,  we can directly apply the MUSIC algorithm as in \eqref{MUSIC_stageI} and choose the first $K$ solutions that maximize the objective function of \eqref{MUSIC_stageI}   rather than one, which corresponds to $K$  DoAs from the BS to $K$ targets.
\subsubsection{IRS-target DoA Estimation} 
There are two on-grid beam scanning methods to estimate multi-target DoA. The first method is 
similar to \eqref{direction_stageII}, we can choose the first $K$  solutions that maximize the  objective function of \eqref{direction_stageII}   rather than one, which corresponds to $K$  DoAs  from the IRS to $K$ targets. However, there are $K!$ possible solutions. To distinguish the true one, we should substitute each one (note that each solution consists of $K$ target DoAs) into ${\left| {\tilde y_{{\rm{BS}}}^{{\rm{II}}}\left[ {i,j} \right]} \right|^2}$ and compute each target's corresponding value and then sum $K$ results. Then, we choose the largest one which corresponds to $K$ true target DoAs.  The second method is directly applying joint azimuth DoA and elevation DoA estimation as   $\left( {t_{{\rm{index}}}^{y,{\rm{opt}}},t_{{\rm{index}}}^{z,{\rm{opt}}}} \right) = \mathop {\arg \max }\limits_{i \in T_2^y,j \in T_2^z} {\left| {\tilde y_{{\rm{BS}}}^{{\rm{II}}}\left[ {i,j} \right]} \right|^2}$ but at the cost of substantial samples, i.e., $T_2 = T_2^yT_2^z$.

\subsubsection{Direction Pair Matching and Location Estimation} 
To determine multi-target locations uniquely, at least $2$ IRSs are required.  In the following, we will provide a heuristic algorithm to uniquely determine multi-target locations. Since there are  $M$ IRSs, we have  $\frac{{M!}}{{2!\left( {M - 2} \right)!}}$  pairs of two IRSs. 
For each IRS pair, there are $K!$ possible multi-target locations that can be obtained based on \eqref{angletolocation}-\eqref{single_location}. 
Let ${{\bf{q}}_{{\rm T},j}^m} = {\left[ {{\bf{q}}_{{\rm{T,}}{t_1}}^T, \ldots ,{\bf{q}}_{{\rm{T,}}{t_K}}^T} \right]^T}\in {{\mathbb C}^{3K \times 1}}$  denote the $j$-th candidate multi-target locations observed by the $m$-th IRS pair, $j=1,\ldots, K!, m = 1, \ldots ,\frac{{M!}}{{2!\left( {M - 2} \right)!}}$. Based on multi-target locations ${{\bf{q}}_{{\rm T},j}^m}$ and BS location, we can calculate DoAs from the BS to the targets, denoted by ${{\bm{\omega }}_j^m}\in {{\mathbb C}^{2K \times 1}}$.  Then, we  compare the calculated DoAs  ${{\bm{\omega }}_j}\in {{\mathbb C}^{2K \times 1}}$ with estimated DoAs, denoted by ${{\bm{\bar \omega }}}\in {{\mathbb C}^{2K \times 1}}$ (note that there are also $K!$ candidates for ${{\bm{\bar \omega }}}$), at the BS, and  choose  the optimal candidate  DoAs from $K!$ candidate DoAs  based on the following least squares criterion:
\begin{align}
{\bm{\omega }}^m = \mathop {\arg \min }\limits_{j,\forall {\bm{\bar \omega }}} \left| {{{\bm{\omega }}_j^m} - {\bm{\bar \omega }}} \right|.
\end{align}
After obtaining ${\bm{\omega }}^m$, we can reconstruct the locations of targets, denoted by ${{\bf{q}}_{{\rm T},j}^m}$. Then, we average  $\frac{{M!}}{{2!\left( {M - 2} \right)!}}$ reconstructed target locations as 
\begin{align}
{\bf{\hat q}}_{\rm{T}}^{} = \frac{{2!\left( {M - 2} \right)!}}{{M!}}\sum\limits_{m = 1}^{\frac{{M!}}{{2!\left( {M - 2} \right)!}}} {{\bf{q}}_{\rm{T}}^m}.
\end{align}

\section{Numerical Results}
In this section, we evaluate the performance of our proposed IRS-aided 3D localization scheme via numerical results.   The system carrier frequency is set as $750~{\rm MHz}$.
The  estimation  performance of  DoA direction is evaluated by ${\rm{RMS}}{{\rm{E}}_\theta } \overset{\triangle}{=} \sqrt {{\mathbb E}\left\{ {{{\left| {\theta  - \hat \theta } \right|}^2}} \right\}} $, where ${\hat \theta }$ stands for the estimate of $\theta $,   and that of target location is  evaluated by ${\rm{RMS}}{{\rm{E}}_{{{\bf{q}}_{\rm{T}}}}} \overset{\triangle}{=} \sqrt {\frac{1}{K}{\mathbb E}\left\{ {\sum\limits_{k = 1}^K {{{\left| {{{\bf{q}}_{{\rm{T,}}k}} - {{{\bf{\hat q}}}_{{\rm{T,}}k}}} \right|}^2}} } \right\}} $, where ${{{{\bf{\hat q}}}_{{\rm T},k}}}$ stands for the estimate of ${{{\bf{q}}_{{\rm T},k}}}$. Unless specified otherwise, the other system parameters are given as follows:  ${{\kappa _{{\rm{RCS}},k}}}=7~{\rm dBsm}, \forall k$, $T_2^y = T_2^z = T_2/2$, $N_{{\rm{BS}}}^y = N_{{\rm{BS}}}^z = 20$, ${{\bar N}_{\rm{r}}} = N_{\rm{r}}^y = N_{\rm{r}}^z$, and ${\sigma ^2} =  - 80~ {\rm dBm}$

\begin{figure}[!t]
	\centerline{\includegraphics[width=3in]{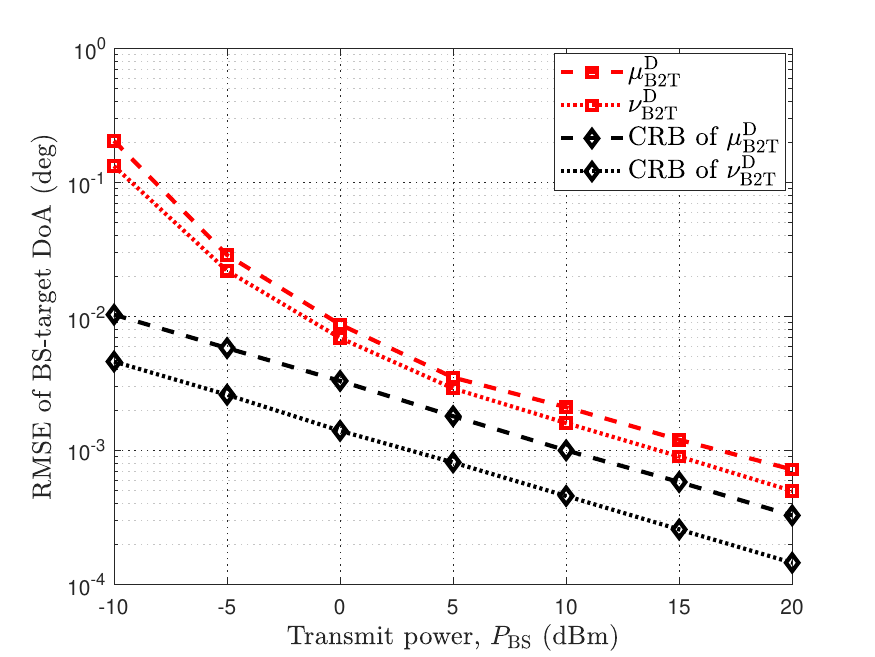}}
	\caption{RMSE of BS-target DoA versus  $P_{\rm BS}$ under $T_1=24$.} \label{fig6}
		\vspace{-0.4cm}
\end{figure}
\begin{figure}[!t]
	\centerline{\includegraphics[width=3in]{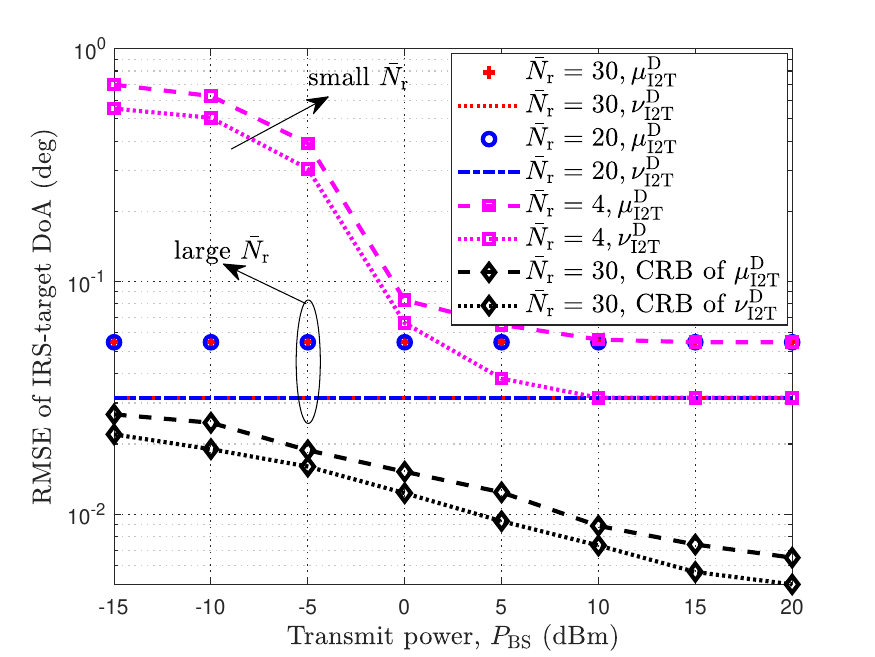}}
	\caption{RMSE of IRS-target DoA versus  $P_{\rm BS}$ under $T_2=20$ and different ${\bar N}_{\rm r}$.} \label{fig7}
		\vspace{-0.4cm}
\end{figure}

\begin{figure}[!t]
	\centerline{\includegraphics[width=3in]{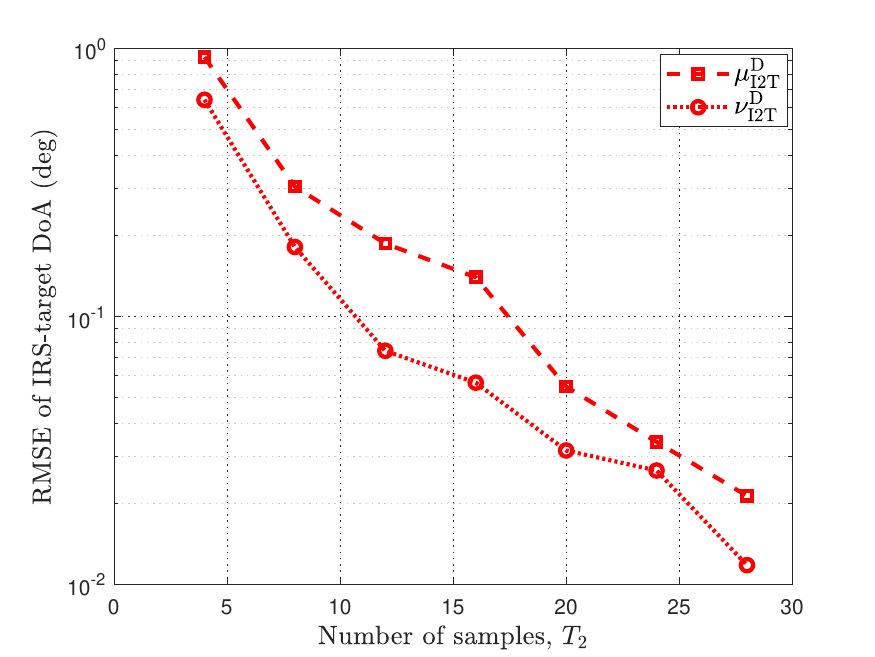}}
	\caption{RMSE of  IRS-target DoA versus  $T_2$ under  $P_{\rm BS}=20~{\rm dBm}$ and ${\bar N}_{\rm r}=30$.} \label{fig8}
	\vspace{-0.4cm}
\end{figure}

\subsection{Single Target Localization}
In this subsection, we consider the single-target localization case, where  the BS, IRS, and  target 
 are located at ${{\bf{q}}_{{\rm{BS}}}} = {\left[ {0,0,5} \right]^T}$ (m), ${{\bf{q}}_{\rm{I}}} = {\left[ { - 20,0,3} \right]^T}$ (m), and ${{\bf{q}}_{\rm{T}}} = {\left[ { - 20,2,0} \right]^T}$ (m), respectively.

In Fig.~\ref{fig6}, we illustrate the RMSE of BS-target DoA versus transmit power $P_{\rm BS}$. It is observed that the RMSEs of two DoAs, namely,  ${\mu _{{\rm{B2T}}}^{\rm{D}}}$ and ${\nu _{{\rm{B2T}}}^{\rm{D}}}$, are monotonically decreasing with the increase of $P_{\rm BS}$. In addition, the derived CRB in Lemma 1 is   plotted. It is observed that although there is a large gap between the  MUSIC estimation error and CRB in a low-power region, the performance obtained by  MUSIC is approaching the CRB as $P_{\rm BS}$ increases. This demonstrates the effectiveness of the MUSIC method for 3D sensing.

In Fig.~\ref{fig7}, we study the RMSE of IRS-target DoA versus  $P_{\rm BS}$ under $T_2=20$ and different number of IRS reflecting elements ${\bar N}_{\rm r}=30$,  ${\bar N}_{\rm r}=20$, and ${\bar N}_{\rm r}=4$. It is observed that the RMSEs of ${\mu _{{\rm{I2T}}}^{\rm{D}}}$ and ${\mu _{{\rm{I2T}}}^{\rm{D}}}$ first decrease with the increase of $P_{\rm BS}$ and then remain constant as $P_{\rm BS}\ge 10~{\rm dBm}$  for ${\bar N}_{\rm r}=4$. This is because the increase of power in a low-power region could combat the impact of background noise, thereby reducing the RMSE. While as $P_{\rm BS}$ further increases, the impact of background noise could be neglected and the RMSE is only related to the on-grid beam scanning refinement. In addition, in a low power region, i.e., $P_{\rm BS}\le 10~{\rm dBm}$, the proposed scheme with ${\bar N}_{\rm r}=20$  achieves a lower RMSE than that with  ${\bar N}_{\rm r}=4$. This is because the IRS can provide a larger passive beamforming gain to combat the noise power.  Moreover, 
it is observed that  the RMSEs of ${\mu _{{\rm{I2T}}}^{\rm{D}}}$ and ${\mu _{{\rm{I2T}}}^{\rm{D}}}$ remain constant regardless of   $P_{\rm BS}$ for a large ${\bar N}_{\rm r}$, e.g., ${\bar N}_{\rm r}=20$. This is because a larger ${\bar N}_{\rm r}$ provides enough passing beamforming gain to against the noise. Furthermore, we observe that increasing ${\bar N}_{\rm r}$ from $20$ to $30$, the RMSE keeps constant. This is because the RMSE only depends on 
 the on-grid beam scanning refinement when the passive gain (or received power) is large. To see it more clearly, we plot the RMSE of  IRS-target DoA versus $T_2$ in Fig.~\ref{fig8}. We can see that a larger $T_2$  leads to a smaller RMSE due to a higher on-grid beam scanning refinement.

\begin{figure}[!t]
 	\centerline{\includegraphics[width=3in]{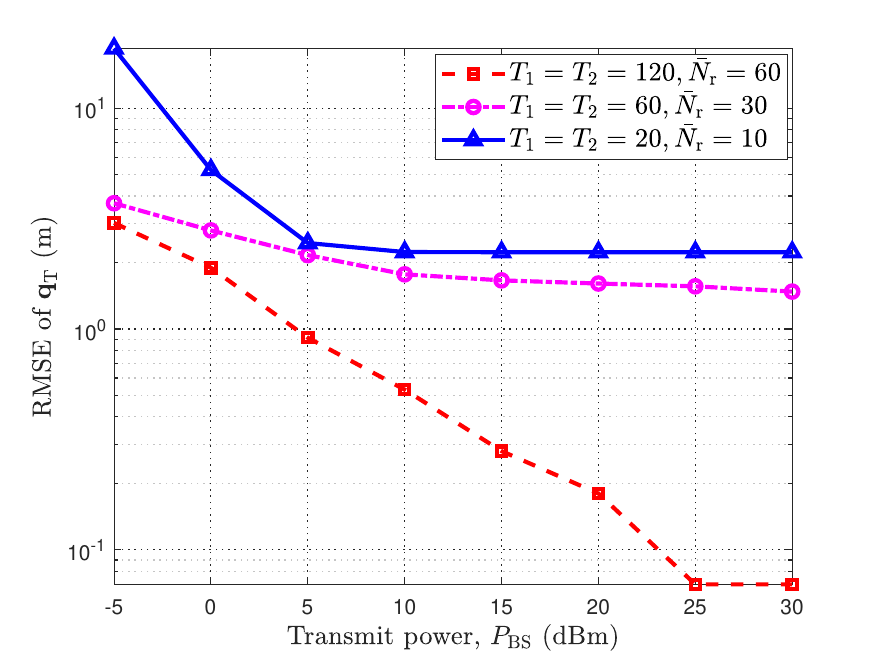}}
 	\caption{Single target location estimation error versus  $P_{\rm BS}$ under different $T_1$, $T_2$, and  ${\bar N}_{\rm r}=30$.} \label{fig9}
 	\vspace{-0.4cm}
 \end{figure}

In Fig.~\ref{fig9}, we study the single-target location estimation error versus  $P_{\rm BS}$ under different numbers of samples and IRS reflecting elements. It is observed that a larger  $T_1$   and ${\bar N}_{\rm r}=30$ leads to a smaller target location estimation error due to the refined on-grid beam scanning resolution as well as improved passive beamforming gain. In addition, we observe that the target location estimation error first decreases with the increase of $P_{\rm BS}$ and then remains constant as $P_{\rm BS}$ becomes larger. This is because the target location estimation error is affected by two factors: received power at the BS and on-grid beam scanning resolution. Specifically,  at the low-power region, the increase of power could combat the background noise, thus increasing the localization accuracy. As the power further improves, the impact of noise can be neglected and the system performance is only related to the on-grid beam scanning resolution. This observation is consistent with Fig.~\ref{fig7}. Furthermore, with a larger  $T_1$  (or $T_2$) and ${\bar N}_{\rm r}=30$, the sub-meter   accuracy for the single-target case can be achieved.
 
 \begin{figure}[!t]
 	\centerline{\includegraphics[width=3in]{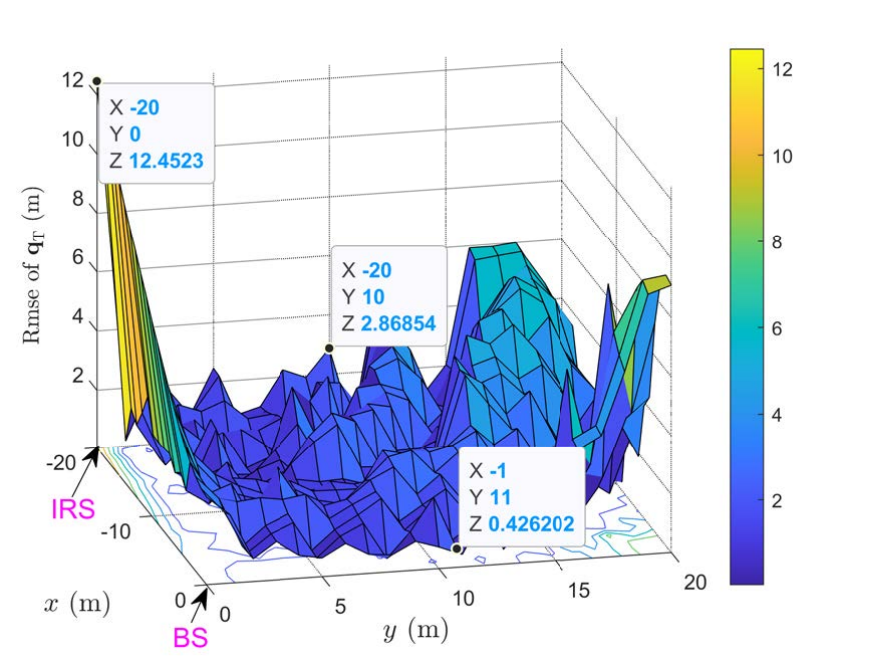}}
 	\caption{Location estimation error for different target locations in the considered area $20~ {\rm m} \times 20~{\rm m}$ under  $P_{\rm BS}=30~{\rm dBm}$, $T_1=T_2=60$, and ${\bar N}_{\rm r}=30$.} \label{fig10}
 	\vspace{-0.4cm}
 \end{figure}
 In Fig.~\ref{fig10}, we further study the location estimation error of the target located in a   $20~ {\rm m} \times 20~{\rm m}$ area. We can observe that for the target  located in the line of $x$-axis, i.e., $y=0$, the location estimation error is very high. This is because both BS and IRS are located in $x$-axis,  the azimuth DoA between the target and the IRS/BS is zero, which indicates that the target is hard to illuminate by the IRS and the BS, thus resulting in a large location estimation error. In the other regions that are not far from both the BS and  IRS, the  error is as small as expected. Furthermore, we can see that in the regions that are far from both the BS and IRS, the location estimation error is large due to the high signal path loss attenuation.

\subsection{Multiple Target Localization}
 In this subsection, we consider the multi-target localization case, where  we consider one BS, three IRSs, and three targets, which are located at  ${{\bf{q}}_{{\rm{BS}}}} = {\left[ {0,0,5} \right]^T}$ (m), ${{\bf{q}}_{{\rm I},1}} = {\left[ { - 20,0,3} \right]^T}$ (m), ${{\bf{q}}_{{\rm I},2}} = {\left[ { - 10,0,3} \right]^T}$ (m), ${{\bf{q}}_{{\rm I},3}} = {\left[ { - 5,0,3} \right]^T}$ (m), ${{\bf{q}}_{{\rm T},1}} = {\left[ { - 10,10,0} \right]^T}$ (m), ${{\bf{q}}_{{\rm T},2}} = {\left[ { - 20,2,0} \right]^T}$ (m), and ${{\bf{q}}_{{\rm T},3}} = {\left[ { - 5,10,0} \right]^T}$ (m).
 
 \begin{figure}[thbp!]
 	\centering
 	\begin{minipage}[t]{1.0\linewidth}
	\centering
	\begin{tabular}{@{\extracolsep{\fill}}c@{}@{\extracolsep{\fill}}}
		\includegraphics[width=3in]{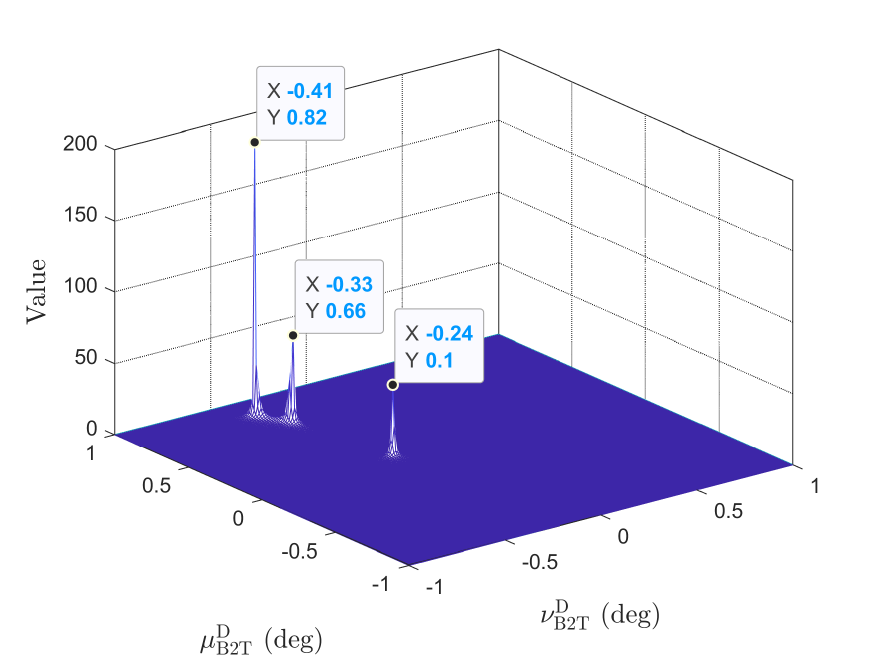}\\
		(a) BS-target DoA \\
	\end{tabular}
\end{minipage}
 	\begin{minipage}[t]{1.0\linewidth}
 		\centering
 		\begin{tabular}{@{\extracolsep{\fill}}c@{}@{\extracolsep{\fill}}}
 			\includegraphics[width=3in]{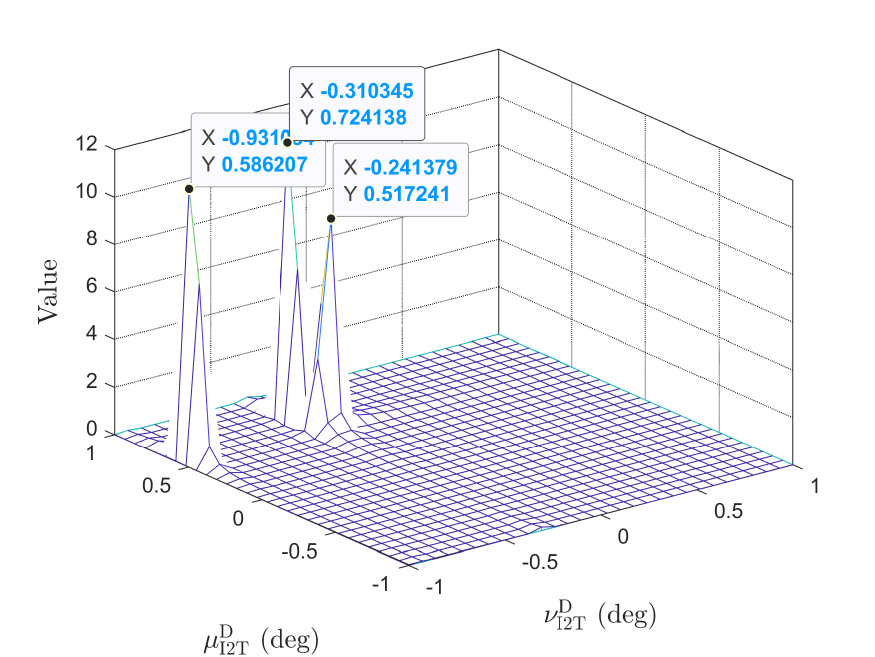}\\
 			(b) IRS-target DoA\\
 		\end{tabular}
 	\end{minipage}
 	\caption{Effectiveness of  two proposed algorithms for multi-target DoA estimation.}
 	\label{fig11}
 \end{figure}

In Fig.~\ref{fig11}, we study the DoA estimation for both the BS-target link and IRS-target link by using the MUSIC algorithm and the on-grid beam scanning algorithm, respectively. Based on the locations of the  BS, IRS, and targets, the   three BS-target DoAs are  $\left( {{\rm{ - 0}}{\rm{.4082}},{\rm{0}}{\rm{.8165}}} \right)$ (deg), $\left( {{\rm{ - 0}}{\rm{.3333}},{\rm{0}}{\rm{.6667}}} \right)$ (deg), and $\left( {{\rm{ - 0}}{\rm{.2414}},{\rm{0}}{\rm{.0966}}} \right)$ (deg). It can be seen that the estimated three BS-target DoAs in Fig.~\ref{fig10}(a) agree well with the true DoAs, demonstrating the effectiveness of the MUSIC algorithm. In addition, the  three IRS-target DoAs are  $\left( {{\rm{ - 0}}{\rm{.2075,0}}{\rm{.6917}}} \right)$ (deg), $\left( {{\rm{ - 0}}{\rm{.8321}},{\rm{0}}{\rm{.5547}}} \right)$ (deg), and $\left( {{\rm{ - 0}}{\rm{.1642}},{\rm{0}}{\rm{.5472}}} \right)$ (deg). Note that Fig. 2 (b) depicts the IRS-target DoA, i.e., $\mu  = \mu _{{\rm{B2I}}}^{\rm{A}}{\rm{ + }}\mu _{{\rm{I2T}}}^{\rm{D}}$  and $\nu  = \nu _{{\rm{B2I}}}^{\rm{A}}{\rm{ + }}\nu _{{\rm{I2T}}}^{\rm{D}}$,  containing the BS-IRS DoA $\left( {{\rm{ - 0}}{\rm{.0995}},0} \right)$ (deg). We   see that the estimated DoAs also match well with the true DoAs, which demonstrates the effectiveness of the on-grid beam scanning algorithm.

 \begin{figure}[htbp]
 	\centerline{\includegraphics[width=3in]{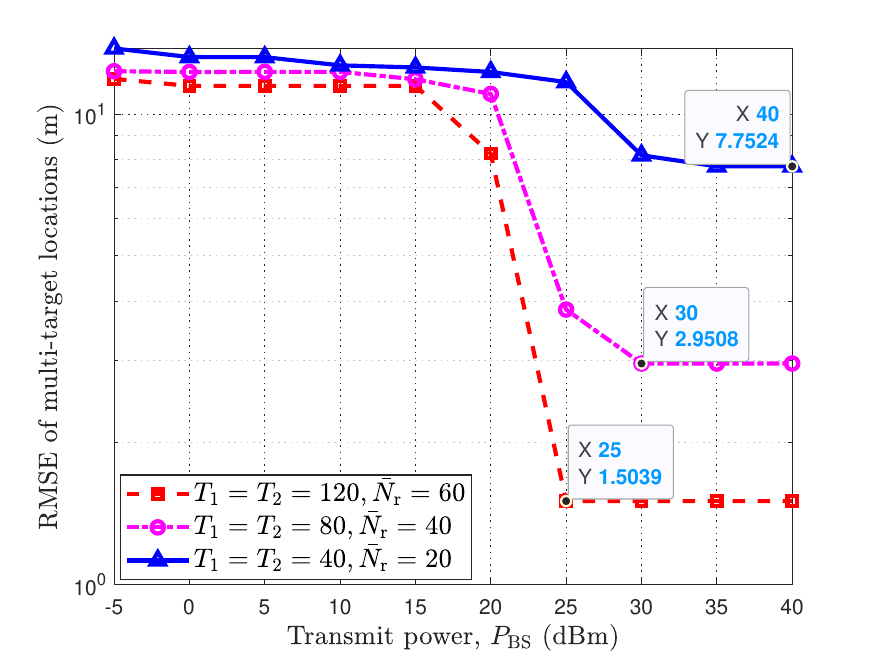}}
 	\caption{Multi-target location estimation error versus  $P_{\rm BS}$ under different samples and number of IRS reflecting elements.} \label{fig12}
 	\vspace{-0.4cm}
 \end{figure}

In Fig.~\ref{fig12}, we show the multi-target location estimation error versus  $P_{\rm BS}$ under different numbers of samples and IRS reflecting elements. It is observed that the RMSE of multi-target locations firstly decreases and then remains unchanged as $P_{\rm BS}$ increases for three schemes. The reason is that the location estimation error for the multiple targets is also affected by the received power at the BS and on-grid beam scanning resolution. Thus, Fig.~\ref{fig12} presents a similar behaviour as in  Fig.~\ref{fig9} for the single-target case. Furthermore, we can see that with a larger  $T_1$, $T_2$, and ${\bar N}_{\rm r}=30$,  meter  positioning accuracy for the multi-target case can be achieved.
\section{Conclusion}
In this paper, we leveraged one BS and multiple IRSs to estimate the 3D locations of multiple targets by exploiting two different DoAs of the BS-target and the IRS-target links. We presented an IRS adaptive protocol by alternately turning on/off the IRS to eliminate the interference reflected by the other IRSs. To fully unveil the fundamental limit of IRS for localization, we first studied the single IRS to estimate the location of a single target.  The MUSIC algorithm was applied to estimate the DoA of the BS-target link and the on-grid beam scanning algorithm was proposed to estimate the IRS-target link.  Particularly, the impact of echo signals reflected by IRS from different paths on sensing performance was analyzed. Besides, we unveiled that the single-beam IRS is not capable of sensing, while a multi-beam IRS is required. Then, we extended the single-target case to the multi-target case, and an effective matching algorithm for multi-target localization was proposed.  Simulation results demonstrated that the proposed scheme can achieve   sub-meter  accuracy.

\appendices
\section{Proof of Lemma 1} \label{appendix_lemma1}
Recall that ${\bf{R}}_{{\rm{BS}}}^{\rm{I}}$ in \eqref{covariance} is irrelevant to estimation parameters ${{\bm{\eta }}_{{\rm{BS}}}^{\rm{I}}}$, thus we have $\frac{{\partial {\bf{R}}_{{\rm{BS}}}^{\rm{I}}}}{{\partial {\bm{\eta }}_{{\rm{BS,}}i}^{\rm{I}}}} = {{\bf{0}}_{{N_{{\rm{BS}}}}{T_1}}},i = 1,2,3$. Then, 
Substituting it  into \eqref{Fisherexpresion}, we have 
\begin{align}
{\left[ {{\bf{F}}_{{\rm{BS}}}^{\rm{I}}} \right]_{ij}}=\frac{2}{{{\sigma ^2}}}{\mathop{\rm Re}\nolimits} \left\{ {\frac{{\partial {\bf{u}}_{{\rm{BS}}}^{{\rm{I,}}H}}}{{\partial {\bm{\eta }}_{{\rm{BS,}}i}^{\rm{I}}}}\frac{{\partial {\bf{u}}_{{\rm{BS}}}^{\rm{I}}}}{{\partial {\bm{\eta }}_{{\rm{BS,}}j}^{\rm{I}}}}} \right\}.
\end{align}
In addition, we have the following identifies:
\begin{align}
&\frac{{\partial {\bf{u}}_{{\rm{BS}}}^{\rm{I}}}}{{\partial \mu _{{\rm{B2T}}}^{\rm{D}}}}={\beta _{{\rm{BTB}}}}{\rm{vec}}\left( {{{{\bf{\dot A}}}_{\mu _{{\rm{B2T}}}^{\rm{D}}}}{\bf{W}}_{{\rm{BS}}}^{\rm{I}}} \right),\\
&\frac{{\partial {\bf{u}}_{{\rm{BS}}}^{\rm{I}}}}{{\partial \nu _{{\rm{B2T}}}^{\rm{D}}}}={\beta _{{\rm{BTB}}}}{\rm{vec}}\left( {{{{\bf{\dot A}}}_{\nu _{{\rm{B2T}}}^{\rm{D}}}}{\bf{W}}_{{\rm{BS}}}^{\rm{I}}} \right),\\
&\frac{{\partial {\bf{u}}_{{\rm{BS}}}^{\rm{I}}}}{{\partial {\bm{\beta }}_{{\rm{BTB}}}^T}} = \left[ {1{\kern 1pt} {\kern 1pt} {\kern 1pt} {\kern 1pt} {\kern 1pt} {\kern 1pt} j} \right] \otimes {\rm{vec}}\left( {{{\bf{A}}_{\mu _{{\rm{B2T}}}^{\rm{D}}}}{\bf{W}}_{{\rm{BS}}}^{\rm{I}}} \right).
\end{align}
Based on these, $f_{\mu _{{\rm{B2T}}}^{\rm{D}}\mu _{{\rm{B2T}}}^{\rm{D}}}^{\rm{I}}$, $f_{\mu _{{\rm{B2T}}}^{\rm{D}}\nu _{{\rm{B2T}}}^{\rm{D}}}^{\rm{I}}$, ${\bf{f}}_{\mu _{{\rm{B2T}}}^{\rm{D}}{\bm{\beta }}_{{\rm{BTB}}}^T}^{\rm{I}}$,
$f_{\nu _{{\rm{B2T}}}^{\rm{D}}\nu _{{\rm{B2T}}}^{\rm{D}}}^{\rm{I}}$, ${\bf{f}}_{\nu _{{\rm{B2T}}}^{\rm{D}}{\bm{\beta }}_{{\rm{BTB}}}^T}^{\rm{I}}$, and ${\bf{F}}_{{\bm{\beta }}_{{\rm{BTB}}}^T{\bm{\beta }}_{{\rm{BTB}}}^T}^{{\rm{I}}}$ can be calculated as 
\begin{align}
&f_{\mu _{{\rm{B2T}}}^{\rm{D}}\mu _{{\rm{B2T}}}^{\rm{D}}}^{\rm{I}}{\rm{ = }}\frac{2}{{{\sigma ^2}}}{\mathop{\rm Re}\nolimits} \left\{ {\frac{{\partial {\bf{u}}_{{\rm{BS}}}^{{\rm{I,}}H}}}{{\partial \mu _{{\rm{B2T}}}^{\rm{D}}}}\frac{{\partial {\bf{u}}_{{\rm{BS}}}^{\rm{I}}}}{{\partial \mu _{{\rm{B2T}}}^{\rm{D}}}}} \right\}\notag\\
&= \frac{{2{{\left| {{\beta _{{\rm{BTB}}}}} \right|}^2}}}{{{\sigma ^2}}}{\mathop{\rm Re}\nolimits} \left\{ {{\rm{ve}}{{\rm{c}}^H}\left( {{{{\bf{\dot A}}}_{\mu _{{\rm{B2T}}}^{\rm{D}}}}{\bf{W}}_{{\rm{BS}}}^{\rm{I}}} \right){\rm{vec}}\left( {{{{\bf{\dot A}}}_{\mu _{{\rm{B2T}}}^{\rm{D}}}}{\bf{W}}_{{\rm{BS}}}^{\rm{I}}} \right)} \right\}\notag\\
& = \frac{{2{{\left| {{\beta _{{\rm{BTB}}}}} \right|}^2}{T_1}}}{{{\sigma ^2}}}{\rm{tr}}\left( {{{{\bf{\dot A}}}_{\mu _{{\rm{B2T}}}^{\rm{D}}}}{{\bf{R}}_{{{\bf{w}}_{{\rm{BS}}}}}}{\bf{\dot A}}_{\mu _{{\rm{B2T}}}^{\rm{D}}}^H} \right),\\
&f_{\mu _{{\rm{B2T}}}^{\rm{D}}\nu _{{\rm{B2T}}}^{\rm{D}}}^{\rm{I}} = \frac{2}{{{\sigma ^2}}}{\mathop{\rm Re}\nolimits} \left\{ {\frac{{\partial {\bf{u}}_{{\rm{BS}}}^{{\rm{I,}}H}}}{{\partial \mu _{{\rm{B2T}}}^{\rm{D}}}}\frac{{\partial {\bf{u}}_{{\rm{BS}}}^{\rm{I}}}}{{\partial \nu _{{\rm{B2T}}}^{\rm{D}}}}} \right\}\notag\\
& = \frac{{2{{\left| {{\beta _{{\rm{BTB}}}}} \right|}^2}}}{{{\sigma ^2}}}{\mathop{\rm Re}\nolimits} \left\{ {{\rm{ve}}{{\rm{c}}^H}\left( {{{{\bf{\dot A}}}_{\mu _{{\rm{B2T}}}^{\rm{D}}}}{\bf{W}}_{{\rm{BS}}}^{\rm{I}}} \right){\rm{vec}}\left( {{{{\bf{\dot A}}}_{\nu _{{\rm{B2T}}}^{\rm{D}}}}{\bf{W}}_{{\rm{BS}}}^{\rm{I}}} \right)} \right\}\notag\\
&= \frac{{2{{\left| {{\beta _{{\rm{BTB}}}}} \right|}^2}}}{{{\sigma ^2}}}{\mathop{\rm Re}\nolimits} \left\{ {{\rm{tr}}\left( {{{{\bf{\dot A}}}_{\nu _{{\rm{B2T}}}^{\rm{D}}}}{{\bf{R}}_{{{\bf{w}}_{{\rm{BS}}}}}}{\bf{\dot A}}_{\mu _{{\rm{B2T}}}^{\rm{D}}}^H} \right)} \right\},\\
&{\bf{f}}_{\mu _{{\rm{B2T}}}^{\rm{D}}{\bm{\beta }}_{{\rm{BTB}}}^T}^{\rm{I}}=\frac{2}{{{\sigma ^2}}}{\mathop{\rm Re}\nolimits} \left\{ {\frac{{\partial {\bf{u}}_{{\rm{BS}}}^{{\rm{I,}}H}}}{{\partial \mu _{{\rm{B2T}}}^{\rm{D}}}}\frac{{\partial {\bf{u}}_{{\rm{BS}}}^{\rm{I}}}}{{\partial {\bm{\beta }}_{{\rm{BTB}}}^T}}} \right\}\qquad \qquad \qquad \qquad \notag\\
&  = \frac{2}{{{\sigma ^2}}}{\rm{Re}}\left\{ {\beta _{{\rm{BTB}}}^H{\rm{ve}}{{\rm{c}}^H}\left( {{{{\bf{\dot A}}}_{\mu _{{\rm{B2T}}}^{\rm{D}}}}{\bf{W}}_{{\rm{BS}}}^{{\rm{Full - I}}}} \right)} \right.\notag\\
&\quad \times \left. {\left( {\left[ {1{\kern 1pt} {\kern 1pt} {\kern 1pt} {\kern 1pt} j} \right] \otimes {\rm{vec}}\left( {{\bf{AW}}_{{\rm{BS}}}^{\rm{I}}} \right)} \right)} \right\}\notag\\
& = \frac{2}{{{\sigma ^2}}}{\rm{Re}}\Big\{ {\beta _{{\rm{BTB}}}^H\left[ {1{\kern 1pt} {\kern 1pt} {\kern 1pt} {\kern 1pt} {\kern 1pt} j} \right] }  \notag\\
&\quad \otimes \left. {\left( {{\rm{ve}}{{\rm{c}}^H}\left( {{{{\bf{\dot A}}}_{\mu _{{\rm{B2T}}}^{\rm{D}}}}{\bf{W}}_{{\rm{BS}}}^{\rm{I}}} \right){\rm{vec}}\left( {{\bf{AW}}_{{\rm{BS}}}^{\rm{I}}} \right)} \right)} \right\}\notag\\
& = \frac{{2{T_1}}}{{{\sigma ^2}}}{\mathop{\rm Re}\nolimits} \left\{ {\beta _{{\rm{BTB}}}^H\left[ {1{\kern 1pt} {\kern 1pt} {\kern 1pt} {\kern 1pt} {\kern 1pt} {\kern 1pt} j} \right]{\rm{tr}}\left( {{\bf{A}}{{\bf{R}}_{{{\bf{w}}_{{\rm{BS}}}}}}{\bf{\dot A}}_{\mu _{{\rm{B2T}}}^{\rm{D}}}^H} \right)} \right\},
\end{align}
\begin{align}
&f_{\nu _{{\rm{B2T}}}^{\rm{D}}\nu _{{\rm{B2T}}}^{\rm{D}}}^{\rm{I}}{\rm{ = }}\frac{2}{{{\sigma ^2}}}{\mathop{\rm Re}\nolimits} \left\{ {\frac{{\partial {\bf{u}}_{{\rm{BS}}}^{{\rm{I,}}H}}}{{\partial \nu _{{\rm{B2T}}}^{\rm{D}}}}\frac{{\partial {\bf{u}}_{{\rm{BS}}}^{\rm{I}}}}{{\partial \nu _{{\rm{B2T}}}^{\rm{D}}}}} \right\}\notag\\
& = \frac{{2{{\left| {{\beta _{{\rm{BTB}}}}} \right|}^2}}}{{{\sigma ^2}}}{\mathop{\rm Re}\nolimits} \left\{ {{\rm{ve}}{{\rm{c}}^H}\left( {{{{\bf{\dot A}}}_{\nu _{{\rm{B2T}}}^{\rm{D}}}}{\bf{W}}_{{\rm{BS}}}^{\rm{I}}} \right){\rm{vec}}\left( {{{{\bf{\dot A}}}_{\nu _{{\rm{B2T}}}^{\rm{D}}}}{\bf{W}}_{{\rm{BS}}}^{\rm{I}}} \right)} \right\}\notag\\
&= \frac{{2{{\left| {{\beta _{{\rm{BTB}}}}} \right|}^2}{T_1}}}{{{\sigma ^2}}}{\rm{tr}}\left( {{{{\bf{\dot A}}}_{\nu _{{\rm{B2T}}}^{\rm{D}}}}{{\bf{R}}_{{{\bf{w}}_{{\rm{BS}}}}}}{\bf{\dot A}}_{\nu _{{\rm{B2T}}}^{\rm{D}}}^H} \right),\\
&{\bf{f}}_{\nu _{{\rm{B2T}}}^{\rm{D}}{\bm{\beta }}_{{\rm{BTB}}}^T}^{\rm{I}}=\frac{2}{{{\sigma ^2}}}{\mathop{\rm Re}\nolimits} \left\{ {\frac{{\partial {\bf{u}}_{{\rm{BS}}}^{{\rm{Full - I,}}H}}}{{\partial \nu _{{\rm{B2T}}}^{\rm{D}}}}\frac{{\partial {\bf{u}}_{{\rm{BS}}}^{{\rm{Full - I}}}}}{{\partial {\bm{\beta }}_{{\rm{BTB}}}^T}}} \right\}\qquad\qquad\qquad\qquad\qquad \notag \\
& = \frac{2}{{{\sigma ^2}}}{\rm{Re}}\Big\{ {\beta _{{\rm{BTB}}}^H\left[ {1{\kern 1pt} {\kern 1pt} {\kern 1pt} {\kern 1pt} {\kern 1pt} j} \right] }  \notag\\
&\quad \otimes \left. {\left( {{\rm{ve}}{{\rm{c}}^H}\left( {{{{\bf{\dot A}}}_{\nu _{{\rm{B2T}}}^{\rm{D}}}}{\bf{W}}_{{\rm{BS}}}^{\rm{I}}} \right){\rm{vec}}\left( {{\bf{AW}}_{{\rm{BS}}}^{\rm{I}}} \right)} \right)} \right\}\notag\\
& = \frac{{2{T_1}}}{{{\sigma ^2}}}{\mathop{\rm Re}\nolimits} \left\{ {\beta _{{\rm{BTB}}}^H\left[ {1{\kern 1pt} {\kern 1pt} {\kern 1pt} {\kern 1pt} {\kern 1pt} {\kern 1pt} j} \right]{\rm{tr}}\left( {{\bf{A}}{{\bf{R}}_{{{\bf{w}}_{{\rm{BS}}}}}}{\bf{\dot A}}_{\nu _{{\rm{B2T}}}^{\rm{D}}}^H} \right)} \right\},\\
&{\bf{F}}_{{\bm{\beta }}_{{\rm{BTB}}}^T{\bm{\beta }}_{{\rm{BTB}}}^T}=\frac{2}{{{\sigma ^2}}}{\mathop{\rm Re}\nolimits} \left\{ {\frac{{\partial {\bf{u}}_{{\rm{BS}}}^{{\rm{I,}}H}}}{{\partial {\bm{\beta }}_{{\rm{BTB}}}^T}}\frac{{\partial {\bf{u}}_{{\rm{BS}}}^{\rm{I}}}}{{\partial {\bm{\beta }}_{{\rm{BTB}}}^T}}} \right\}\qquad \qquad \qquad \qquad \notag\\
& = \frac{2}{{{\sigma ^2}}}{\mathop{\rm Re}\nolimits} \left\{ {\left( {{{\left[ {1{\kern 1pt} {\kern 1pt} {\kern 1pt} {\kern 1pt} {\kern 1pt} {\kern 1pt} j} \right]}^H}\left[ {1{\kern 1pt} {\kern 1pt} {\kern 1pt} {\kern 1pt} {\kern 1pt} {\kern 1pt} j} \right]} \right)} \right.\notag\\
&\quad \left. { \otimes \left( {{\rm{ve}}{{\rm{c}}^H}\left( {{\bf{AW}}_{{\rm{BS}}}^{\rm{I}}} \right){\rm{vec}}\left( {{\bf{AW}}_{{\rm{BS}}}^{\rm{I}}} \right)} \right)} \right\}\notag\\
&= \frac{{2{T_1}}}{{{\sigma ^2}}}{\mathop{\rm Re}\nolimits} \left\{ {\left( {{{\left[ {1{\kern 1pt} {\kern 1pt} {\kern 1pt} {\kern 1pt} {\kern 1pt} {\kern 1pt} j} \right]}^H}\left[ {1{\kern 1pt} {\kern 1pt} {\kern 1pt} {\kern 1pt} {\kern 1pt} {\kern 1pt} j} \right]} \right){\rm{tr}}\left( {{\bf{A}}{{\bf{R}}_{{{\bf{w}}_{{\rm{BS}}}}}}{{\bf{A}}^H}} \right)} \right\}\notag\\
&= \frac{{2{T_1}}}{{{\sigma ^2}}}{\rm{tr}}\left( {{\bf{A}}{{\bf{R}}_{{{\bf{w}}_{{\rm{BS}}}}}}{{\bf{A}}^H}} \right){{\bf{I}}_2}.
\end{align}
This completes the proof of Lemma 1.

\section{Proof of Theorem 1} \label{appendix_Theorem1}
Before proceeding with Theorem 1, we first derive some equations to facilitate the proof. 
 Define $\frac{{\partial {{\bf{u}}}\left( {\mu _{{\rm{B2T}}}^{\rm{D}},N_{{\rm{BS}}}^y} \right)}}{{\partial \mu _{{\rm{B2T}}}^{\rm{D}}}} = {{{\bf{\dot u}}}_{\mu _{{\rm{B2T}}}^{\rm{D}}}}\left( {\mu _{{\rm{B2T}}}^{\rm{D}},N_{{\rm{BS}}}^y} \right)$ and $\frac{{\partial {\bf{u}}\left( {\nu _{{\rm{B2T}}}^{\rm{D}},N_{{\rm{BS}}}^z} \right)}}{{\nu _{{\rm{B2T}}}^{\rm{D}}}} = {\bf{\dot u}}_{\nu _{{\rm{B2T}}}^{\rm{D}}}^{}\left( {\nu _{{\rm{B2T}}}^{\rm{D}},N_{{\rm{BS}}}^z} \right)$.
Based on \eqref{response_vector}, we have the following identities:
\begin{align}
&{\bf{\dot u}}_{\mu _{{\rm{B2T}}}^{\rm{D}}}^H\left( {\mu _{{\rm{B2T}}}^{\rm{D}},N_{{\rm{BS}}}^y} \right){{\bf{\dot u}}_{\mu _{{\rm{B2T}}}^{\rm{D}}}}\left( {\mu _{{\rm{B2T}}}^{\rm{D}},N_{{\rm{BS}}}^y} \right) = \qquad\qquad\quad\qquad\qquad \notag\\
&\qquad\qquad\qquad\qquad\qquad\quad  \frac{{{\pi ^2}}}{4}\sum\limits_{n = 1}^{N_{{\rm{BS}}}^y} {{{\left( {N_{{\rm{BS}}}^y - 2n + 1} \right)}^2}}, \\
&{\bf{\dot u}}_{\nu _{{\rm{B2T}}}^{\rm{D}}}^H\left( {\nu _{{\rm{B2T}}}^{\rm{D}},N_{{\rm{BS}}}^z} \right){{\bf{\dot u}}_{\nu _{{\rm{B2T}}}^{\rm{D}}}}\left( {\nu _{{\rm{B2T}}}^{\rm{D}},N_{{\rm{BS}}}^z} \right) = \notag\\
& \qquad\qquad\qquad\qquad\qquad\quad  \frac{{{\pi ^2}}}{4}\sum\limits_{n = 1}^{N_{{\rm{BS}}}^z} {{{\left( {N_{{\rm{BS}}}^z - 2n + 1} \right)}^2}},\\
&{\bf{\dot u}}_{\mu _{{\rm{B2T}}}^{\rm{D}}}^H\left( {\mu _{{\rm{B2T}}}^{\rm{D}},N_{{\rm{BS}}}^y} \right){\bf{u}}\left( {\mu _{{\rm{B2T}}}^{\rm{D}},N_{{\rm{BS}}}^y} \right) = 0,\\
&{\bf{\dot u}}_{\nu _{{\rm{B2T}}}^{\rm{D}}}^H\left( {\nu _{{\rm{B2T}}}^{\rm{D}},N_{{\rm{BS}}}^z} \right){\bf{u}}\left( {\nu _{{\rm{B2T}}}^{\rm{D}},N_{{\rm{BS}}}^z} \right) = 0. 
\end{align}
Define ${{{\bf{\dot a}}}_{\mu _{{\rm{B2T}}}^{\rm{D}}}}\left( {\mu _{{\rm{B2T}}}^{\rm{D}},\nu _{{\rm{B2T}}}^{\rm{D}}} \right) = {{{\bf{\dot u}}}_{\mu _{{\rm{B2T}}}^{\rm{D}}}}\left( {\mu _{{\rm{B2T}}}^{\rm{D}},N_{{\rm{BS}}}^y} \right) \otimes {\bf{u}}\left( {\nu _{{\rm{B2T}}}^{\rm{D}},N_{{\rm{BS}}}^z} \right)$ and ${{{\bf{\dot a}}}_{\nu _{{\rm{B2T}}}^{\rm{D}}}}\left( {\mu _{{\rm{B2T}}}^{\rm{D}},\nu _{{\rm{B2T}}}^{\rm{D}}} \right) = {\bf{u}}\left( {\mu _{{\rm{B2T}}}^{\rm{D}},N_{{\rm{BS}}}^y} \right) \otimes {{{\bf{\dot u}}}_{\nu _{{\rm{B2T}}}^{\rm{D}}}}\left( {\nu _{{\rm{B2T}}}^{\rm{D}},N_{{\rm{BS}}}^z} \right)$. Accordingly, it follows that 
\begin{align}
	&{\bf{\dot a}}_{\mu _{{\rm{B2T}}}^{\rm{D}}}^H\left( {\mu _{{\rm{B2T}}}^{\rm{D}},\nu _{{\rm{B2T}}}^{\rm{D}}} \right){{{\bf{\dot a}}}_{\mu _{{\rm{B2T}}}^{\rm{D}}}}\left( {\mu _{{\rm{B2T}}}^{\rm{D}},\nu _{{\rm{B2T}}}^{\rm{D}}} \right)\qquad\qquad\qquad\qquad\qquad \notag\\
	& \qquad\qquad\qquad\qquad= \frac{{{\pi ^2}N_{{\rm{BS}}}^z}}{4}\sum\limits_{n = 1}^{N_{{\rm{BS}}}^y} {{{\left( {N_{{\rm{BS}}}^y - 2n + 1} \right)}^2}}, \\
	&{\bf{\dot a}}_{\nu _{{\rm{B2T}}}^{\rm{D}}}^H\left( {\mu _{{\rm{B2T}}}^{\rm{D}},\nu _{{\rm{B2T}}}^{\rm{D}}} \right){{{\bf{\dot a}}}_{\nu _{{\rm{B2T}}}^{\rm{D}}}}\left( {\mu _{{\rm{B2T}}}^{\rm{D}},\nu _{{\rm{B2T}}}^{\rm{D}}} \right)\notag\\
	&\qquad\qquad\qquad\qquad = \frac{{{\pi ^2}N_{{\rm{BS}}}^y}}{4}\sum\limits_{n = 1}^{N_{{\rm{BS}}}^z} {{{\left( {N_{{\rm{BS}}}^z - 2n + 1} \right)}^2}}, \\
	&{\bf{\dot a}}_{\mu _{{\rm{B2T}}}^{\rm{D}}}^H\left( {\mu _{{\rm{B2T}}}^{\rm{D}},\nu _{{\rm{B2T}}}^{\rm{D}}} \right){{{\bf{\dot a}}}_{\nu _{{\rm{B2T}}}^{\rm{D}}}}\left( {\mu _{{\rm{B2T}}}^{\rm{D}},\nu _{{\rm{B2T}}}^{\rm{D}}} \right)=0,\\
	&{\bf{\dot a}}_{\mu _{{\rm{B2T}}}^{\rm{D}}}^H\left( {\mu _{{\rm{B2T}}}^{\rm{D}},\nu _{{\rm{B2T}}}^{\rm{D}}} \right){{\bf{a}}_{}}\left( {\mu _{{\rm{B2T}}}^{\rm{D}},\nu _{{\rm{B2T}}}^{\rm{D}}} \right)=0,\\
	&{{\bf{a}}^H}\left( {\mu _{{\rm{B2T}}}^{\rm{D}},\nu _{{\rm{B2T}}}^{\rm{D}}} \right){{{\bf{\dot a}}}_{\nu _{{\rm{B2T}}}^{\rm{D}}}}\left( {\mu _{{\rm{B2T}}}^{\rm{D}},\nu _{{\rm{B2T}}}^{\rm{D}}} \right)=0,\\
	&{{\bf{a}}^H}\left( {\mu _{{\rm{B2T}}}^{\rm{D}},\nu _{{\rm{B2T}}}^{\rm{D}}} \right){{\bf{a}}_{}}\left( {\mu _{{\rm{B2T}}}^{\rm{D}},\nu _{{\rm{B2T}}}^{\rm{D}}} \right)= {N_{{\rm{BS}}}}.
\end{align}
Furthermore, we have 
\begin{align}
	{{{\bf{\dot A}}}_{\mu _{{\rm{B2T}}}^{\rm{D}}}}& = {{{\bf{\dot a}}}_{\mu _{{\rm{B2T}}}^{\rm{D}}}}\left( {\mu _{{\rm{B2T}}}^{\rm{D}},\nu _{{\rm{B2T}}}^{\rm{D}}} \right){{\bf{a}}^T}\left( {\mu _{{\rm{B2T}}}^{\rm{D}},\nu _{{\rm{B2T}}}^{\rm{D}}} \right) \notag\\
	&+{{\bf{a}}_{}}\left( {\mu _{{\rm{B2T}}}^{\rm{D}},\nu _{{\rm{B2T}}}^{\rm{D}}} \right){\bf{\dot a}}_{\mu _{{\rm{B2T}}}^{\rm{D}}}^T\left( {\mu _{{\rm{B2T}}}^{\rm{D}},\nu _{{\rm{B2T}}}^{\rm{D}}} \right),\\
	{{{\bf{\dot A}}}_{\nu _{{\rm{B2T}}}^{\rm{D}}}} &= {{{\bf{\dot a}}}_{\nu _{{\rm{B2T}}}^{\rm{D}}}}\left( {\mu _{{\rm{B2T}}}^{\rm{D}},\nu _{{\rm{B2T}}}^{\rm{D}}} \right){{\bf{a}}^T}\left( {\mu _{{\rm{B2T}}}^{\rm{D}},\nu _{{\rm{B2T}}}^{\rm{D}}} \right) \notag\\
	& + {{\bf{a}}_{}}\left( {\mu _{{\rm{B2T}}}^{\rm{D}},\nu _{{\rm{B2T}}}^{\rm{D}}} \right){\bf{\dot a}}_{\nu _{{\rm{B2T}}}^{\rm{D}}}^T\left( {\mu _{{\rm{B2T}}}^{\rm{D}},\nu _{{\rm{B2T}}}^{\rm{D}}} \right).
\end{align}
Then, we can obtain the following identities:
 \begin{align}
&\!\!\!\!{\bf{\dot A}}_{\mu _{{\rm{B2T}}}^{\rm{D}}}^H{{{\bf{\dot A}}}_{\mu _{{\rm{B2T}}}^{\rm{D}}}} \! =\! \frac{{{\pi ^2}N_{{\rm{BS}}}^z}}{4}\sum\limits_{n = 1}^{N_{{\rm{BS}}}^y} {{{\left( {N_{{\rm{BS}}}^y \!- 2n\! + 1} \right)}^2}} \Big( {{{\bf{a}}^*}\left( {\mu _{{\rm{B2T}}}^{\rm{D}},\nu _{{\rm{B2T}}}^{\rm{D}}} \right)} \notag\\ \label{appendxi2_1}
&\!\! {{{\bf{a}}^T}\left( {\mu _{{\rm{B2T}}}^{\rm{D}},\nu _{{\rm{B2T}}}^{\rm{D}}} \right) + {\bf{a}}\left( {\mu _{{\rm{B2T}}}^{\rm{D}},\nu _{{\rm{B2T}}}^{\rm{D}}} \right){{\bf{a}}^H}\left( {\mu _{{\rm{B2T}}}^{\rm{D}},\nu _{{\rm{B2T}}}^{\rm{D}}} \right)} \Big),\\
&\!\!\!\!{\bf{\dot A}}_{\nu _{{\rm{B2T}}}^{\rm{D}}}^H{{{\bf{\dot A}}}_{\nu _{{\rm{B2T}}}^{\rm{D}}}} \! =\! \frac{{{\pi ^2}N_{{\rm{BS}}}^y}}{4}\sum\limits_{n = 1}^{N_{{\rm{BS}}}^z} {{{\left( {N_{{\rm{BS}}}^z \!- 2n\! + 1} \right)}^2}} \Big( {{{\bf{a}}^*}\left( {\mu _{{\rm{B2T}}}^{\rm{D}},\nu _{{\rm{B2T}}}^{\rm{D}}} \right)} \notag\\
& \!\!{{{\bf{a}}^T}\left( {\mu _{{\rm{B2T}}}^{\rm{D}},\nu _{{\rm{B2T}}}^{\rm{D}}} \right) + {\bf{a}}\left( {\mu _{{\rm{B2T}}}^{\rm{D}},\nu _{{\rm{B2T}}}^{\rm{D}}} \right){{\bf{a}}^H}\left( {\mu _{{\rm{B2T}}}^{\rm{D}},\nu _{{\rm{B2T}}}^{\rm{D}}} \right)} \Big),\\
&\!\!{\bf{\dot A}}_{\mu _{{\rm{B2T}}}^{\rm{D}}}^H{{{\bf{\dot A}}}_{\nu _{{\rm{B2T}}}^{\rm{D}}}} \!\!= {N_{{\rm{BS}}}}{\bf{\dot a}}_{\mu _{{\rm{B2T}}}^{\rm{D}}}^*\left( {\mu _{{\rm{B2T}}}^{\rm{D}},\nu _{{\rm{B2T}}}^{\rm{D}}} \right){\bf{\dot a}}_{\nu _{{\rm{B2T}}}^{\rm{D}}}^T\left( {\mu _{{\rm{B2T}}}^{\rm{D}},\nu _{{\rm{B2T}}}^{\rm{D}}} \right),\\
&\!\!{\bf{\dot A}}_{\mu _{{\rm{B2T}}}^{\rm{D}}}^H{\bf{A}} = {N_{{\rm{BS}}}}{\bf{\dot a}}_{\mu _{{\rm{B2T}}}^{\rm{D}}}^*\left( {\mu _{{\rm{B2T}}}^{\rm{D}},\nu _{{\rm{B2T}}}^{\rm{D}}} \right){{\bf{a}}^T}\left( {\mu _{{\rm{B2T}}}^{\rm{D}},\nu _{{\rm{B2T}}}^{\rm{D}}} \right),\\
&\!\!{\bf{\dot A}}_{\nu _{{\rm{B2T}}}^{\rm{D}}}^H{\bf{A}} = {N_{{\rm{BS}}}}{\bf{\dot a}}_{\nu _{{\rm{B2T}}}^{\rm{D}}}^*\left( {\mu _{{\rm{B2T}}}^{\rm{D}},\nu _{{\rm{B2T}}}^{\rm{D}}} \right){{\bf{a}}^T}\left( {\mu _{{\rm{B2T}}}^{\rm{D}},\nu _{{\rm{B2T}}}^{\rm{D}}} \right),\\
&\!\!{\bf{A}}_{}^H{\bf{A}} = {N_{{\rm{BS}}}}{{\bf{a}}^*}\left( {\mu _{{\rm{B2T}}}^{\rm{D}},\nu _{{\rm{B2T}}}^{\rm{D}}} \right){{\bf{a}}^T}\left( {\mu _{{\rm{B2T}}}^{\rm{D}},\nu _{{\rm{B2T}}}^{\rm{D}}} \right).\label{appendxi2_6}
 \end{align}
To prove Theorem 1, we should prove two points: 1) ${{\bf{R}}_{{{\bf{w}}_{{\rm{BS}}}}}} = \frac{{{P_{{\rm{BS}}}}}}{{{N_{{\rm{BS}}}}}}{{\bf{I}}_{{N_{{\rm{BS}}}}}}$ makes FIM ${{\bf{F}}_{{\bm{\eta }}_{{\rm{BS}}}^{\rm{I}}}}$ diagonal; 2) ${{\bf{R}}_{{{\bf{w}}_{{\rm{BS}}}}}} = \frac{{{P_{{\rm{BS}}}}}}{{{N_{{\rm{BS}}}}}}{{\bf{I}}_{{N_{{\rm{BS}}}}}}$ makes each  diagonal  entry of FIM ${{\bf{F}}_{{\bm{\eta }}_{{\rm{BS}}}^{\rm{I}}}}$ maximized.
 
First, we will verify that as ${{\bf{R}}_{{{\bf{w}}_{{\rm{BS}}}}}} = \frac{{{P_{{\rm{BS}}}}}}{{{N_{{\rm{BS}}}}}}{{\bf{I}}_{{N_{{\rm{BS}}}}}}$, the FIM ${{\bf{F}}_{{\bm{\eta }}_{{\rm{BS}}}^{\rm{I}}}}$ is diagonal. Together with  \eqref{appendxi2_1}-\eqref{appendxi2_6},   \eqref{FIM_1}-\eqref{FIM_6} can be simplified to 
 \begin{align}
& f_{\mu _{{\rm{B2T}}}^{\rm{D}}\mu _{{\rm{B2T}}}^{\rm{D}}}^{\rm{I}} = \frac{{{{\left| {{\beta _{{\rm{BTB}}}}} \right|}^2}{T_1}{P_{{\rm{BS}}}}{\pi ^2}N_{{\rm{BS}}}^z}}{{{\sigma ^2}}}\sum\limits_{n = 1}^{N_{{\rm{BS}}}^y} {{{\left( {N_{{\rm{BS}}}^y - 2n + 1} \right)}^2}} ,\\
& f_{\mu _{{\rm{B2T}}}^{\rm{D}}\nu _{{\rm{B2T}}}^{\rm{D}}}^{\rm{I}} = 0,\\
& {\bf{f}}_{\mu _{{\rm{B2T}}}^{\rm{D}}{\bm{\beta }}_{{\rm{BTB}}}^T}^{\rm{I}} = \left[ {0{\kern 1pt} {\kern 1pt} {\kern 1pt} 0} \right],\\
&f_{\nu _{{\rm{B2T}}}^{\rm{D}}\nu _{{\rm{B2T}}}^{\rm{D}}}^{\rm{I}} = \frac{{{{\left| {{\beta _{{\rm{BTB}}}}} \right|}^2}{T_1}{P_{{\rm{BS}}}}{\pi ^2}N_{{\rm{BS}}}^y}}{{{\sigma ^2}}}\sum\limits_{n = 1}^{N_{{\rm{BS}}}^z} {{{\left( {N_{{\rm{BS}}}^z - 2n + 1} \right)}^2}},\\
&{\bf{F}}_{{\bm{\beta }}_{{\rm{BTB}}}^T{\bm{\beta }}_{{\rm{BTB}}}^T}^{{\rm{I,}}T} = \frac{{2{T_1}{N_{{\rm{BS}}}}{P_{{\rm{BS}}}}}}{{{\sigma ^2}}}{{\bf{I}}_2},
 \end{align}
and FIM ${{\bf{F}}_{{\bm{\eta }}_{{\rm{BS}}}^{\rm{I}}}}$ becomes  a diagonal matrix, which indicates that   ${{\bf{R}}_{{{\bf{w}}_{{\rm{BS}}}}}} = \frac{{{P_{{\rm{BS}}}}}}{{{N_{{\rm{BS}}}}}}{{\bf{I}}_{{N_{{\rm{BS}}}}}}$ is optimal  to problem \eqref{problem1}. 

Next, we will prove that ${{\bf{R}}_{{{\bf{w}}_{{\rm{BS}}}}}} = \frac{{{P_{{\rm{BS}}}}}}{{{N_{{\rm{BS}}}}}}{{\bf{I}}_{{N_{{\rm{BS}}}}}}$ makes each  diagonal  entry maximized. Considering the following optimization problem:
 \begin{subequations} \label{max_min}
	\begin{align}
&\mathop {\max }\limits_{{{\bf{R}}_{{{\bf{w}}_{{\rm{BS}}}}}}} \mathop {\min }\limits_{\bf{E}} {\rm{tr}}\left( {{\bf{RE}}} \right)\\
&{\rm{s}}{\rm{.t}}{\rm{.}}{\kern 1pt} {\kern 1pt} {\kern 1pt} {\kern 1pt} {\rm{tr}}\left( {{{\bf{R}}_{{{\bf{w}}_{{\rm{BS}}}}}}} \right) = {P_{{\rm{BS}}}},\\
&\qquad  {{\bf{R}}_{{{\bf{w}}_{{\rm{BS}}}}}} \succeq {\bf{0}}, {\bf{E}} \succeq {\bf{0}},
	\end{align}
\end{subequations} 
where ${\bf{E}} \in \left\{ {{\bf{\dot A}}_{\mu _{{\rm{B2T}}}^{\rm{D}}}^H{{{\bf{\dot A}}}_{\mu _{{\rm{B2T}}}^{\rm{D}}}},{\bf{\dot A}}_{\nu _{{\rm{B2T}}}^{\rm{D}}}^H{{{\bf{\dot A}}}_{\nu _{{\rm{B2T}}}^{\rm{D}}}},{{\bf{A}}^H}{\bf{A}}} \right\}$. Problem \eqref{max_min} can be interpreted as choosing ${{\bf{R}}_{{{\bf{w}}_{{\rm{BS}}}}}}$ such that  ${\rm{tr}}\left( {{\bf{RE}}} \right)$ (${\bf{E}}$ is unknown but a Hermitian semi-definite matrix) is maximized in  the worst case.
According to \cite{Stoica2002Maximum}, the optimal solution to problem \eqref{max_min} is ${{\bf{R}}_{{{\bf{w}}_{{\rm{BS}}}}}} = \frac{{{P_{{\rm{BS}}}}}}{{{N_{{\rm{BS}}}}}}{{\bf{I}}_{{N_{{\rm{BS}}}}}}$. This indicates that ${{\bf{R}}_{{{\bf{w}}_{{\rm{BS}}}}}} = \frac{{{P_{{\rm{BS}}}}}}{{{N_{{\rm{BS}}}}}}{{\bf{I}}_{{N_{{\rm{BS}}}}}}$ makes diagonal entries, i.e., \eqref{FIM_1}, \eqref{FIM_4}, and \eqref{FIM_6}, maximized.
Based on these two results,  the proof of  Theorem 1 is completed.

\section{Proof of Lemma 4}\label{appendix_lemma3}
Define the following notations:
\begin{align}
&{{{\bf{\dot q}}}_\mu } = \frac{{\partial {\bf{q}}}}{{\partial \mu }} = {\bf{\dot u}}\left( {\mu ,N_{\rm{r}}^y} \right) \otimes {\bf{u}}\left( {\nu ,N_{\rm{r}}^z} \right),\\
&{{{\bf{\dot q}}}_\nu } = \frac{{\partial {\bf{q}}}}{{\partial \nu }} = {\bf{u}}\left( {\mu ,N_{\rm{r}}^y} \right) \otimes {\bf{\dot u}}\left( {\nu ,N_{\rm{r}}^z} \right),
\end{align}
with ${\bf{\dot u}}\left( {\mu ,N_{\rm{r}}^y} \right) = \frac{{\partial {\bf{u}}\left( {\mu ,N_{\rm{r}}^y} \right)}}{{\partial \mu }}$ and ${\bf{\dot u}}\left( {\nu ,N_{\rm{r}}^z} \right) = \frac{{\partial {\bf{u}}\left( {\nu ,N_{\rm{r}}^y} \right)}}{{\partial \nu }}$. Thus, $\frac{{\partial {\bf{u}}_{{\rm{IRS}}}^{{\rm{II}}}}}{{\partial \mu }}$, $\frac{{\partial {\bf{u}}_{{\rm{IRS}}}^{{\rm{II}}}}}{{\partial \nu }}$, and $\frac{{\partial {\bf{u}}_{{\rm{IRS}}}^{{\rm{II}}}}}{{\partial {\bm{\alpha }}}}$ can be  calculated as 
\begin{align}
&\frac{{\partial {\bf{u}}_{{\rm{IRS}}}^{{\rm{II}}}}}{{\partial \mu }} = \alpha \left[ {{\bf{w}}_{{\rm{IRS}}}^T\left[ {{T_1} + 1} \right]\left( {{{{\bf{\dot q}}}_\mu }{{\bf{q}}^T} + {\bf{q\dot q}}_\mu ^T} \right){{\bf{w}}_{{\rm{IRS}}}}\left[ {{T_1} + 1} \right]} \right.\notag\\
&{\left. {, \ldots ,{\kern 1pt} {\bf{w}}_{{\rm{IRS}}}^T\left[ {{T_1} + {T_2}} \right]\left( {{{{\bf{\dot q}}}_\mu }{{\bf{q}}^T} + {\bf{q\dot q}}_\mu ^T} \right){{\bf{w}}_{{\rm{IRS}}}}\left[ {{T_1} + {T_2}} \right]} \right]^T},\label{fisherelement_derivative_1}\\ 
&\frac{{\partial {\bf{u}}_{{\rm{IRS}}}^{{\rm{II}}}}}{{\partial \nu }} = \alpha \left[ {{\bf{w}}_{{\rm{IRS}}}^T\left[ {{T_1} + 1} \right]\left( {{{{\bf{\dot q}}}_\nu }{{\bf{q}}^T} + {\bf{q\dot q}}_\nu ^T} \right){{\bf{w}}_{{\rm{IRS}}}}\left[ {{T_1} + 1} \right]} \right.\notag\\
&{\left. {, \ldots ,{\kern 1pt} {\bf{w}}_{{\rm{IRS}}}^T\left[ {{T_1} + {T_2}} \right]\left( {{{{\bf{\dot q}}}_\nu }{{\bf{q}}^T} + {\bf{q\dot q}}_\nu ^T} \right){{\bf{w}}_{{\rm{IRS}}}}\left[ {{T_1} + {T_2}} \right]} \right]^T},\\
&\frac{{\partial {\bf{u}}_{{\rm{IRS}}}^{{\rm{II}}}}}{{\partial {\bm{\alpha }}^T}} = \left[ {1{\kern 1pt} {\kern 1pt} {\kern 1pt} {\kern 1pt} j} \right] \otimes \Big[ {{\bf{w}}_{{\rm{IRS}}}^T\left[ {{T_1} + 1} \right]{\bf{q}}{{\bf{q}}^T}{{\bf{w}}_{{\rm{IRS}}}}\left[ {{T_1} + 1} \right]} \notag\\
&\qquad \quad { {, \ldots ,{\kern 1pt} {\bf{w}}_{{\rm{IRS}}}^T\left[ {{T_1} + {T_2}} \right]{\bf{q}}{{\bf{q}}^T}{{\bf{w}}_{{\rm{IRS}}}}\left[ {{T_1} + {T_2}} \right]} \Big]^T}.\label{fisherelement_derivative_3}
\end{align}
Based on \eqref{fisherelement_derivative_1}-\eqref{fisherelement_derivative_3}, ${f_{\mu \mu }^{{\rm{II}}}}$, ${f_{\mu \nu }^{{\rm{II}}}}$, ${{\bf{f}}_{\mu {{\bm{\alpha }}^T}}^{{\rm{II}}}}$, ${f_{\nu \nu }^{{\rm{II}}}}$, ${{\bf{f}}_{\nu {{\bm{\alpha }}^T}}^{{\rm{II}}}}$, and  ${{\bf{F}}_{{{\bm{\alpha }}^T}{{\bm{\alpha }}^T}}^{{\rm{II}}}}$ can be   calculated as
\begin{align}
&f_{\mu \mu }^{{\rm{II}}} = \frac{2}{{{\sigma ^2}{N_{{\rm{BS}}}}}}{\mathop{\rm Re}\nolimits} \left\{ {\frac{{\partial {\bf{u}}_{{\rm{IRS}}}^{{\rm{II,}}H}}}{{\partial \mu }}\frac{{\partial {\bf{u}}_{{\rm{IRS}}}^{{\rm{II}}}}}{{\partial \mu }}} \right\}\quad\qquad\qquad\qquad\qquad\qquad\notag\\
&  = \frac{{2{{\left| \alpha  \right|}^2}}}{{{\sigma ^2}{N_{{\rm{BS}}}}}}\sum\limits_{i = 1}^{{T_2}} {{{\left| {{\bf{w}}_{{\rm{IRS}}}^T\left[ {{T_1} + i} \right]{{\bf{Q}}_\mu }{{\bf{w}}_{{\rm{IRS}}}}\left[ {{T_1} + i} \right]} \right|}^2}}, \label{fisher_matrix_stageII_1}\\
&f_{\mu \nu }^{{\rm{II}}} = \frac{2}{{{\sigma ^2}{N_{{\rm{BS}}}}}}{\mathop{\rm Re}\nolimits} \left\{ {\frac{{\partial {\bf{u}}_{{\rm{IRS}}}^{{\rm{II,}}H}}}{{\partial \mu }}\frac{{\partial {\bf{u}}_{{\rm{IRS}}}^{{\rm{II}}}}}{{\partial \nu }}} \right\}\quad\qquad\qquad\qquad\qquad\qquad\notag\\
&  = \frac{{2{{\left| \alpha  \right|}^2}}}{{{\sigma ^2}{N_{{\rm{BS}}}}}}{\mathop{\rm Re}\nolimits} \left\{ {\sum\limits_{i = 1}^{{T_2}} {{\bf{w}}_{{\rm{IRS}}}^H\left[ {{T_1} + i} \right]{\bf{Q}}_\mu ^H{\bf{w}}_{{\rm{IRS}}}^*\left[ {{T_1} + i} \right]} } \right.\notag\\
&\quad\times {{\bf{w}}_{{\rm{IRS}}}^T\left[ {{T_1} + i} \right]{{\bf{Q}}_\nu }{{\bf{w}}_{{\rm{IRS}}}}\left[ {{T_1} + i} \right]} \bigg\},\\
&{\bf{f}}_{\mu {{\bm{\alpha }}^T}}^{{\rm{II}}} = \frac{2}{{{\sigma ^2}{N_{{\rm{BS}}}}}}{\mathop{\rm Re}\nolimits} \left\{ {\frac{{\partial {\bf{u}}_{{\rm{IRS}}}^{{\rm{II,}}H}}}{{\partial \mu }}\frac{{\partial {\bf{u}}_{{\rm{IRS}}}^{{\rm{II}}}}}{{\partial {{\bm{\alpha }}^T}}}} \right\}\notag\\
& = \frac{2}{{{\sigma ^2}{N_{{\rm{BS}}}}}}{\mathop{\rm Re}\nolimits} \left\{ {\left[ {1{\kern 1pt} {\kern 1pt} {\kern 1pt} {\kern 1pt} {\kern 1pt} j} \right]{\alpha ^*}\sum\limits_{i = 1}^{{T_2}} {{\bf{w}}_{{\rm{IRS}}}^H\left[ {{T_1} + i} \right]{\bf{Q}}_\mu ^H{\bf{w}}_{{\rm{IRS}}}^*\left[ {{T_1} + i} \right]} } \right.\notag\\
&\quad \times {{\bf{w}}_{{\rm{IRS}}}^T\left[ {{T_1} + i} \right]{{\bf{Q}} }{{\bf{w}}_{{\rm{IRS}}}}\left[ {{T_1} + i} \right]} \bigg\},\\
&f_{\nu \nu }^{{\rm{II}}} = \frac{2}{{{\sigma ^2}{N_{{\rm{BS}}}}}}{\mathop{\rm Re}\nolimits} \left\{ {\frac{{\partial {\bf{u}}_{{\rm{IRS}}}^{{\rm{II,}}H}}}{{\partial \nu }}\frac{{\partial {\bf{u}}_{{\rm{IRS}}}^{{\rm{II}}}}}{{\partial \nu }}} \right\}\quad \qquad\qquad\qquad\qquad\qquad\notag\\
& = \frac{{2{{\left| \alpha  \right|}^2}}}{{{\sigma ^2}{N_{{\rm{BS}}}}}}\sum\limits_{i = 1}^{{T_2}} {{{\left| {{\bf{w}}_{{\rm{IRS}}}^T\left[ {{T_1} + i} \right]{{\bf{Q}}_\nu }{{\bf{w}}_{{\rm{IRS}}}}\left[ {{T_1} + i} \right]} \right|}^2}},\\
&{\bf{f}}_{\nu {{\bm{\alpha }}^T}}^{{\rm{II}}} = \frac{2}{{{\sigma ^2}{N_{{\rm{BS}}}}}}{\mathop{\rm Re}\nolimits} \left\{ {\frac{{\partial {\bf{u}}_{{\rm{IRS}}}^{{\rm{II,}}H}}}{{\partial \nu }}\frac{{\partial {\bf{u}}_{{\rm{IRS}}}^{{\rm{II}}}}}{{\partial {{\bm{\alpha }}^T}}}} \right\}\notag\\
& = \frac{2}{{{\sigma ^2}{N_{{\rm{BS}}}}}}{\mathop{\rm Re}\nolimits} \left\{ {\left[ {1{\kern 1pt} {\kern 1pt} {\kern 1pt} {\kern 1pt} {\kern 1pt} j} \right]{\alpha ^*}\sum\limits_{i = 1}^{{T_2}} {{\bf{w}}_{{\rm{IRS}}}^H\left[ {{T_1} + i} \right]{\bf{Q}}_\nu ^H{\bf{w}}_{{\rm{IRS}}}^*\left[ {{T_1} + i} \right]} } \right.\notag\\
&\quad \times {{\bf{w}}_{{\rm{IRS}}}^T\left[ {{T_1} + i} \right]{{\bf{Q}} }{{\bf{w}}_{{\rm{IRS}}}}\left[ {{T_1} + i} \right]} \bigg\},\\
&{\bf{f}}_{{{\bm{\alpha }}^T}{{\bm{\alpha }}^T}}^{{\rm{II}}} = \frac{2}{{{\sigma ^2}{N_{{\rm{BS}}}}}}{\mathop{\rm Re}\nolimits} \left\{ {\frac{{\partial {\bf{u}}_{{\rm{IRS}}}^{{\rm{II,}}H}}}{{\partial {{\bm{\alpha }}^T}}}\frac{{\partial {\bf{u}}_{{\rm{IRS}}}^{{\rm{II}}}}}{{\partial {{\bm{\alpha }}^T}}}} \right\}\qquad\qquad\qquad\qquad\qquad\notag\\
& = \frac{2}{{{\sigma ^2}{N_{{\rm{BS}}}}}}\sum\limits_{i = 1}^{{T_2}} {{{\left| {{\bf{w}}_{{\rm{IRS}}}^T\left[ {{T_1} + 1} \right]{\bf{Q}}{{\bf{w}}_{{\rm{IRS}}}}\left[ {{T_1} + 1} \right]} \right|}^2}} {{\bf{I}}_2},  \label{fisher_matrix_stageII_6}
\end{align}
where ${\bf{Q}} = {\bf{q}}{{\bf{q}}^T}$, ${{\bf{Q}}_\mu } = {{{\bf{\dot q}}}_\mu }{{\bf{q}}^T} + {\bf{q\dot q}}_\mu ^T$, and ${{\bf{Q}}_\nu } = {{{\bf{\dot q}}}_\nu }{{\bf{q}}^T} + {\bf{q\dot q}}_\nu ^T$.

\section{Proof of Theorem 2} \label{appendix_Theorem2}
To prove ${\rm{CRB}}\left( {{\bm \eta} _{{\rm{IRS}}}^{{\rm{II}}}} \right)=+\infty$, we can equivalently prove by verifying $\left| {{{\bf{F}}_{\eta _{{\rm{IRS}}}^{{\rm{II}}}}}} \right| = 0$. Let  ${{\bf{w}}_{{\rm{IRS}}}} = {{\bf{w}}_{{\rm{IRS}}}}\left[ {{T_1} + 1} \right] = , \ldots ,{{\bf{w}}_{{\rm{IRS}}}}\left[ {{T_1} + {T_2}} \right]$. Then, \eqref{fisher_matrix_stageII_1}-\eqref{fisher_matrix_stageII_6} can be reduced to 
\begin{align}
&f_{\mu \mu }^{{\rm{II}}} = \frac{{2{{\left| \alpha  \right|}^2}{T_2}}}{{{\sigma ^2}{N_{{\rm{BS}}}}}}{\left| {{\bf{w}}_{{\rm{IRS}}}^T{{\bf{Q}}_\mu }{{\bf{w}}_{{\rm{IRS}}}}} \right|^2},\label{fisher_matrix_stageII_new_1}\\
&f_{\mu \nu }^{{\rm{II}}} = \frac{{2{{\left| \alpha  \right|}^2}{T_2}}}{{{\sigma ^2}{N_{{\rm{BS}}}}}}{\mathop{\rm Re}\nolimits} \left\{ {{\bf{w}}_{{\rm{IRS}}}^H{\bf{Q}}_\mu ^H{\bf{w}}_{{\rm{IRS}}}^*{\bf{w}}_{{\rm{IRS}}}^T{{\bf{Q}}_\nu }{{\bf{w}}_{{\rm{IRS}}}}} \right\},\\
&{\bf{f}}_{\mu {{\bm{\alpha }}^T}}^{{\rm{II}}} = \frac{{2{T_2}}}{{{\sigma ^2}{N_{{\rm{BS}}}}}}{\mathop{\rm Re}\nolimits} \left\{ {\left[ {1{\kern 1pt} {\kern 1pt} {\kern 1pt} {\kern 1pt} {\kern 1pt} j} \right]{\alpha ^*}{\bf{w}}_{{\rm{IRS}}}^H{\bf{Q}}_\mu ^H{\bf{w}}_{{\rm{IRS}}}^*{\bf{w}}_{{\rm{IRS}}}^T{\bf{Q}}{{\bf{w}}_{{\rm{IRS}}}}} \right\},\\
&f_{\nu \nu }^{{\rm{II}}} = \frac{{2{{\left| \alpha  \right|}^2}{T_2}}}{{{\sigma ^2}{N_{{\rm{BS}}}}}}{\left| {{\bf{w}}_{{\rm{IRS}}}^T{{\bf{Q}}_\nu }{{\bf{w}}_{{\rm{IRS}}}}} \right|^2},
\end{align}
\begin{align}
&{\bf{f}}_{\nu {{\bm{\alpha }}^T}}^{{\rm{II}}} = \frac{{2{T_2}}}{{{\sigma ^2}{N_{{\rm{BS}}}}}}{\mathop{\rm Re}\nolimits} \left\{ {\left[ {1{\kern 1pt} {\kern 1pt} {\kern 1pt} {\kern 1pt} {\kern 1pt} j} \right]{\alpha ^*}{\bf{w}}_{{\rm{IRS}}}^H{\bf{Q}}_\nu ^H{\bf{w}}_{{\rm{IRS}}}^*{\bf{w}}_{{\rm{IRS}}}^T{\bf{Q}}{{\bf{w}}_{{\rm{IRS}}}}} \right\},\\
&{\bf{f}}_{{{\bm{\alpha }}^T}{{\bm{\alpha }}^T}}^{{\rm{II}}} = \frac{{2{T_2}}}{{{\sigma ^2}{N_{{\rm{BS}}}}}}{\left| {{\bf{w}}_{{\rm{IRS}}}^T{\bf{Q}}{{\bf{w}}_{{\rm{IRS}}}}} \right|^2}{{\bf{I}}_2}.\label{fisher_matrix_stageII_new_6}
\end{align}
We rewrite ${{{\bf{F}}_{\eta _{{\rm{IRS}}}^{{\rm{II}}}}}}$ in a block  matrices form given by 
\begin{align}
{{\bf{F}}_{{\bm \eta} _{{\rm{IRS}}}^{{\rm{II}}}}} = \left[ {\begin{array}{*{20}{c}}
		{\bf{F}}_{11}&{\bf{F}}_{12}\\
		{\bf{F}}_{21}&{\bf{F}}_{22}
\end{array}} \right],
\end{align}
where ${\bf{F}}_{11} = \left[ {\begin{array}{*{20}{c}}
		{f_{\mu \mu }^{{\rm{II}}}}&{f_{\mu \nu }^{{\rm{II}}}}\\
		{f_{\mu \nu }^{{\rm{II}},T}}&{f_{\nu \nu }^{{\rm{II}}}}
\end{array}} \right]$, ${\bf{F}}_{12} = \left[ {\begin{array}{*{20}{c}}
{{\bf{f}}_{\mu {{\bm \alpha} ^T}}^{{\rm{II}}}}\\
{{\bf{f}}_{\nu {{\bm \alpha} ^T}}^{{\rm{II}}}}
\end{array}} \right]$, ${\bf{F}}_{21} = {{\bf{F}}_{12}^T}$, and ${\bf{F}}_{22} = {\bf{F}}_{{{\bm \alpha} ^T}{{\bm \alpha} ^T}}^{{\rm{II}}}$. It can be readily checked that ${\bf F}_{21}$  and ${\bf F}_{22}$ commute, i.e., ${\bf F}_{21}{\bf F}_{22}={\bf F}_{22}{\bf F}_{21}$. As a result,  $\left| {{{\bf{F}}_{\eta _{{\rm{IRS}}}^{{\rm{II}}}}}} \right| = \left| {\bf F}_{11}{\bf F}_{22}-{\bf F}_{12}{\bf F}_{21} \right|$.
Substituting \eqref{fisher_matrix_stageII_new_1}-\eqref{fisher_matrix_stageII_new_6} into ${\bf F}_{11}{\bf F}_{22}$ yields   \eqref{Fisher_determinant_1} at the top of  next page. 
\newcounter{mytempeqncnt0}
\begin{figure*}
	\normalsize
	\setcounter{mytempeqncnt0}{\value{equation}}
	\begin{align}
{\bf F}_{11}{\bf F}_{22} = \frac{{4{{\left| \alpha  \right|}^2}T_2^2}}{{{\sigma ^4}N_{{\rm{BS}}}^2}}{\left| {{\bf{w}}_{{\rm{IRS}}}^T{\bf{Q}}{{\bf{w}}_{{\rm{IRS}}}}} \right|^2}\left[ {\begin{array}{*{20}{c}}
		{{{\left| {{\bf{w}}_{{\rm{IRS}}}^T{{\bf{Q}}_\mu }{{\bf{w}}_{{\rm{IRS}}}}} \right|}^2}}&{{\mathop{\rm Re}\nolimits} \left\{ {{\bf{w}}_{{\rm{IRS}}}^H{\bf{Q}}_\mu ^H{\bf{w}}_{{\rm{IRS}}}^*{\bf{w}}_{{\rm{IRS}}}^T{{\bf{Q}}_\nu }{{\bf{w}}_{{\rm{IRS}}}}} \right\}}\\
		{{\mathop{\rm Re}\nolimits} \left\{ {{\bf{w}}_{{\rm{IRS}}}^H{\bf{Q}}_\mu ^H{\bf{w}}_{{\rm{IRS}}}^*{\bf{w}}_{{\rm{IRS}}}^T{{\bf{Q}}_\nu }{{\bf{w}}_{{\rm{IRS}}}}} \right\}}&{{{\left| {{\bf{w}}_{{\rm{IRS}}}^T{{\bf{Q}}_\nu }{{\bf{w}}_{{\rm{IRS}}}}} \right|}^2}}
\end{array}} \right]. \label{Fisher_determinant_1}
	\end{align}
	\hrulefill 
	\vspace*{4pt} 
\end{figure*}
Then, substituting \eqref{fisher_matrix_stageII_new_1}-\eqref{fisher_matrix_stageII_new_6} into ${\bf F}_{12}{\bf F}_{21}$, we can obtain the same results as in \eqref{Fisher_determinant_1}, i.e, ${\bf F}_{11}{\bf F}_{22}={\bf F}_{12}{\bf F}_{21}$,  which indicates that $\left| {{{\bf{F}}_{\eta _{{\rm{IRS}}}^{{\rm{II}}}}}} \right| = 0$. This completes the proof of Theorem 2.

\bibliographystyle{IEEEtran}
\bibliography{ExploitingIRSforsensing}
\end{document}